# Two-dimensional Superconductors with Atomic-scale Thicknesses

Takashi Uchihashi

International Center for Materials Nanoarchitectonics (MANA), National Institute for Materials Science, 1-1, Namiki, Tsukuba, Ibaraki 305-0044, Japan

**Abstract**

Recent progress in two-dimensional superconductors with atomic-scale thicknesses is reviewed mainly from the experimental point of view. The superconducting systems treated here involve a variety of materials and forms: elemental-metal ultrathin films and atomic layers on semiconductor surfaces; interfaces and superlattices of heterostructures made of cuprates, perovskite oxides, and rare-earth metal heavy-fermion compounds; interfaces of electric-double-layer transistors; graphene and atomic sheets of transition-metal dichalcogenide; iron selenide and organic conductors on oxide and metal surfaces, respectively. Unique phenomena arising from the ultimate two-dimensionality of the system and the physics behind them are discussed.

**Contents**

**List of acronyms**

| | |
|---|---|
| 1D, 2D, 3D | one-, two-, three- dimension (dimensional) |
| KTB | Kosterlitz-Thouless-Berezinskii |
| BCS | Bardeen-Cooper-Schrieffer |
| S-I | superconductor-insulator |
| MBE | molecular beam epitaxy |
| PLD | pulsed laser deposition |
| UHV | ultrahigh vacuum |
| LT | low temperature |
| STM | scanning tunneling microscopy (microscope) |
| STS | scanning tunneling spectroscopy |
| ML | monolayer |
| FET | field-effect transistors |
| EDL | electric double layer |
| UC | unit cell |
| GL | Ginzburg-Landau |
| AL | Aslamazov-Larkin |
| MT | Maki-Thompson |
| LEED | low energy electron diffraction |
| QWS | quantum well states |
| SQUID | superconducting quantum interference device |
| ARPES | Angle-resolved photoemission spectroscopy |
| FFLO | Fulde-Ferrell-Larkin-Ovchinnikov |
| SOI | spin-orbit interaction |
| FET | field-effect transistor |
| RHEED | reflection high energy electron diffraction |
| CDW | charge density wave |
| SDW | spin density wave |

# 1. Introduction

Superconductivity arises from the Cooper pair formation of a huge number of conduction electrons in a metal at sufficiently low temperatures. Since it is a representative order-disorder phase transition, the dimensionality of the system, i.e., one, two, or three dimension (1D, 2D, 3D), can have a crucial influence on its characteristics [1-3]. Generally, the lower the dimension, the more difficult for the phase transition to take place, because the interaction between microscopic constituents of the system (in this case, electrons) becomes spatially limited and a less number of partners are available for the interaction with a particular constituent. This means that, even in an ordered phase below the transition temperature ($T_c$), each subset of the system has a tendency to behave more independently and the order parameter suffers from larger spatial and temporal fluctuations. In the extreme case, the coherence throughout the system is completely lost and the phase transition itself is destroyed. Now let us ask a simple question: Does superconductivity survive in a 2D system, especially when one of the material dimensions is reduced to a truly atomic-scale size? If this is the case, what are the unique characteristics of 2D superconductivity and what kind of new phenomena are expected to occur? These are obviously relevant to the modern state-of-the-art nanotechnology and will be crucial issues when the present superconducting devices are shrunk towards the atomic-scale limit in the future [4-6].

There is a long history regarding this problem. First of all, we should note that, for a 2D system, the famous Mermin-Wager theory prohibits the superconducting phase transition that accompanies a symmetry breaking and a long-range correlation of the order parameter [7, 8]. This does not mean, however, that 2D superconductivity is unrealistic. The Kosterlitz-Thouless-Berezinskii (KTB) transition, which is compatible with the Mermin-Wager theory, can occur in a 2D system and allows the establishment of a *quasi*-long-range correlation of the order parameter [9-11]. In this case, the zero resistance state is retained for infinitesimally small external perturbation and the Meissner effect can also be well-defined [12, 13]. Even without the KTB transition, Cooper pairs can condense at the mean-field level due to the Bardeen-Cooper-Schrieffer (BCS) mechanism (i.e., $T_{c0} > T_{KTB}$, where $T_{c0}$ is the Cooper-pair condensation temperature and $T_{KTB}$ is the KTB transition temperature). In a practical sense, the system may be considered superconducting if the correlation of the order parameter is sufficiently developed at low temperatures. Nevertheless, 2D superconductivity is on the verge of transition to a metallic or an insulating state and thus could be fragile. Indeed, introduction of disorder into a 2D superconductor can readily induce a superconductor-insulator (S-I) transition [14], in clear contrast to the 3D counterpart where superconductivity is robust against disorder [15]. Extensive experimental and theoretical efforts have clarified that the S-I transition always occurs when disorder is introduced to such a level that the sheet resistance (2D resistivity) of the sample is on the order of quantum resistance of Coopers, $R_Q$ ($\equiv h/4e^2$ = 6.45 k$\Omega$) [14]. Usually, superconductivity is lost when the thickness of a metal film approaches 1-2 nm and disorder becomes significant [16]. Coming back to the earlier question, the answer should be "yes" in principle, but realization of a 2D superconductor in the atomic-scale limit has been a technical challenge for a long time. This is because, in this limit, the whole system consists entirely of surfaces and interfaces, which are vulnerable to structural and chemical disorder in general. For this purpose, therefore, it is demanded that samples with highly ordered and controlled surfaces and interfaces should be fabricated and their superconducting properties probed through advanced techniques.

During this decade, the studies on ultrathin 2D superconductors have seen remarkable progress beyond the traditional experimental framework in various fields, and the existence of 2D superconductors with truly atomic-scale thicknesses has also been established. This was mostly driven by rapid advancements on nanotechnology in recent years, including molecular beam epitaxy (MBE), pulsed laser deposition (PLD), *in-situ* ultrahigh vacuum (UHV)-low temperature (LT) measurement, scanning tunneling microscopy/spectroscopy (STM/STS), etc. These technologies now allow various types of superconducting materials to be fabricated with the atomic-scale precision and to be characterized in unprecedented details, which will be the subject of the present review article. For example, ultrathin elemental metal films grown on silicon surfaces in a layer-by-layer fashion were found to exhibit robust superconductivity down the monolayer (ML) thickness regime [17, 18]. This system also exhibits $T_c$ oscillation as a function of film thickness due to the electron quantum confinement effects [19]. $La_{2-x}Sr_xCuO_4$, a representative cuprate high-$T_c$ superconductor, also showed superconductivity at one-unit-cell (1-UC) thickness and was electrically tuned to an insulating state, revealing a S-I transition driven by quantum phase fluctuations [20]. Furthermore, the interface between $LaAlO_3$ and $SrTiO_3$, both of them being perovskite oxide insulators, was found to exhibit 2D superconductivity that can coexist or compete with ferromagnetism [21]. Recent technological breakthrough of field-effect transistors (FET) using an electric-double-layer (EDL) gate have enabled carrier doping with an unprecedentedly high level ($n_{2D}$ ~$10^{14}$ cm$^{-2}$) at the subsurface region [22]. This led to the successful realization of field-induced 2D superconductivity in various insulating materials including $SrTiO_3$ and $ZrNCl$ [23, 24]. Another great invention concerning 2D materials was the exfoliation of atomic sheets from a piece of a layered material, as was first demonstrated for graphene [25]. Search for superconductivity in atomic sheets such as graphene and transition-metal dichalcogenide monolayers is now under rapid progress [26, 27]. One of the most surprising findings in 2D superconductors is the strong enhancement of $T_c$ in 1-UC-thick FeSe layers that are epitaxially grown on $SrTiO_3$ substrates; $T_c$ was found to increase remarkably up to 40 ~ 100 K from 8 K of a bulk FeSe crystal by reaching the atomic-scale limit [28, 29]. This phenomenon suggests a possible generic scheme for realizing high-$T_c$ superconductors based on atomic-scale film thinning and a strong interaction with an appropriate substrate. Finally, investigations into 2D superconductors consisting of organic conductors and of rare-earth based heavy-fermion compounds are also under progress, although the number of available publications have been limited so far [30, 31].

The study on 2D superconductors has evolved into one of the most active fields of superconductivity, which is highly relevant for material science, device physics, and instrumental technologies as well as for fundamental physics. In the present article, the recent progress in 2D superconductors will be reviewed. The main interest here is to overview the topics in various research fields in a comparative way and is to find commonalities and differences among them. This will help us grasp what is unique in 2D superconductors in general and to understand how new physics and novel phenomena emerge as a consequence of the atomic-scale thicknesses of the materials. It is my great hope that this holistic and comprehensive approach will stimulate researches of individual fields, promote interactions among them, and help create new interdisciplinary realm of research. While the topics treated in this paper share the atomic-scale two-dimensionality, they are so extensive and diverse in terms of material that it is impossible to cover all literatures in each field. Nevertheless, I believe that even the selected set of work introduced here allow us to achieve the present goal to a considerable degree. The paper is organized as follows. Section 2 briefly reviews the 2D superconductors made of granular and amorphous metal thin films, which have been studied since 1970s mainly from the viewpoint of the KTB and S-I transitions. This will serve as a basis for understanding the progress in more modern 2D materials. Section 3 constitutes the main body of the paper and describes the recent topics of 2D superconductors in individual fields. These 2D superconductors features high crystallinity and well-defined, sharp interfaces in general. Figure 1 shows a schematic chart where materials dealt here are roughly categorized in terms of two axes: the horizontal axis reflecting the material and electronic complexity and the vertical axis the location of 2D superconducting layer, i.e., top-exposed surface or buried interface. The former may also be labeled with the strength of electron correlations, while the latter with the required experimental technique, i.e., UHV environment or device configuration. Note that the positioning of each field within the chart is only qualitative and subjective. Readers are also referred to Table 1 where representative 2D superconductors treated here are listed along with their categories and important attributes. The paper is summarized in Sec. 4.

**2. Granular and amorphous 2D superconductors**

The Ginzburg-Landau (GL) theory of superconductivity allows the definition of the GL coherence length $\xi_{GL}$, which is the characteristic length scale for spatial variation of the amplitude of the order parameter $\Psi$. A superconducting film with a thickness $d << \xi_{GL}$ is considered to be 2D because $\Psi$ becomes uniform in the out-of-plane direction to avoid a cost in free energy [32]. For conventional (BCS-type) superconductors, $\xi_{GL}$ at zero temperature, $\xi_{GL}(0)$, is in the order of a few hundred nanometers. Therefore, the thickness of a 2D superconductor may still be much larger than the atomic-scale thickness. It should also be noted that $\xi_{GL}$ of an unconventional superconductor with a high $T_c$ can be as small as a few nanometer, which is often in the order of the 1-UC thickness because of large unit-cell sizes in these materials.

Straightforward way of preparing a superconducting thin film is the vacuum evaporation of an elemental metal such as Pb and Al on an insulating substrate. This can be followed by deposition of a capping layer (*e.g.* SiO$_x$) for the purpose of avoiding post-oxidation and contamination. If glass or surface-oxidized silicon is used as a substrate, the metal overlayer often grows in the form of a granular film when the thickness is reduced because of a poor wettability of metal on these substrates. This naturally introduces disorder into the metal film, the degree of which can be controlled by the film thickness and is represented typically by the sheet resistance $R_{\text{sheet}}$ [14]. The degree of disorder (and hence $R_{\text{sheet}}$) can also be controlled by intentional surface oxidation of individual metal granules and by repeating this process several times [33]. Thus prepared granular superconducting film may be modeled as a Josephson-junction coupled island network in the case of strong disorder, where the junctions are of either tunneling or weak-link type. Within individual metal islands, relatively high crystallinity can be obtained. More uniform thin films can be prepared by quench condensation of an evaporated material on a liquid-He cooled substrate, which strongly suppresses diffusion of adsorbed atoms and/or molecules and formation of large clusters [16]. Sputter deposition of certain alloys can also be used for this purpose. To enhance the uniformity of the film, the substrate is often precoated with a buffer layer. The resulting films have amorphous-like microscopic structures with atomic-scale disorder and defects, and are believed to be spatially homogeneous at larger scales [34]. For the quench-condensed films, electron transport measurements should be performed *in-situ* at the sample-preparation cold stage to avoid morphological changes of the film. This experimental setup is advantageous in terms of precise tuning of the film thickness, which in this case is the dominant parameter representing disorder. A disadvantage is that it is difficult to perform detailed structural analysis of the film. Regarding the superconducting material, elemental metals such as Pb and Bi [34] as well as metal composites and alloys like In$_2$O$_3$ and MoGe have been used [35, 36]. Note that thin films of amorphous Bi become superconducting at $T \sim 6$ K although bulk crystals of Bi do not.

The effect of the low dimensionality manifests itself as an enhancement of fluctuations as mentioned earlier. This can be seen first of all as a precursor of the superconducting transition. Namely, a 2D superconductor exhibits a sizable decrease in conductivity even above $T_c$ due to the temporal formation of Cooper pairs and the inertia of Cooper pairs after decaying into quasiparticles [37, 38]. Within the microscopic theory of superconductivity, the former is expressed by the Aslamazov-Larkin (AL) term and the latter the Maki-Thompson (MK) term. The contribution of the AL term to 2D conductivity, which is added to the normal conductivity, is given by

$$\Delta\sigma_{2D} = \frac{1}{R_0}\frac{T}{T - T_c}, \qquad (1)$$

where $R_0 = 16\hbar/e^2 = 65.8$ k$\Omega$ is a universal constant. The contribution of the MK term is expressed by a similar form that includes material dependent parameters. These effects should appear commonly in all 2D superconductors.

In a 2D system, conduction electrons can be readily localized due to the quantum interference effect in the presence of disorder, which is known as the Anderson localization [39]. Since the resulting insulating state is incompatible with a superconducting state, a superconductor-insulator (S-I) transition is expected to take place as the degree of disorder in the system is increased. In this viewpoint, the transition is driven by the suppression of amplitude of the superconducting order parameter $\Psi$. Alternatively, superconductivity can also be destroyed when the phase of $\Psi$ fluctuates strongly and its coherence is lost. This leads to a quantum phase transition at zero temperature [14, 33, 40]. A number of experiments have been performed in this respect since late 1970s, a representative result of which is shown in Fig. 2 [34]. Here, the sheet resistance $R_{sheet}$ of quench-condensed amorphous Bi films was measured as a function of temperature $T$ while increasing the film thickness. The data clearly show a S-I transition around $R_{sheet} \cong R_Q \equiv h/4e^2$ (= 6.45 k$\Omega$). The presence of this universal constant suggests that the transition is actually a quantum phase transition and can be described by the so-called dirty-boson model at least near the critical point [41, 42]. In this model, Cooper pairs are treated as composite bosons and the possibility of their breaking into fermionic quasiparticles is neglected. Cooper pairs (bosons) are localized due to the strong disorder in the system, but their very existence is assumed even in the resultant insulating state. In addition to Cooper pairs, superconducting vortices are treated as another kind of competing boson in this model. Because of the duality between the two bosons, i.e., Cooper pairs and vortices, the universal constant of $R_Q \equiv h/4e^2$ is deduced for the critical sheet resistance $R_c$. A number of experiments on disorder-induced S-I transitions in different systems revealed that $R_c$ was close to $R_Q$, but the deviation from this value was often noticed [43]. The failure of the dirty-boson picture may be due to extrinsic effects and/or to oversimplification in the theoretical modeling. Otherwise, it suggests that Cooper pairs are at least partially broken at the critical point as a result of the Anderson localization [44]. How satisfactorily this model applies to a real experiment seems to be dependent on the details of the system.

Apart from the particular model, the possibility of a quantum phase transition has also been tested using the finite-size scaling analysis [40, 41]. Here the sheet resistance $R_{sheet}$ of the sample has the following form:

$$R_{\text{sheet}}(x, T) = R_c F(|x - x_c| T^{-1/z\nu}) \ . \tag{2}$$

Here, $x$ is a control parameter for the phase transition (e.g., disorder and film thickness) and $R_c$ is the critical resistance that separates the superconducting and the insulating phases, corresponding to $x = x_c$. $F(u)$ is a universal function of $u$ that goes to unity for $u \to 0$. The parameters $z$ and $\nu$ characterize the critical behaviors near the quantum phase transition, which gives crucial information on the universality class of the system. The experiment cited above concluded that this scaling analysis worked fine, with the extracted parameters of $z\nu$ = 2.4 and 1.2 for the insulating and superconducting sides of the transition, respectively [45]. The success of this analysis was found in many analogous experiments. In addition to disorder, magnetic field induces an S-I transition with $R_c \sim R_Q$, for which the scaling analysis was also successfully applied [35, 36]. All these results evidence the existence of quantum phase transitions in these systems. With technological advancements, the study on the S-I transition has seen new developments. For example, carrier density of amorphous Bi layer grown on a thin $SrTiO_3$ substrate was continuously tuned by a gate voltage to investigate this phenomenon quantitatively [46]. High-$T_c$ cuprate superconductors have also become the target of this study. Remarkably, a S-I transition with $R_c = R_Q$ and a nearly perfect scaling behavior were demonstrated using a 1-UC-thick $La_{2-x}Sr_xCuO_4$ and an ionic liquid EDL gate (see Sec. 3.5) [20].

Another important phenomenon that is relevant for 2D superconductors in general is the Kosterlitz-Thouless-Berezinskii (KTB) transition [9, 11]. As mentioned earlier, the Mermin-Wagner theory prohibits the emergence of the superconducting phase transition in a 2D system in the strict sense, i.e., the establishment of a long-range correlation at a finite temperature. In a more physical picture, 2D superconductors suffer from phase fluctuations of the order parameter $\Psi$ due to the thermally excited free vortices, even if the amplitude of $\Psi$ is well developed below a Cooper-pair condensation temperature, $T_{c0}$. When these vortices are moved by an external current in the transverse direction, the motion gives rise to a voltage drop in the longitudinal direction, resulting in finite energy dissipation. This precludes the realization of the true zero-resistance state [32]. However, a vortex in a 2D superconductor can form a bound state with an antivortex (a vortex with the opposite supercurrent circulation) to form a "neutral" pair. If the vortices interact logarithmically as a function of their separation $r$ like $E_{\text{int}} \propto \log r$ ($E_{\text{int}}$: interaction energy), all vortices and antivortices form pairs below a certain critical temperature, thus leaving no free vortices. This is called the KTB transition, the transition temperature of which is denoted as $T_{KTB}$. For $T > T_{KTB}$ (and $T < T_{c0}$), the average separation $\xi$ between free vortices diverges like $\xi \propto |T - T_{KTB}|^{-1/2}$ as $T$ approaches $T_{KTB}$, and the zero-bias sheet resistance $R_{\text{sheet}}$ decreases to zero according to a relation

$$R_{\text{sheet}} \propto \exp\left\{-\left(\frac{T_{c0} - T}{T - T_{KTB}}\right)^{1/2}\right\}. \tag{3}$$

Since no free vortices exist for $T < T_{KTB}$, the true zero resistance state can be realized. However, the vortex-antivortex pairs become unbound under a finite external current. This unbinding occurs progressively as the current is increased, leading to a current-voltage (I-V) characteristics of a power-law dependence, i.e., $V \propto I^a$. As $T$ is lowered from above $T_{KTB}$, exponent $a$ jumps universally from 1 to 3 at $T = T_{KTB}$ and increases further at lower temperatures, which is one of the hallmarks of the KTB transition. The phase transition is unique in the sense that it does not lead to a true long-range order, but rather to a quasi-long-range order where the spatial correlation of the order decays as a function of distance according to a power-law dependence. The general features of the KTB transition have been observed using 2D superconductors with moderately high sheet resistances, which were prepared in the similar method as described above [47-49]. Figure 3 shows a representative result of the KTB transition, which was observed for an amorphous homogeneous film made of In/InO$_x$ composites with a thickness of 10 nm. The left panel is a logarithmic plot of the sheet resistance as a function of $(T-T_{KTB})^{-1/2}$ where the relation of Eq.(3) holds over four orders of magnitude in resistance. The right panel displays a I-V characteristics in the log-log plot. The exponent $a$ of the relation $V \propto I^a$ was found to cross 3 around $T_{KTB}$ = 1.94 K. It should be noted that the occurrence of the KTB transition requires that the logarithmic interaction between vortices be retained up to a sufficiently (ideally, infinitely) long distance. Cutoff of the logarithmic interaction stemming from a finite sample/domain size, a finite perpendicular penetration depth, a residual magnetic field, etc. leads to a non-vanishing zero-bias resistance down to $T$ = 0 [50]. In recent experiments on various 2D superconductors, the KTB transition has been discussed based on the non-linear characteristics $V \propto I^a$, but this relation was found to apply only at relatively high bias regions [21, 51-55]. Whether its origin is ascribed to the interaction cutoff inevitable in real experiments is not clear in many studies.

## 3. Recent progress in 2D superconductors with atomic-scale thicknesses

### 3.1. Ultrathin metal films and islands on semiconductor surfaces

One of the first experiments on superconducting ultrathin metal films with atomically well-defined thicknesses and high crystallinity was performed using Pb films grown on a clean Si(111)-(7×7) surface with the molecular beam epitaxy (MBE) method in UHV [56]. An advantage of using these materials is that Pb and Si does not form an alloy so that their interface is atomically sharp. Furthermore, Pb films can grow on a clean Si surface in a layer-by-layer fashion at low temperatures [57], which makes it possible to fabricate ultrathin films with nearly uniform thicknesses. Thus obtained ultrathin Pb films can be an ideal system to study 2D superconductivity. In Ref. [56], the films were characterized using the spot-profile analysis of low energy electron diffraction (LEED) and the obtained structural information was utilized for the analysis of the transport data. The transport measurements were performed *in-situ* in UHV to exclude contamination and oxidation of the samples. The $T_c$ of the Pb film was found to remain around 5~7 K down to the thickness of 4 ML, which was close to the bulk $T_c$ value of 7.2 K. However, the $T_c$ decreased substantially to 1~2 K in the sub-4 ML regime, where structural disorder in the macroscopic scale was unavoidable.

A particular interest in ultrathin metal films with the atomically smooth interfaces arises from the fact that conduction electrons are quantized in the out-of-plane direction to form quantum well states (QWS). The quantum confinement occurs when the Bohr-Sommerfeld quantization rule is satisfied [58]:

$$\frac{4\pi d}{\lambda(E)} + \Phi(E) = 2n\pi, \qquad (4)$$

where $\lambda(E)$ is the electron wavelength at an energy $E$, $d$ is the film thickness, $\Phi(E)$ is the total phase shift at the boundaries, and $n$ is an integer. Since $\lambda(E_F) = 1.06$ nm for Pb ($E_F$: Fermi energy), $d$ must be both in the order of nanometers and atomically uniform for this effect to be observed. The formation of QWS leads to an oscillation in the electron density of states at the Fermi level, $\rho(E_F)$, as a function of $d$, with the periodicity of $\lambda(E_F)/2$. Since the $T_c$ of a BCS-type superconductor is proportional to $\exp[-1/\rho(E_F)V]$ [32], it can lead to an oscillation of $T_c$ as a function of $d$ (here, $V$ denotes the effective interaction between electrons). The first clear observation of this phenomenon was reported using ultrathin Pb films on Si(111) surfaces in the thickness regime above 21 ML (see Fig. 4) [19]. The observed periodicity in $d$ was 2 ML, approximately equal to $\lambda(E_F)/2$ for Pb. The interpretation as a quantum oscillation was supported by the corresponding oscillatory behavior for $\rho(E_F)$, which was determined from the temperature dependence of out-of-plane upper critical magnetic field $H_{c2\perp}$ near $T_c$. In addition, the electron-phonon coupling constant $\lambda$ appearing in the Eliashberg-McMillan theory [59] was estimated from the temperature dependence of the quasiparticle lifetime through photoemission spectroscopy. This also showed a similar oscillation as a function of $d$. In this experiment, the transport measurements were performed *ex-situ* with Au capping layers to protect the superconducting Pb layers from air exposure. The presence of the capping layer may shift the energy positions of the quantum well states due to the change in the boundary condition.

Another important consequence of the QWS in atomically thin Pb films is the manifestation of "magic thicknesses" during the growth. When the energy positions of QWS are located far from $E_F$ for a certain thickness, the film becomes stable due to a decrease in electronic energy. This occurs with a periodicity of $\lambda(E_F)/2$ and the resulting magic thicknesses are $d$ = 4, 5, 7, 9, ... ML for Pb films [60]. This effect helps to make the film thickness constant at one of these values, and when the nominal coverage of Pb is slightly less than it, flat voids with a lower local magic thickness are created on an otherwise uniform film. The unique morphology leads to a strong pinning of magnetic fluxes at these voids and to realization of the so-called hard superconductor [32] albeit the atomic-scale thicknesses. This unexpected robustness of superconductivity was observed for Pb ultrathin films grown on a Si surface by *ex-situ* magnetization measurements using a superconducting quantum interference device (SQUID) [17, 61]. Figures 5(a) and (c) display a STM image of a 7-ML-thick Pb film with 2-ML deep voids and the corresponding d.c. magnetization with a "hard" hysteresis loop. In contrast, a 7-ML-thick Pb film with 2-ML tall mesas only exhibited a "soft" hysteresis loop (Figs. 5(b) and (d)). The analysis of the real and imaginary diamagnetism susceptibility showed that it had a nearly perfect Bean-like critical state corresponding to a very hard superconductivity. A critical current density as large as $\sim 2\times 10^6$ A/cm$^2$ was obtained at 2 K, which was about 10% of the depairing supercurrent density. It was also shown that the structural stability and the superconducting properties could be tuned by adding Bi into Pb to make ultrathin alloy films [62]. Just below $T_c$, the temperature dependence of $H_{c2\perp}$ was found to markedly deviate from the prediction of the GL theory, i.e., $H_{c2\perp} \propto (1 - T/T_c)$.

This indicated that, in this 2D geometry, the superconducting order was anomalously suppressed by scattering in violation of Anderson's theorem.

Ultrathin metal films on semiconductor surfaces have been studied not only by the conventional electron transport and magnetic measurements but also by LT-STM/STS. In this case, superconductivity is detected by observing the superconducting energy gap through the measurement of bias-voltage dependent differential conductance (*dI/dV*). This technique allows direct and real-space investigations of spatially inhomogeneous superconducting properties at atomic scales, which are hardly accessible by the conventional macroscopic experiments. Since the surfaces of the samples treated here are atomically clean and well-defined, this system is highly suited for STM investigation in UHV. Particularly interesting is the observation of superconducting vortices under application of magnetic field [63]; the vortex core, where the superconducting energy gap is suppressed due to an excessive supercurrent density, can be imaged as a region with high local density of states (i.e., high *dI/dV*). Figure 6(a)(b) shows an example of such experiments; 12-14 ML-thick Pb islands were imaged in (a) the topographic mode and (b) the *dI/dV* mapping mode with $E = E_F$ [64]. In the *dI/dV* image, the superconducting regions appeared dark due to the presence of energy gap while the normal-like vortex cores bright. The vortices were located at the voids, indicating void-induced trapping as discussed above (see also Ref. [65]). Furthermore, hysteretic behaviors of vortex dynamics were observed under varying magnetic fields using a Pb island whose size is about several times the coherence length $\xi_{GL}$ (~30 nm at 2 K) [66]. The experiment showed that a vortex penetrated into and escaped from the superconducting island at different magnetic fields, revealing the presence of the Bean-Livingston energy barrier at the periphery of the island [67]. In a 2D superconductor with a sufficiently large area, vortices are usually quantized into those with a vorticity of unity and form the Abrikosov lattice, but they can be merged into a giant vortex with a multiple vorticity when squeezed into a narrow region. Such anomalous behaviors were observed for vortices within an ultrathin Pb island using a LT-STM [65, 68].

When a superconductor is in close contact with a normal metal, the superconducting correlations may propagate into the neighboring region [69] where a superconducting energy gap can be detected by a spectroscopic method. This proximity effect is a spatially inhomogeneous phenomenon and was successfully studied in real space with a LT-STM using superconducting ultrathin Pb islands on a Si(111) surface [64, 70, 71]. Here, 2D atomic layers of Pb (amorphous or crystalline, depending on the sample preparation) on the Si(111) surface played the role of the normal metal region. In a simple configuration, the length scale with which superconductivity penetrate into the normal-metal region is solely determined by the proximity length $\xi_M \equiv \sqrt{\hbar D / 2\pi k_B T}$, where *D* is the diffusion constant in the normal metal. However, a LT-STM study revealed the presence of a strong geometric effect on the proximity effect [70]. The normal-metal region surrounded by superconductors exhibited an enhanced proximity effect, *i.e.*, the energy gap was found to survive up to distances of several times $\xi_M$ from the boundary. This is due to the quantum interference originating from the multiple Andreev reflection in the confined geometry [72, 73]. At a superconductor-normal metal (S-N) junction, an electron in the N region with an energy $-\Delta < E < \Delta$ ($\Delta$: energy gap) is reflected as a hole (and vice versa) to allow a Cooper pair injection into the S region, and this Andreev reflection can be multiplied and enhanced in the case of a narrow S-N-S junction. Since the proximity effect is caused by the Andreev reflection, the former is also enhanced by this mechanism. The essentially same phenomenon was observed when the proximity region was terminated and confined by a surface atomic step [74]. Such an N region in the proximity of S can host vortices when out-of-plane magnetic field was applied, which were successfully imaged with a LT-STM [64]. Figure 6(c)(d) shows the superconducting energy gap and its spatial mapping within the proximity N region where a vortex core is visible. The vortex was called a Josephson vortex, since the proximity effect and the multiple Andreev reflection are the origin of the Josephson effect for a S-N-S junction [75]. The locations of these vortices were found to be determined by the spatial distributions of supercurrents within the Pb islands.

*3.2. Metal atomic layers on semiconductor surfaces*

It has been generally accepted that metal thin films fabricated on a substrate should lose superconductivity when they approach the atomic-scale limit in thickness. In the case of granular and amorphous films, this phenomenon can be understood in terms of the disorder-driven superconductor-insulator (S-I) transition as discussed in Sec. 2. However, crystalline thin films with atomically uniform thicknesses for which the disorder is minimized also tend to lose superconductivity in this limit. For example, *ex-situ* magnetic measurements revealed that the $T_c$ of a Pb ultrathin film decreased as a function of thickness $d$ according to the relation:

$$T_c(d) = T_{c0}(1 - d_c/d) \tag{5}$$

where $T_{c0}$ is the asymptotic value of $T_c(d)$ for $d \to \infty$ (the bulk limit) and $d_c$ is the critical thickness for the disappearance of superconductivity [17, 61, 62]. Experimentally, $d_c$ was determined to be 0.43-0.615 nm for Pb, which corresponds to 1.5-2 ML. Within the framework of the GL theory, this behavior can be naturally explained by introduction of surface energy term into the GL free energy [76]. While this term originates from the decrease in density of states near the surface in Ref. [76], it could also be attributed to other effects such as surface/interface roughness or disorder, which is hardly avoidable in real experiments (for example, because of a capping layer). Whatever the origin is, superconductivity may persist until the truly monatomic thickness if this effect is negligibly small. *In-situ* UHV-LT-STM experiments on 2D Pb islands on a Si(111) surface showed that this was actually the case [77, 78]. The $T_c$ values, obtained from the analysis of temperature-dependent superconducting energy gap $\Delta(T)$, remained between 6.0 K and 6.7 K down to $d$ = 4 ML, while $T_c$ = 7.2 K for bulk Pb . $T_c$ was also found to oscillate with a periodicity of 2 ML due to the formation of QWS as discussed in Sec. 3.1. Even 2-ML-thick Pb film, for which only a single QWS channel exists, retained superconductivity with $T_c$ = 3.65 K or 4.9 K depending on the type of the Pb-Si interface (i.e., Pb-induced surface reconstructions, see below). It should be noted that a UHV-LT-STM study by another group on the same Pb/Si(111) system found that $T_c$ steadily decreased with decreasing $d$ according to Eq.(5), where $d_c$ was determined to be 1.88 ML [79]. The reason for this discrepancy is not clear, but may be attributed to a subtle difference in the resulting Pb-Si interface structure originating from the experimental conditions.

The naturally grown ultrathin film of Pb on a clean Si(111) surface retains the atomic structure of a bulk crystal while an amorphous-like wetting layer exists at the interface to the silicon substrate, which is the result of the first stage of the Stranski-Krastanov growth mode [80]. This wetting layer can be transformed into a crystalline atomic layer of Pb through an appropriate annealing process. The structural change is driven by formation of covalent bonding between the metal adatoms and the silicon substrate and leads to a unique lattice periodicity and electronic states distinct from those of bulk Pb. The resulting surface structure is called a Pb-induced (generally, metal-induced) silicon surface reconstruction, and is considered an ultimately thin form of the 2D electron system consisting only of the surface and the interface [81]. It might seem that these surface reconstructions are unlikely to become superconducting due to their ultimate 2D character, but recent experiments have changed this view; UHV-LT-STM observations of the energy gap revealed that three kinds of metal-induced silicon surface reconstructions, Si(111)-SIC-Pb, Si(111)-($\sqrt{7}\times\sqrt{3}$)-Pb, and Si(111)-($\sqrt{7}\times\sqrt{3}$)-In, exhibited superconductivity with $T_c$ =1.83 K, 1.52 K, 3.18 K, respectively (see Fig. 7 and Figs. 8(a)(b)) [18]. (Here, SIC stands for "striped incommensurate" and $\sqrt{7}\times\sqrt{3}$ for the periodicity against the ideally terminated Si(111) surface.) Compared to $T_c$ = 7.2 K of bulk Pb, those of the two Pb-induced surface reconstructions (Si(111)-SIC-Pb, Si(111)-($\sqrt{7}\times\sqrt{3}$)-Pb) are much lower, while that of the In-induced surface (Si(111)-($\sqrt{7}\times\sqrt{3}$)-In) is comparable to $T_c$ = 3.4 K of bulk In. For the Si(111)-SIC-Pb surface, the vortices were also observed using STM under out-of-plane magnetic field. The estimated upper critical field is $\mu_0 H_{c2\perp}$ = 1.45 T, giving the coherence length of $\xi_{GL}(0)$ = 49 nm. Angle-resolved photoemission spectroscopy (ARPES) measurements on these surface atomic layers revealed highly dispersed metallic band structures, which can be understood based on 2D free-electron-like states [18, 82]. Furthermore, the temperature dependence of the electronic band width showed that the electron-phonon coupling constant $\lambda$ was ~1 for both Si(111)-SIC-Pb and Si(111)-($\sqrt{7}\times\sqrt{3}$)-In, being enhanced from the bulk values. This indicates that the Pb-Si and In-Si interfaces play an important role for the occurrence of the BCS-type superconductivity in this atomically thin limit.

The above finding was based on the tunneling spectroscopy experiments using STM, but more direct evidence of superconductivity was obtained through *in-situ* electron transport measurements under the UHV environment (see Figs. 8(c)(d)) [83, 84]. The experiments revealed that the emergence of zero resistance, i.e., the existence of supercurrents through atomic layers and of the superconducting coherence at macroscopic scales. The observed $T_c$ was 1.1 K for Si(111)-SIC-Pb and 2.4 K or 2.8 K for Si(111)-(√7×√3)-In. The two different values of $T_c$ for Si(111)-(√7×√3)-In were attributed to the presence of two kinds of phases that have different nominal coverages of In for this surface reconstruction [85]. It should be noted that the resistance changes were remarkably sharp just at $T_c$ although the systems were in the atomic-scale limit. The decrease in resistance above $T_c$ was accelerated as the temperature approached $T_c$, which was well described by the 2D fluctuation theories (see Sec.2) [84, 86]. Analysis of magnetic field dependence of $T_c$ gave $\xi_{GL}(0) = 74$ nm for Si(111)-SIC-Pb and $\xi_{GL}(0) = 19$-$29$ nm for Si(111)-(√7×√3)-In [84]. Unexpectedly, the robustness of the superconductivity at this 2D limit was evidenced in terms of a high 2D critical supercurrent density $J_{2D}$. For Si(111)-(√7×√3)-In, current-voltage (*I-V*) characteristics measurements gave $J_{2D} = 1.8 \times 10^{-2}$ A/cm at 1.8 K (correspondingly, 3D supercurrent density of $3$-$6 \times 10^5$ A/cm$^2$) [83]. The temperature dependence of $J_{2D}$ was described by the Ambegaokar-Baratoff relation [87], suggesting that $J_{2D}$ was determined by Josephson junctions that must exist within the current path on the surface. The most likely locations for the junctions are atomic steps since they terminate and separate the surface terraces. This assignment was corroborated by STM observation of Josephson vortices, which in this case were trapped at the atomic steps on the Si(111)-(√7×√3)-In surface [88]. The Josephson vortices here are anomalously elongated along the steps and superconductivity is significantly recovered within their core regions. These features are distinct from those of Josephson vortices found in the normal region in the proximity of superconducting Pb islands [64]. The presence of Josephson junctions was also indicated for Si(111)-(√7×√3)-Pb from the observation of suppressed superconductivity near atomic steps [89].

There are two unique features that should be considered for atomically thin 2D superconductors in general. First, when magnetic field is applied in the in-plane direction, there exists no orbital pair-breaking effect that usually dominates the upper critical magnetic field $H_{c2}$. This is because electrons cannot move in the out-of-plane direction (see Fig. 9(a)). In the simplest case, the in-plane critical magnetic field $H_{c2//}$ is determined by the Pauli pair-breaking effect, leading to $\mu_0 H_{c2//} = \Delta(0)/\sqrt{2}\mu_B \, (\equiv \mu_0 H_P)$ at $T = 0$ where $\Delta(0)$ is the BCS energy gap and $H_P$ is called the Pauli paramagnetic limit [32]. The low dimensionality of the system and the dominance of the Pauli effect may also lead to a realization of the Fulde-Ferrell-Larkin-Ovchinnikov (FFLO) state at a high magnetic field around $H_P$ [90, 91]. This unusual superconducting state is the result of the competition between an energy gain due to a partial Cooper pair formation and a paramagnetic energy gain due to the remaining normal electrons, both occurring on the Zeeman-split Fermi surfaces. The Cooper pairs have (multiple) non-zero momentums in striking contrast to the BCS state. Consequently, the order parameter $\Psi(\vec{r})$ becomes spatially modulated with wave vectors of $\vec{q}_m$ even at equilibrium, and is generally expressed by [92]

$$\Psi(\vec{r}) = \sum_{m=1}^{M} \Psi_m \exp(i\vec{q}_m \cdot \vec{r}) \ . \tag{6}$$

The second important feature is the manifestation of the Rashba effect due to the space-inversion symmetry breaking of the system and the spin-orbit interaction (SOI) [93]. The Hamiltonian of the Rashba effect is expressed as

$$\mathcal{H}_R(\vec{k}) = \alpha_R(\vec{\epsilon} \times \vec{k}) \cdot \vec{\sigma} \quad , \tag{7}$$

where $\vec{k} = (k_x, k_y, 0)$ is the kinetic momentum, $\alpha_R$ is the strength of the Rashba SOI, $\vec{\epsilon}$ is the unit vector perpendicular to the 2D xy plane, and $\vec{\sigma}$ is the Pauli matrices. The Rashba effect leads to a spin splitting and the spin-momentum locking where electron spins are chirally polarized in the directions perpendicular to both the electron momentum and the electric field (see Fig. 9(b)). Due to the space-inversion symmetry breaking at surface, this effect has been observed by ARPES for the metal-induced surface reconstructions with a strong SOI [94, 95]. In terms of superconductivity, Rashba effect may lead to exotic phenomena under high in-plane magnetic fields. As schematically depicted in Fig. 9(c), the two Rashba-split Fermi surfaces move in the opposite directions by ±q/2, perpendicular to the magnetic field $H_{//}$ (note that this is a simplified picture and the Fermi surfaces are actually deformed under a high field). This is due to a magnetoelectric effect based on the reconfiguration of momentum-locked spins under magnetic field [96]. Then the electrons on each Fermi surface can form Cooper pairs with non-zero momentum ±q, leading to $\Psi(y) \propto \exp(iqy) + \exp(-iqy) \propto \cos(qy)$ similarly to Eq. (6) [97] (see Fig. 9(d); here the x direction is set parallel to $H_{//}$). The resulting superconducting state is often called a FFLO-like state, while the original FFLO state is caused purely by the Zeeman effect and does not require the Rashba effect. Another possible consequence of the magnetoelectric effect is induction of *unidirectional* supercurrents perpendicular to the magnetic field (note that magnetic field usually induces *circular* supercurrents as a vortex; see also Fig. 9(d)). Since such stationary supercurrents cannot run in an isolated object, $\Psi$ becomes spatially modulated as $\Psi(y) \propto \exp(iq'y)$ to compensate the internal supercurrent (note that $q'$ is different from q in the above equation). This state is called the helical state [98]. The FFLO-like state and the helical state compete with each other [99], but the helical state is more robust against the impurity scattering. These exotic superconducting states have unusually high values of $H_{c2//}$, and have been studied within the context of heavy fermion systems (see Sec.3.8) [92, 100]. Since metal atomic layers on semiconductor surfaces (metal-induced surface reconstructions) treated here are much simpler systems, they may provide a new platform where complexities such as strong electron-correlation effects can be avoided.

Experimentally, there have already been reports on surface 2D superconductivity related to these topics. Notably, the Si(111)-(√3×√3)-(Tl,Pb) surface reconstruction consisting of 1 ML of Tl and 1/3 ML of Pb on Si(111) was found to exhibit the Rashba effect based on ARPES measurements (see Fig 10(a)) [52]. The same surface also showed a superconducting transition at 2.25 K, opening a route for investigating exotic superconducting state as stated above. In Ref.[89], the STS observation of the superconducting Si(111)-($\sqrt{7}\times\sqrt{3}$)-Pb surface and the analysis of its energy gap structure indicated the presence of a large pair breaking parameter. This was attributed to the scattering of the triplet part of the spin-triplet mixed Cooper pairs, which can be caused by the space-inversion symmetry breaking and the resultant Rashba effect [96]. Furthermore, *in-situ* transport experiments on Pb monatomic layers grown on a vacuum-cleaved GaAs surface showed a surprising robustness against in-plane magnetic field (see Fig. 10(b) [101]. The $T_c$ was found to decease only by ~2% under in-plane field of 15T. This phenomenon was interpreted as a manifestation of the helical state described above, based on the analysis of elastic and spin-orbit scattering rates deduced from the transport data.

Finally, 2 ML-thick Ga atomic layer on GaN(0001) was found to exhibit superconductivity. The measured sheet resistance showed an onset of transition at $T_c^{onset}$ = 5.4 K and appearance of the zero resistance at $T_c^{zero}$ = 3.8 K, which were much higher than $T_c$ = 1.08 K for the bulk stable phase of α-Ga [102]. This $T_c$ enhancement was attributed to a possible strong interaction at the interface between the Ga layer and the GaN substrate. It is reminiscent of the high-$T_c$ superconductivity of 1-UC-thick FeSe layers on a SrTiO$_3$ substrate, which will be described in Sec. 3.7.

### 3.3. Cuprate: $La_{2-x}Sr_xCuO_4/La_2CuO_4$ interface.

It is widely recognized that high-$T_c$ cuprate superconductors such as La$_{2-x}$Sr$_x$CuO$_4$, YBa$_2$Cu$_3$O$_7$, and Bi$_2$Sr$_2$CaCu$_2$O$_{8+x}$ have layered perovskite structures where hole-doped conducting CuO$_2$ planes are separated from each other by insulating layers [103]. The bulk crystal retains strong 2D characters because the conduction channels of the CuO$_2$ planes are only weakly coupled (often regarded as Josephson-coupled), and the coherence length in the out-of-plane direction (along the *c*-axis) $\xi_{GL\perp}(0)$ is comparable to the unit-cell (UC) length. This naturally leads to the following questions: What is the minimum number of CuO$_2$ planes (or unit cells along the *c*-axis) for the occurrence of superconductivity, and how the coupling between the CuO$_2$ planes helps establish high $T_c$ in these systems? Indeed, soon after the finding of cuprate superconductors, the investigation into this problem started [104-108]. Generally, superconductivity was found to survive even when the number of unit cells in a cuprate thin film was reduced to one. Nevertheless, the temperatures where the zero resistance sets in ($T_c^{zero}$) were substantially suppressed compared to both the bulk $T_c$ values and the temperatures where the resistance starts to decrease ($T_c^{onset}$). For example, a 1-UC-thick YBa$_2$Cu$_3$O$_7$ layer sandwiched between 6-UC-thick PrBa$_2$Cu$_3$O$_7$ layers showed $T_c^{zero}$ ~30 K and $T_c^{onset}$ ~80 K while $T_c \cong$ 90 K was obtained for a 100 nm-thick YBa$_2$Cu$_3$O$_7$ film [107]. Consequently, the superconducting transition was found to be substantially broadened. It was difficult to determine in the early stage of the research whether this is the intrinsic effect due to the ultimate 2D character or is caused by extrinsic factors such as structural and/or stoichiometric disorder at the interface.

Recent remarkable advancements of MBE/PLD growth and surface characterization techniques have now allowed researchers to investigate cuprate superconductors with ultimately thin layers of CuO$_2$ with an unprecedented precision [109, 110]. For example, La$_{2-x}$Sr$_x$CuO$_4$/La$_2$CuO$_4$ heterostructures with atomically sharp interfaces were studied through electron transport and magnetic susceptibility measurements [109]. Here neither of La$_{2-x}$Sr$_x$CuO$_4$ nor La$_2$CuO$_4$ layer is superconducting when isolated, since La$_{2-x}$Sr$_x$CuO$_4$ (x = 0.45) is in the overdoped region of metal and La$_2$CuO$_4$ is a Mott insulator. The resulting heterostructure, however, showed superconductivity with $T_c \cong 30$ K, which should be attributed to the presence of the La$_{2-x}$Sr$_x$CuO$_4$/La$_2$CuO$_4$ interface. When the top La$_2$CuO$_4$ insulating layer was transformed into a La$_2$CuO$_{4+\delta}$ superconducting layer through annealing in ozone atmosphere, $T_c$ reached $\cong 50$ K. This value was 25% higher than $T_c$ of a La$_2$CuO$_{4+\delta}$ single phase. Furthermore, the transitions were sufficiently sharp so that $T_c^{zero}$ was nearly equal to $T_c^{onset}$. The observed sharp transition strongly suggested the completeness of the interface at the atomic scale. The temperature dependence of critical current density $J_c$ obtained from analysis of diamagnetic response showed that the enhanced superconducting region was spatially limited within 1-2 UC thicknesses around the interface. A subsequent resonant soft x-ray scanning measurement on the La$_{2-x}$Sr$_x$CuO$_4$/La$_2$CuO$_4$ heterostructure ($x = 0.36$, $T_c = 38$ K) also revealed that the conducting hole channel was located at the interface over a characteristic distance of ~0.6 nm [111]. This is half the UC length of La$_2$CuO$_4$ (= 1.3 nm) where only single CuO$_2$ plane is included. The existence of the superconductivity within a nominally single CuO$_2$ plane was further demonstrated using δ-doped heterostructures (see Fig. 11) [110]. Here one of the CuO$_2$ planes of La$_{2-x}$Sr$_x$CuO$_4$/La$_2$CuO$_4$ ($x = 0.45$) was selectively doped with Zn to suppress superconductivity, identifying the location where superconductivity occurred. Transport and diamagnetic induction measurements revealed that only the second CuO$_2$ plane from the interface located on the La$_2$CuO$_4$ side was responsible for the phenomenon. The conducting holes were found to arise from charge transfer over the interface due to a chemical potential difference, but its exact location was also influenced by the Sr spatial profile.

The same group also demonstrated that superconductivity of a 1-UC-thick La$_{2-x}$Sr$_x$CuO$_4$ ($x$=0.06 to 0.2) can be electrically modulated [20]. The samples were fabricated through epitaxial growth on insulating La$_2$CuO$_4$/LaSrAlO$_4$ substrates, and the EDL gating technique based on polymer electrolyte or ionic liquid was applied (see Sec. 3.5) to successfully tune the $T_c$ by up to 30 K. This has become possible due to the facts that the thickness of the conduction layer was only ~1 nm and that a huge local electric field of > $10^9$ Vm$^{-1}$ was obtained in this configuration. Using underdoped samples in the same setup, they also investigated the S-I transition by modulating the carrier density to reveal its nature as a quantum phase transition (see Fig. 12). Accumulated data on the sample sheet resistance $R_{sheet}$ as a function of temperature $T$ and the number of mobile holes per unit cell, $x$, were found to follow perfectly the scaling equation function given in Eq. (2) (see Sec. 2). The analysis gave $R_c$ = 6.45±0.10 kΩ and $zv$ = 1.5±0.1. Remarkably, the critical resistance $R_c$ obtained here is equal to the quantum resistance for Cooper pair, $R_Q$ = 6.45 kΩ, within an experimental error. This strongly suggests that Cooper pairs exist in the form of localized bosons in the insulating region near the boundary and that the transition is driven by quantum phase fluctuation [42]. The value of $z\nu = 1.5$ is clearly different from those of previously investigated systems, such as amorphous MoGe films ($z\nu \cong 1.3$) [35], LaAlO$_3$/SrTiO$_3$ interface ($z\nu \cong 2/3$) (see Sec.3.4) [112], and amorphous Bi films ($z\nu \cong 0.7$) [46]. This means that they belong to different universality classes and that the observed quantum phase transitions are governed by different physics (for example, $z\nu = 4/3$ and $2/3$ correspond to the classical percolation and the 3D XY models, respectively). Also, related experiments on hole-doped YBaCuO$_{7-x}$ in the similar configuration revealed $R_c = 6.0$ k$\Omega$ and $z\nu = 2.2$ [113]. Here, the result $R_c \cong R_Q = 6.45$ k$\Omega$ was also indicative of a phase transition by quantum phase fluctuation of Cooper pairs, but the failure of the scaling analysis at the lowest temperatures suggested a possible presence of an intermediate phase near the transition. For electron-doped Pr$_{2-x}$Ce$_x$CuO$_4$ ($x = 0.04, 0.1$), $R_c = 2.88$ k$\Omega$ and $z\nu = 2.4$ were found [114]. The large deviation of $R_c$ from $R_Q$ indicated that fermionic quasiparticle excitations, arising from the suppression of Cooper pair formation, played an import role in the transition. These results give an important clue for understanding the physics of high-$T_c$ cuprates. It should also be noted that the superfluid density $n_s(0)$ of 2-UC-thick Y$_{1-x}$Ca$_x$Ba$_2$Cu$_3$O$_{7-\delta}$ was determined from the ac conductivity measured with the two-coil mutual inductance method [115]. The $T_c$ was found to be proportional to $n_s(0)$ near the S-I transition, which was attributed to the quantum fluctuation near a 2D critical point.

Preparation of cuprate thin layers with an atomic-scale precision based on MBE or PLD methods as shown above requires a highly advanced instrumentation. However, the recent progress in mechanical exfoliation of atomic sheets, which was first demonstrated for graphene [25, 116], has now allowed preparation of atomic sheets of Bi$_2$Sr$_2$CaCu$_2$O$_{8+x}$ in a simple and cost-effective way [117]. The prepared atomic sheets with thicknesses down to half-UC (including two CuO$_2$ planes) were protected by graphene sheets from degradation and transport measurements were performed. While the sheet resistance $R_{sheet}$ of the sample increased from a few $\Omega$ to 5 k$\Omega$ as the sheet thickness was reduced from 270 to 0.5 UC, sharp superconducting transitions were consistently observed, with nearly constant $T_c$ of ~82 K. This indicates that the interlayer coupling does not play an important role for the superconductivity of Bi$_2$Sr$_2$CaCu$_2$O$_{8+x}$. Superconductivity of exfoliated atomic sheets of other layered materials will be described in Sec.3.6.

*3.4. LaAlO$_3$/SrTiO$_3$ interface*

As seen in Sec. 3.3, recent technological advancements have allowed researchers to grow complex oxides in a layer-by-layer fashion and to fabricate oxide heterostructures with atomically sharp and well-defined interfaces. Since many oxide materials have unique properties originating from strong electron correlations and orbital/spin degrees of freedom, the interface of oxide heterostructure have a huge potential for realizing exotic and functional artificial 2D materials [118]. One of the most famous examples is the interface between two perovskite transition-metal oxides, LaAlO$_3$ and SrTiO$_3$, where LaAlO$_3$ layers are epitaxially grown on TiO$_2$-teminated (100) surface of SrTiO$_3$ due to a small lattice mismatch [119, 120]. Although LaAlO$_3$ and SrTiO$_3$ are both wide-bandgap insulators, their interface is known to possess a 2D metallic conduction channel with a high carrier mobility. This has been widely ascribed to a mechanism called "polar catastrophe", which arises from the discontinuity of ionic characters of oxide layers [121]. While the layers consisting of SrTiO$_3$ are formally charge-neutral [(SrO)$^0$-(TiO$_2$)$^0$], those of LaAlO$_3$ are ionic [(LaO)$^+$-(AlO$_2$)$^-$] (see Fig. 13(a)). Within the LaO-AlO$_2$ stacking layers, an electrostatic voltage is accumulated and becomes linearly divergent as the number of layers increases, which is apparently unrealistic. This difficulty can be solved by the polar catastrophe where half of the charge of the top-surface LaO layer is transferred to the TiO$_2$ layer of the interface. Consequently, in the simplest case, the TiO$_2$ sheets of the LaAlO$_3$-SrTiO$_3$ interface are doped by 1/2 electron per 2D unit cell to have a carrier density of $n_{2D} = 3.4 \times 10^{14}$ cm$^{-2}$. The actual interface is much more complex, due to effects such as structural and electronic reconstructions, extrinsic electron doping from oxygen defects, and stoichiometric inhomogeneity. Nevertheless, it generally has a high carrier density (~$10^{13}$ cm$^{-2}$) and a low sheet resistance (~$10^3$ Ω) even when extrinsic electron doping from oxygen defects is minimized by choosing the appropriate growth condition. [119]. The interface also has a high carrier mobility (~$10^4$ cmV$^{-1}$s$^{-1}$), indicating the electron scatterings due to impurities and defects are weak. Generally speaking, LaAlO$_3$ growth at a lower oxygen pressure leads to a higher density of oxygen deficiency defects in the SrTiO$_3$ substrate and the electron conduction becomes more bulk-like due to the doping [122].

Remarkably, electron transport measurements have revealed that the LaAlO$_3$/SrTiO$_3$ interface become superconducting at low temperatures (see Fig. 13(b)-(e)) [21]. For a representative result using a sample with a 8-UC-thick LaAlO$_3$ layer, important physical quantities related to superconductivity were given as follows: transition temperature $T_c \cong 200$ mK, 2D critical supercurrent density $J_{c,2D} = 98$ μA/cm, out-of-plane critical magnetic field $\mu_0 H_{c2\perp} = 65$ mT. The observed $T_c$ was within the range of those previously reported for oxygen-defect-doped bulk SrTiO$_3$ [123]. The GL coherence length at zero temperature was estimated to be $\xi_{GL}(0)$ ~70 nm from the temperature dependence of $H_{c2\perp}$. Other parameters regarding the electron transport were as follows: sheet resistance $R_{sheet} \cong 300$ Ω, Hall carrier density $\cong 4 \times 10^{13}$ cm$^{-2}$, carrier mobility $\cong 350$ cmV$^{-1}$s$^{-1}$ (all at $T = 4.2$ K). The significant contribution of the interface conduction channel to the transport phenomena and the 2D character of superconductivity were discussed from the viewpoint of the KTB transition. $T_{KTB} = 188$ mK was obtained from the analysis of the temperature dependence of the $I$-$V$ characteristics, where the exponent $a$ of the relation $V \propto I^a$ was found to cross the universal constant of 3. In addition, the zero bias resistance was found to follow Eq. (3), giving $T_{KTB} = 190$ mK in consistency with the above value. It should be noted that the KTB transition considered here is not of the conventional type where vortex-antivortex bound pairs are formed from thermally excited free vortices in a 2D superconductor; rather it is analogous to the dislocation-induced melting of a 2D crystal treated in the original paper by Kosterlitz and Thouless [9]. Just below $T_c$, a high density of vortices and antivortices can form an ionic-like crystal if the energy required for thermal excitation of a vortex is sufficiently small. In this case, the KTB transition is characterized by the freezing of vortex-lattice defect motion, which leads to the occurrence of true zero resistance in the limit of an infinitely small bias current [124, 125]. In Ref. [21], the residual ohmic regime at small currents observed even below $T_{KTB}$ was attributed to the finite sample size effect [50].

The KTB transition observed here indicates that superconductivity in this system is of 2D character, thus limiting the maximum thickness of the conducting layer sufficiently below the coherence length of $\xi_{GL}(0)$ ~70 nm. A subsequent experiment on the same system revealed a strong anisotropy of critical magnetic field ($H_{c2//}/H_{c2\perp} \cong 20$) and showed that the LaAlO$_3$/SrTiO$_3$ superconductivity had indeed a strong 2D character [126]. The thickness of the conduction layer was estimated to be 11±3 nm from the analysis of temperature dependence of the critical magnetic fields. More direct information on the conducting layer thickness was obtained with a microscopic imaging technique based on atomic force microscopy (AFM) [122]. The local resistance was mapped over the cross-section cut of a LaAlO$_3$/SrTiO$_3$ heterostructure using an AFM tip to investigate the extent of the conducting region near the interface. For samples prepared by annealing around 750 °C at an oxygen pressure of 300 mbar after the LaAlO$_3$ growth, the spatial extension of the conducting layer was found to be smaller than 7 nm. In contrast, it could spread up to as large as ~500 μm when a different annealing process was taken due to the oxygen-deficiency doping. Furthermore, hard x-ray photoelectron spectroscopy taken on similarly prepared samples revealed that doped electrons are located at Ti sites of the TiO$_2$ layers within a distance of 4 nm from the interface [127].

Thanks to an extremely high dielectric constant of SrTiO$_3$ ($\varepsilon_r$ ~300 at RT and ~20,000 at LT) and a carrier density much lower than those of usual metals, the 2D conduction can be electrically tuned using a back gate on the substrate [120]. Superconductivity in this system was electrically enhanced based on this method where the maximum KTB transition temperature $T_{KTB}$ reached 310 mK [112]. Clear S-I transitions were also demonstrated through electrostatic gate control. The critical sheet resistance for the transition was $R_c \cong 4.5$ kΩ, which was close to but lower than the quantum resistance of Cooper pairs, $R_Q = 6.45$ kΩ [14]. The transition temperature identified with $T_{KTB}$ was found to scale with the carrier density n$_{2D}$ according to a relation $T_{KTB} \propto |n_{2D} - n_{2D,c}|^{z\nu}$, where $n_{2D,c}$ is the critical carrier density for the S-I transition and $z\nu$ is a parameter related to the critical behavior near the transition (see Eq. (2) of Sec. 2). The successful application of the scaling theory and the value of $z\nu = 2/3$ indicate that the transition is driven by the quantum fluctuations of Cooper pairs in a clean 2D system. Furthermore, a large negative magnetoresistance observed for the insulating phase in the same experiment was attributed to the weak localization of electrons.

Intriguingly, the LaAlO$_3$/SrTiO$_3$ interface exhibits not only superconductivity but also ferromagnetism, as in a theoretical prediction of ferromagnetic alignment of local spins at Ti$^{3+}$ sites of the interface TiO layer [128]. Experimentally, the existence of ferromagnetic phase was first reported using samples prepared in high oxygen pressures (i.e., with small oxygen deficiency doping) nearly simultaneously as superconductivity was discovered for the same system [129]. A clear magnetic-field hysteresis of sample sheet resistance $R_{sheet}$ was found as an evidence for the formation of ferromagnetic domains, and a logarithmic temperature dependence of $R_{sheet}$ and a large negative magnetoresistance observed at low temperatures were attributed to the presence of local magnetic moments. Subsequent studies have revealed that superconductivity and ferromagnetism can coexist within the *same* sample. Magnetic-field dependence of superconducting transition temperature $T_c$, which was electrically tuned by a back gate, was found to have hysteresis within the range of -30 mT < $\mu_0 H$ < -30 mT [130]. This clearly indicated that two conduction channels, corresponding to superconducting and ferromagnetic regions, existed within the interface. More direct evidence for this unexpected phenomenon was soon reported by two independent groups. A combined experiment based on high-resolution magnetic torque magnetometry and electron transport measurement revealed the presence of magnetic ordering in a superconducting LaAlO$_3$/SrTiO$_3$ interface [131]. Also reported was the real-space 2D mappings of magnetization and diamagnetic susceptibility using a SQUID, demonstrating the existence of nanoscale phase separation of ferromagnetic and superconducting regions (see Fig. 14) [132]. Ferromagnetic domains were observed as separate dipoles and persisted over the whole temperature range of the experiment, while inhomogeneous superconducting domains appeared below 100 mK. A control experiment using Nb δ-doped samples showed that magnetism in this system could only arise from the interface. It should be noted that the space inversion symmetry is broken at the interface, which is important for the emergence of the Rashba effect (see Sec. 3.2). The spin-orbit interaction (SOI) needed for the Rashba effect was also detected and was successfully controlled by a gate voltage [133]. Together with the electric-field tunability and the coexistence with ferromagnetism, the LaAlO$_3$/SrTiO$_3$ system offers a platform for exploring exotic superconducting phenomena.

*3.5. Electric-field induced interface superconductivity*

In the preceding two sections, we have seen that the interface between two non-superconducting (metallic or insulating) materials may spontaneously host a conduction channel with 2D superconductivity. Realization of such an interface superconductivity in an artificial and controlled way, preferably in a field-effect transistor (FET) configuration, should provide great opportunities for basic physics, materials science, and device applications [134]. In the case of the SrTiO$_3$ substrate, the back-gate control was possible due to its high dielectric constant [112], but the real power of this approach was demonstrated based on an electrochemical method, i.e., using an electric-double-layer (EDL) as a gate electrode (also see Sec. 3.3) [23, 24]. The conventional FET device uses a gate insulator made of solid-state dielectric material such as SiO$_2$ and high-$k$ complex oxides, but the accumulation of carriers at the interface is limited up to the level of $n_{2D}$ ~1×10$^{13}$cm$^{-2}$ because of current leakage and dielectric breakdown of the insulator. In an EDL transistor, however, the solid-state gate insulator is replaced by a liquid electrolyte, which allows free cations or anions to be assembled at the liquid-substrate interface under the application of gate voltage [22, 135-137]. The resulting EDL works as a subnanometer-gap capacitor and can induce carriers with an areal density up to $\sim 1\times 10^{15}$ cm$^{-2}$ at the interface (more precisely, at the subsurface) of the target material.

In the early stage of the EDL device, a polymer electrolyte consisting of KClO$_4$ and polyethylene oxide was used to fabricate a n-type FET with a 2D channel at the SrTiO$_3$ interface [23]. This device successfully induced superconductivity at $T_c \sim 0.4$ K under the application of the gate voltage above 2.5 V. This is the first report on electric-field-induced superconductivity in an insulator without chemical doping. A substantially high density of electron carriers with $n_{2D} = 1\times 10^{13} \sim 1\times 10^{14}$ cm$^{-2}$ was confirmed through Hall resistance measurement. This corresponds to a 3D carrier density as high as $3\times 10^{18} \sim 1\times 10^{20}$ cm$^{-3}$, which is roughly equal to that of superconducting Nb-doped SrTiO$_3$ bulk samples with $T_c = 0.4$ K. The effective thickness of the conduction channel was estimated to be 5~15 nm based on the calculated spatial distribution of carriers. This relatively large value is due to the large dielectric constant of SrTiO$_3$ that is on the verge of transition to ferroelectricity. The analysis showed that several 2D quantized subbands were involved within the conduction channel; in this sense, the system is not considered a superconductor in the 2D limit.

Substitution of the polymer electrolyte with ionic liquid, which is an organic salt in the liquid state even at room temperature, enables even an stronger gating function and accumulation of a higher density of carriers at the interface [138]. This has allowed investigation of field-induced superconductivity using various kinds of insulators including ZrNCl, KTaO$_3$, and 1T-TiSe$_2$. Some of the insulators also exhibited many-body quantum states featuring, e.g., magnetic ordering and charge density waves (CDW) [24, 139-141]. The experiment on thin layers of ZrNCl, obtained by mechanical micro-cleavage, realized for the first time superconducting transitions with $T_c = 12 \sim 15.2$ K under the gate voltages of 4~5 V using a EDL-FET device (see Fig. 15) [24]. The measured net carrier density amounted up to $1.7\times 10^{14}$ cm$^{-2}$. This seminal work was followed by a more elaborate experiment that revealed many intriguing phenomena (see Fig. 16) [142]. Generally, quasi-2D bulk superconductors made of weakly coupled conducting layers have an angular dependence of the upper critical magnetic field $H_{c2}(\theta)$, where $\theta$ is the angle of magnetic field relative to the in-plane direction (note that, in Fig. 15(a), $\theta$ is measured relative to the out-of-plane direction). Within the 3D anisotropic GL model, the anisotropy can be expressed by the following equation [32]:

$$\left[\frac{H_{c2}(\theta)\cos\theta}{H_{c2}(0°)}\right]^2 + \left[\frac{H_{c2}(\theta)\sin\theta}{H_{c2}(90°)}\right]^2 = 1 \quad . \tag{8}$$

This form applies when the coherence length in the out-of-plane direction, $\xi_{GL\perp}$, is larger than the interlayer distance. In contrast, for an isolated 2D superconductor with a thickness much smaller than the magnetic penetration depth, $H_{c2}(\theta)$ obeys the relation:

$$\left[\frac{H_{c2}(\theta)\cos\theta}{H_{c2}(0°)}\right]^2 + \left[\frac{H_{c2}(\theta)\sin\theta}{H_{c2}(90°)}\right] = 1 \quad , \tag{9}$$

which is called the 2D Tinkham model [143]. The clear difference in the angle dependence between the two model can be found around $\theta = 0°$; for the former, the change in $H_{c2}(\theta)$ is smooth (i.e., $|\partial H_{c2}/\partial\theta|_{\theta=0°} = 0$) while, for the latter, it exhibits a cusp-like shape (i.e., $|\partial H_{c2}/\partial\theta|_{\theta=0°} > 0$). In Ref. [142], $H_{c2}(\theta)$ for the field-induced superconductivity of ZrNCl exhibited a huge anisotropy between the in-plane ($\theta = 0°$) and out-of-plane ($\theta = 90°$) directions. Furthermore, a cusp-like angle dependence of $H_{c2}(\theta)$ was found around $\theta = 0°$, which was well described by Eq. (9) of the 2D Tinkham model. From the analysis of the angle dependence, the thickness of the superconducting region was determined to be 1.8 nm. This value corresponds to the $(ZrNCl)_2$ bilayer thickness and is smaller than the 1-UC-thickness of ZrNCl (= 2.76 nm). Thus it is considered a superconductor in the 2D limit. Correspondingly, the 2D signatures of the KTB transition and the superconducting fluctuation were found from the analysis of temperature dependence of the zero bias resistance (see Sec. 2). This atomically thin 2D superconductor is relatively clean; the electron mean free path was estimated to be 18~35 nm from the normal sheet resistance, and this value is comparable to the Pippard's coherence length $\xi_{Pippard}$ = 43.4 nm ($\xi_{Pippard} \equiv \hbar v_F/\Delta(0)$, $v_F$: Fermi velocity, $\Delta(0)$: BCS energy gap at $T = 0$). These ideal 2D character and high crystallinity led to an almost free vortex motion under magnetic fields of $\mu_0 H_{c2\perp} > 1.3$ T. For 0.05 T < $\mu_0 H_{c2\perp}$ < 1.3 T, a quantum creep motion of vortex lattice was indicated, and the detected finite residual resistances at the lowest temperatures demonstrated the presence of metallic states in this regime [144]. True superconducting states with undetectable residual resistance were obtained only for $\mu_0 H_{c2\perp}$ < 0.05 T. The result is contradictory to the theoretical prediction of the dirty-boson model where there should be no metallic state between the superconducting and the insulating states [41]. Accordingly, the magnetic-field dependence of the $R_{sheet}$-$T$ relations did not follow the scaling equation of Eq.(2), thus clarifying the limitation of the conventional theories of the magnetic-field induced S-I transition when applied to a clean system.

2H-type $MoS_2$, a band insulator and a representative transition-metal dichalcogenide, was also turned into a superconductor using an EDL transistor device with ionic liquid [145, 146]. The induced carriers with an areal density of ~$10^{14}$ cm$^{-2}$ were estimated to concentrate within a thickness of ~0.6 nm, which corresponds to a monolayer of $MoS_2$ (half the unit cell of a $MoS_2$ layered crystal). In addition, a high Hall mobility of ~240 cm$^2$/Vs was obtained at 20 K. Most intriguingly, the superconducting state had a dome-like phase diagram [146] that was reminiscent of those of cuprate high-$T_c$ superconductors [103]. Namely, $T_c$ plotted as a function of 2D carrier density $n_{2D}$ was found to have a maximum value of 10.8 K at $n_{2D}$ = 1.2×10$^{14}$ cm$^{-2}$. Further increase in $n_{2D}$ resulted in a lower $T_c$. The $T_c$ = 10.8 K is ~40% higher than the maximum $T_c$ of alkali-doped bulk $MoS_2$ crystal.

The fact that electron carriers are confined within a $MoS_2$ monolayer has a remarkable consequence as explained in the following (see Fig. 17) [54, 147]. MoS$_2$ as a bulk crystal has the global inversion symmetry due to its $D_{6h}$ symmetry, and its superconducting state is considered conventional. However, monolayer of MoS$_2$ lacks the in-plane inversion symmetry because it has half the unit cell of a bulk crystal and the symmetry is lowered to $D_{3h}$ [148]. Together with a large spin-orbit interaction (SOI) originating from the Mo $d$-orbitals, this causes an effective Zeeman field in the order of ~100 T and a spin polarization in the out-of-plane direction. In the momentum space, carriers induced in the MoSe$_2$ monolayer are located at electron pockets around the K and K' points in the Brillouin zone. The sign of the Zeeman splitting is inverted between the K and K' points due to the time-reversal relation, which is called the spin-valley locking [149]. The Hamiltonian for this effect is expressed as

$$\mathcal{H}_{\text{sv}}(\vec{k} + \epsilon \vec{K}) = \epsilon \beta_{\text{SO}} \sigma_z \quad (10)$$

where $\vec{k} = (k_x, k_y, 0)$ is the kinetic momentum in the K and K' valleys, $\vec{K}$ is the momentum of the K valley, $\epsilon = \pm 1$ is the valley index, $\beta_{\text{SO}}$ is the strength of the intrinsic SOI, and $\sigma_z$ is the z-component of the Pauli matrices. When Cooper pairs are formed by two electrons with opposite spins (up and down) and momenta (at K and K'), the only allowed spin directions are along the out-of-plane direction. The resultant state, called Ising superconductivity, is robust against in-plane magnetic field because the spin polarization direction is perpendicular to the field. Although the spin-valley locking is analogous to the in-plane spin-momentum locking that originates from the Rashba effect (see Eq. (7) in Sec.3.2) [52, 101], they are different in terms of the locking direction. Experimentally, huge in-plane critical magnetic fields $H_{c2//}$ were observed for MoS$_2$ EDL transistors and were attributed to the spin-valley locking phenomena. The observed $\mu_0 H_{c2//}$ = 52 T at 1.5 K (for $T_c$ = 9.7 K) [147] and $\mu_0 H_{c2//}$ > 20 T at 1.46 K (for $T_c$ = 2.37 K) [27] are about 4-5 times larger than the Pauli paramagnetic limit $H_P$. These $H_{c2//}$ values are also an order of magnitude larger than those of alkali-doped bulk MoS$_2$ crystal, for which $H_{c2//}$ is dominated by the orbital pair-breaking effect caused by the interlayer coupling. It should be noted that a huge $H_{c2//}$ value surpassing the Pauli limit was also indicated for the 2D superconductor with Rashba effect as discussed in Sec.3.2, but its origin is different. In the system with in-plane inversion symmetry breaking like the MoS$_2$ monolayer, the presence of the Rashba effect tends to weaken the spin-momentum locking because their spin polarization directions are orthogonal.

*3.6. Atomic sheets: graphene and transition-metal dichalcogenide atomic layers*

In 2004, graphene was mechanically exfoliated from a piece of graphite to fabricate an high-performance ambipolar FET device for the first time [25]. This invention was followed by the discovery of the half-integer quantum hall effect at low temperatures that signified the presence of massless Dirac fermions [150, 151], and quickly by an explosive number of graphene-related work. Finding superconductivity in graphene has been a long-standing goal in this field and should also greatly contribute to the study on 2D superconductivity. Until today, however, superconductivity was not reported in pristine graphene except for graphene-based junctions where Josephson supercurrents were detected to run [152]. This is presumably because of a small density of states at the Fermi level near the Dirac point. Nevertheless, it is well known that graphite intercalation compounds with foreign alkaline or alkaline-earth metal layers exhibit superconductivity, examples of which include $KC_8$ with $T_c$ = 0.14 K [153], $CaC_6$ with $T_c$ = 11.5 K, and $YbC_6$ with $T_c$ = 6.5 K [154, 155]. The metal layers play important roles in terms of the BCS-type superconductivity because they donate electrons into the π*-bands of graphite and modify the electron-phonon interaction. They also form electronic bands by themselves, which affect total properties of graphite. This raises a hope that intercalated or chemically doped graphene may become superconducting.

Recently, metal-doped few-layer graphene was reported to exhibit superconductivity in a variety of experiments. For example, a wet chemistry method allowed preparation of K-doped few-layer graphene from graphite flakes in dimethoxyethane solution, and superconductivity with $T_c$ = 4.5 K was found based on magnetic susceptibility measurement [156]. Similarly prepared Li-intercalated few-layer graphene was found to have $T_c$ = 7.4 K from magnetization measurement, although transport measurement did not show the evidence of resistance decrease around that temperature [157]. In terms of controlling the quality of graphene and the number of its layers, it is preferable to grow graphene in a layer-by-layer fashion on a semiconducting substrate such as SiC under the UHV environment. Thus prepared multi-layer graphene was Ca-intercalated *ex situ* by following the standard intercalation method used for bulk compounds [158]. Both magnetic and transport measurements revealed superconducting transitions down to a thickness of 10 ML, with the maximum $T_c$ of 7 K. However, since the intercalated dopants of alkaline or alkaline earth metal are very reactive in air, the above experiments may have been influenced by sample degradation due to air exposure, particularly when the number of layers is very small. To avoid such an undesirable effect, Ca-intercalated bilayer graphene ($C_6CaC_6$) was fabricated by growing graphene on a 6H-SiC(0001) substrate and subsequently by depositing Ca in UHV [159]. The samples were characterized by STM and ARPES to reveal their atomic-scale structures and free-electron-like interlayer electronic bands. Furthermore, transport measurements were taken on similarly prepared $C_6CaC_6$ samples *in situ* using a micro four-point-probe technique under magnetic fields, which revealed $T_c^{onset} \cong$ 4 K and $T_c^{zero} \cong$ 2 K (see Fig. 18) [160]. Analysis by reflection high energy electron diffraction (RHEED) showed that intercalated Ca atoms formed a (√3×√3)-R30° structure against the C(1×1) surface of the host graphene and that this structural ordering was crucial to have a clear superconducting transition. However, they did not observe a superconducting transition for Li-intercalated bilayer graphene ($C_6LiC_6$). This is puzzling because an ARPES study on Li-intercalated *monolayer* graphene ($LiC_6$) prepared using the similar method indicated the emergence of superconductivity at $T_c$ ~5.9 K through observation of an energy gap opening at the Fermi level [161]. In this experiment, the occurrence of superconductivity was ascribed to the enhancement of electron-phonon coupling constant $\lambda$ to 0.58, which was estimated from the ARPES measurement of the electronic bands. Indeed, first-principle calculations on $LiC_6$ assuming the ($\sqrt{3}\times\sqrt{3}$)-R30° structure of Li adatoms on a monolayer graphene predicted that $T_c$ would be strongly enhanced to 8.1 K from the bulk value of 0.9 K [162]. This enhancement was attributed to an increase in the electron-phonon coupling constant $\lambda$, in line with the experiment. The same theory also predicted $T_c$ = 1.4 K for Ca-intercalated monolayer graphene ($CaC_6$), which is suppressed significantly from the bulk value of 11.5 K. Nevertheless, direct evidence of superconductivity of doped monolayer graphene is so far missing from the viewpoint of electron transport or magnetization measurement. Coming back to the bulk material, superconductivity of intercalated graphene laminates were recently studied [163]. This layered material is similar to bulk graphite, but the coupling between the individual layers is weaker due to the presence of rotational disorder. Thus its electronic structure should be close to that of isolated monolayer graphene. The $T_c$ of Ca-intercalated graphene laminates was found to be ~6 K at maximum, but it was strongly dependent on the sample condition. None of K, Cs, or Li induced superconductivity in the temperature range of $T$ > 1.8 K.

The mechanical exfoliation of graphene from graphite has motivated creation of atomic sheets from layered materials using the same method. Particularly, the studies on atomic sheets of transition-metal dichalcogenides are now very active, in view of valleytronics (electronics combined with the valley degree of freedom in certain semiconductors) and optoelectronics applications [164]. Since the layers of transition-metal dichalcogenides are only weakly bonded through van der Waals force, the technique of mechanical exfoliation is readily applicable. In terms of studies on superconductivity in the 2D limit, $NbSe_2$ is a promising material because it has a relatively high $T_c$ of 7.2 K in the bulk form. It also has a phase transition accompanying charge density waves (CDW) at 33 K, so the competition or coexistence of the two ordered states in the 2D limit would also be interesting to investigate. An early experiment in 1972 already reported on the exfoliation of atomic sheets of $NbSe_2$ from a bulk crystal and the electron transport data indicative of the superconducting transition at a few UC thickness [165]. However, recent studies reported the absence of superconductivity in this regime [166], suggesting that the sample degradation due to air exposure and the lithography process was a serious problem. This was solved by performing the whole sample treatment in a controlled inert atmosphere and by encapsulating the target atomic sheet using a graphene or BN atomic sheet [167]. With this technique, monolayer (half-UC-thick) $NbSe_2$ was found to exhibit superconductivity with $T_c$ $\cong$ 2 K. Similarly prepared monolayer $NbSe_2$ showed a superconducting transition at 3 K, while a strong enhancement of the CDW transition up to $T_c$ = 145 K was detected from the Raman spectroscopy measurement [168]. High-quality monolayers of $NbSe_2$ were also prepared by MBE growth on epitaxial bilayer graphene on a 6H-SiC(0001) substrate under UHV environment [26]. After addition and removal of a Se capping layer, electron transport measurements were taken to reveal a superconducting transition with $T_c^{onset}$ = 1.9 K and $T_c^{zero}$ = 0.46 K. The suppression of $T_c$ compared to that of bulk NbSe$_2$ was partly attributed to the reduction of density of states at the Fermi level, which was caused by changes in electronic band structures in the monolayer of NbSe$_2$. STM/STS observations also clarified the occurrence of the CDW phase in the same sample. The 3×3 CDW superlattice was found to set in around 25 K and to develop fully at 5 K, while there was no signature of the CDW phase at 45 K. This is contradictory to the $T_c$ enhancement observed in Ref. [168].

Exactly like MoS$_2$, NbSe$_2$ monolayer is of half-UC thickness and lacks the in-plane inversion symmetry. Together with a strong SOI of transition-metal Nb, this can lead to the spin-valley locking as observed for the field-induced superconductivity in the MoSe$_2$ EDL transistors [54, 147] (see Sec. 3.5). This expectation was recently demonstrated by using a mechanically exfoliated NbSe$_2$ monolayer that was encapsulated by thin layers of BN, which exhibited superconductivity with $T_c$ = 3.0 K [27]. Magnetotransport measurements showed that superconductivity was remarkably robust against in-plane magnetic field, with an estimated upper critical field at zero temperature $\mu_0 H_{c2//}(0)$ ~ 35 T. This value is more than six times the Pauli paramagnetic limit $\mu_0 H_P$. Analysis of the temperature dependence of $H_{c2//}$ gave a spin splitting energy 2$\Delta_{SO}$ ~76 meV, equivalent to a spin-orbit field $\mu_0 H_{SO}$ ~660 T ($H_{SO} \equiv \Delta_{SO}/\mu_B$). These results indicate that the NbSe$_2$ monolayer is an Ising superconductor where the spins of Cooper pairs are aligned in the out-of-plane direction due to the spin-valley locking. Furthermore, magnetotransport measurement on NbSe$_2$ bilayer with a 1-UC-thickness ($T_c$ = 5.26K) showed that its true superconducting state was easily destroyed by a small out-of-plane magnetic field of 0.175T (see Fig. 19) [169]. At higher fields, the presence of so-called Bose metal phase was indicated, in which uncondensed Cooper pairs and vortices are responsible for the non-zero resistances [170, 171]. In view of the emergence of a metallic ground state under application of small out-of-plane magnetic fields, the obtained $H_\perp$-$T$ phase diagram was similar to that of field-induced superconductivity of a ZrNCl EDL transistor [142]. The absence of the magnetic-field induced S-I transition, unlike the conventional theories on 2D superconductors, should be attributed to the extremely small disorder and weak vortex pinning in these systems.

*3.7. One-unit-cell thick FeSe layer on SrTrO$_3$*

The tetragonal phase α-FeSe with PbO-type structure belongs to the iron-based high-$T_c$ superconductor family (see Fig. 20(a)) and exhibits superconductivity with $T_c$ ~ 8 K at ambient pressure and with $T_c$ ~37 K at a high pressure of 8.9 GP [172, 173]. FeSe has attracted extensive interest because of its compositional and structural simplicity and of a variety of unique physical properties attributable to its strong electron correlation [174]. The bulk crystal of FeSe consists of covalent-bonded 2D layers that are weakly coupled via van der Waals interaction. Together with its simple chemical form, this makes FeSe a promising material for the study of superconductivity in the 2D limit as well.

Surprisingly, 1-UC-thick FeSe layers (monolayers) epitaxially grown on a SrTiO$_3$(001) substrate using the MBE technique were found to exhibit superconductivity at much higher temperatures than for bulk crystals. The reported $T_c$ ranges from 23.5 K to 109 K depending on the type of measurement and the experimental environment. The first indication of a high $T_c$ of the monolayer FeSe was given by UHV-LT-STM experiments with a similar setup described in Sec. 3.1 and 3.2 [28], where degradation of FeSe layers by air exposure was avoided. Atomically well-ordered FeSe monolayer was fabricated on Nb-doped SrTiO$_3$ substrates by optimizing the growth condition (Fig. 20(b)). The STS measurements revealed the presence of a large gap structure with an energy gap $\Delta \cong 20$ meV at 4.2 K (Fig. 20(c)), which persisted at least up to 43 K. Assuming that the relation between the energy gap $\Delta$ and $T_c$ of bulk FeSe ($2\Delta/k_BT_c \cong 5.5$) still holds, it indicates that $T_c$ would be as high as 80 K. This value could break the $T_c$ record of the iron-based superconductors, 56 K, found for Sr$_{1-x}$Sm$_x$FFeAs [175]. STM measurements under an out-of-plane magnetic field of 11 T revealed formation of vortices as expected (Fig. 20(d)), from which the GL coherence length $\xi_{GL}$ was estimated to be a few nm. The result is in striking contrast to a related work on FeSe layers grown on bilayer graphene that was prepared on a SiC(0001) substrate [176]. In this case, the superconducting energy gap was not detected for 1-UC-thick FeSe layers down to $T = 2.2$ K. In addition, $T_c$ was found to decrease with decreasing thickness in accordance with Eq. (5), from which $d_c = 0.7$ nm was deduced as the critical thickness for disappearance of superconductivity. This form of thickness dependence of $T_c$ was also observed for ultrathin Pb films (see Sec. 3.2) [17, 79], suggesting the same physical mechanism behind it. In this case, the FeSe layers should have only weak van der Waals interaction with the bilayer graphene underneath, judging from the observation of frequent displacements and rotations of FeSe islands during STM imaging. On the contrary, for FeSe/SrTrO$_3$, the overlayer should have a strong interaction with the substrate considering the heteroepitaxial growth under a tensile strain. Therefore, the interaction at the interface, whatever it is, should be the origin of the high $T_c$ of monolayer FeSe.

More direct evidence for high-$T_c$ superconductivity of the 1-UC-thick FeSe layer was provided by *ex-situ* electron transport and diamagnetic response measurements using insulating SrTrO$_3$ substrates. The samples were capped with protection layers made of FeTe and amorphous-Si [53, 177]. The transport measurements indeed revealed a high $T_c$, with $T_c^{onset} = 40.2$ K and $T_c^{zero} = 23.5$ K. The latter value is close to $T_c = 21$ K determined from the diamagnetic response. In accordance with the high $T_c$, a large out-of-plane upper critical field $\mu_0 H_{c2\perp} > 52$ T was indicated at 1.4 K from the magnetotransport measurements. A large critical current density $J_c = 1.7 \times 10^6$ A/cm$^2$ was also observed at 2 K, which is two orders of magnitude larger than that of bulk FeSe. Nevertheless, the protection layers used in these experiments may have caused structural and chemical disorder in the FeSe layer and have adversely affected the observed superconducting properties. To avoid this problem, *in-situ* electron transport measurements were performed using a micro four-point-probe technique in UHV without any protection layers [29]. A surprisingly high $T_c$ of 109 K was locally detected from the emergence of the zero resistance state (see Fig. 20(e)). Accordingly, $\mu_0 H_{c2\perp}(0) \cong 116$ T, was deduced from magnetic field dependence of $T_c$. Although the samples suffered from spatial inhomogeneity and hence the high-$T_c$ superconductivity was not found all over the surface, the result suggests that macroscopically uniform 2D superconductors with $T_c > 100$ K may be available if the sample fabrication process is optimized. In terms of spectroscopic measurements, ARPES is a powerful surface-sensitive tool to reveal the electronic band structure and the superconducting energy gap that should appear below $T_c$. All ARPES experiments reported the opening of an energy gap that amounts to 15-20 meV in the gap size at the lowest temperatures [178-181]. $T_c$ determined from the temperature dependence of the energy gap ranges from 55 K to 65 K. These values are higher than that reported from the *ex-situ* macroscopic transport measurement ($T_c^{onset}$ = 40.2K) [53] but is lower than the maximum value in the *in-situ* local transport study ($T_c$ = 109K) [29]. This discrepancy may be explained considering the fact that ARPES is performed *in-situ* but at a macroscopic scale.

The emergence of high-$T_c$ superconductivity driven by the atomic-scale thinning is highly unusual because $T_c$ becomes generally suppressed with decreasing layer thickness as in Eq. (5). This certainly calls for investigations into its mechanism and the physics behind it, which are currently under intense debate. As mentioned above, the interface between the FeSe monolayer and the $SrTiO_3$ substrate must play a crucial role. There are several possibilities for the main mechanism, such as strain effects due to a lattice mismatch, enhancement of electron-phonon coupling, polaronic effects associated with the high dielectric constant of $SrTiO_3$ substrate, and carrier doping from the interface [28]. Some of the ARPES experiments indicated the presence of optical phonon modes at the interface [182] and of spin density waves (SDW) in the FeSe film, both of which may contribute to the high-$T_c$ superconductivity [180]. However, it is becoming clearer from the following experiments that the electron doping from the $SrTiO_3$ substrate plays a central role.

The Fermi surface of a bulk FeSe crystal consists of a hole-like pocket near the $\Gamma$ point and an electron-like pocket near the M point in the Brillouin zone, making FeSe a multiband superconductor. In contrast, the FeSe monolayer on a $SrTiO_3$ substrate has only an electron-like pocket near the M point, featuring a complete change in the topology of the Fermi surface [178-181]. The disappearance of the hole-like Fermi surface is naturally attributed to electron doping from the $SrTiO_3$ substrate with oxygen-vacancies, which was confirmed by step-by-step annealing taken after the growth of a FeSe monolayer [179]. On this Fermi surface, a nearly isotropic (*s*-wave-like) superconducting energy gap was detected below $T_c$ [181]. The band structure change due to the electron doping was also demonstrated by intentional chemical doping on the surface with K atoms, not only for a monolayer but also for a *trilayer* of FeSe (see Fig. 21) [178]. When the doping was tuned to the optimum level, a superconducting energy gap that corresponds to $T_c \sim 50$ K was observed for the trilayer, revealing that the monolayer was not the prerequisite for the emergence of a high $T_c$. The $T_c$ enhancement in a FeSe multilayer due to the surface K doping was also observed by STS when bilayer graphene on SiC(0001) was used as a substrate [183]. The idea was further supported by an experiment using an EDL transistor device [184]. Here, in addition to the field-induced carrier doping, electrochemical etching of FeSe layers by ionic liquid was performed through the gate operation by adjusting the working temperature and voltage. This allowed the layers to be thinned down to atomic-scale thicknesses. After the thinning process was terminated, $T_c^{onset}$ ~40 K was observed even for 10-UC-thickness FeSe multilayers when electrons were doped via gate voltage. Furthermore, the essentially same result was obtained when the substrate material was changed from SrTiO$_3$ to MgO. These results suggest that the main role of the SrTiO$_3$ substrate in $T_c$ enhancement is to work as a source of carrier doping. Nevertheless, in Ref. [178], the maximum available $T_c$ estimated from the energy gap is dependent on the number of FeSe layers and indeed the monolayer had the highest $T_c$. Very recently, the same trend was also observed in STS studies on the thickness-dependent energy gap of FeSe on SrTiO$_3$ [185, 186]. Correspondingly, the contribution of the intrinsic interface effect rather than doping was also noticed in related experiments [186, 187]. Thus it can be concluded that, although the emergence of a high $T_c$ up to ~50 K is possible only from substrate-induced electron doping, the presence of the FeSe-SrTiO$_3$ interface itself contributes to the additional $T_c$ enhancement up to 65~100 K, e.g., through electron-phonon coupling and polaronic effects. More experimental and theoretical investigations are needed to clarify the origin of the high-$T_c$ superconductivity in this system.

*3.8. Monolayer organic conductor and heavy-fermion superlattice*

In this final subsection, we will treat two important superconducting materials: organic conductors and rare-earth based heavy-fermion compounds. These two systems are known to exhibit varieties of exotic phenomena originating from strong electron correlations and/or intrinsic low dimensionality of the crystal. Examples include CDW/SDW formation, non-BCS type superconductivity such as *d*-wave Cooper pairing, the Kondo lattice, and coexistence and competition of superconductivity and magnetism [188, 189]. The studies on 2D superconductors with atomic-scale thicknesses made of these materials have been relatively limited so far, but the reported results are definitely worthwhile to be introduced here.

λ-(BETS)$_2$GaCl$_4$ (BETS=bis(ethylenedithio)tetraselenafulvalene) belongs to the family of charge-transfer complex type organic conductors and exhibits superconductivity at 4.7 K [190, 191]. Within a bulk crystal, 2D conduction layers consisting of hole-doped BETS molecules are stacked in the perpendicular direction while separated by GaCl$_4$ insulating layers, forming a highly anisotropic quasi-2D electron system. This native 2D character of the organic crystal poses a question whether 1-UC-thick layer (monolayer) of (BETS)$_2$GaCl$_4$ can become superconducting or not, as in the case of cuprates discussed in Sec. 3.3. Such an experiment was performed for monolayers of (BETS)$_2$GaCl$_4$ grown on a Ag(111) surface using a UHV-LT-STM (see Fig. 22) [30]. The organic molecule was deposited through thermal evaporation on a clean Ag substrate cooled down to ~120 K and was subsequently imaged with STM. The arrangement of BETS and GaCl$_4$ molecular units on the Ag substrate was found to be identical to that of a bulk crystal, confirming the 1-UC limit of the 2D organic crystal. STS measurements revealed that the monolayer of (BETS)$_2$GaCl$_4$ indeed exhibited an energy gap $\Delta \cong 12$ meV at 6 K that was attributed to the emergence of superconductivity. The $T_c$ was estimated to be ~10 K from the disappearance of the gap. It should be noted that this value is comparable to or higher than the bulk $T_c$ of ~ 8 K. The ratio of the energy gap to $T_c$, $2\Delta/k_B T_c$, was ~13.6, which is much larger than the BCS value of 3.52. Furthermore, the spectral shape of the observed energy gap was best fitted by assuming a $d_{xy}$ symmetry of Cooper pairs. These results showed that the superconductivity detected here is of non-BCS-type as found for many organic superconductors [192]. Surprisingly, the energy gap structure at the Fermi level persisted even in isolated molecular chains and the gap size remained nearly constant until the number of (BETS)$_2$GaCl$_4$ molecule pairs were reduced to 15. A small gap-like structure was noticed even for just four pairs of molecule unit. This means that superconductivity can survive nearly down to the zero-dimensional limit.    A subsequent experiment on the same system revealed that the obtained molecular structure was sensitive to the substrate temperature during the deposition [193]. Kagome-like lattices were formed when the substrate temperature was kept at 125 K, which were found to be insulating from STS measurements. For room temperature deposition, BETS molecules grew on a closely packed GaCl$_4$ layer on a Ag(111) surface. Despite the structural difference, this phase also exhibited a gap structure with $\Delta \cong 12$ meV at 4.6 K as seen in the molecular chains of Ref. [30].

There are many open questions for the monolayer of (BETS)$_2$GaCl$_4$ and its superconductivity. For example, precise determinations of molecular arrangement and of its electronic states are still far from complete. Particularly, charge transfers between the BETS/GaCl$_4$ molecules and the Ag substrate are very likely to take place as a result of their chemical potential difference. This should have a strong influence on the system, since the charge transfer within the molecular unit (2BETS + GaCl$_4$) is the driving force for the emergence of superconductivity in this type of complex organic conductors. Use of a closely related organic conductor λ-(BETS)$_2$FeCl$_4$ would also be very interesting because it is one of the few examples of magnetic-field-induced superconductors [194]. Since there has been no direct evidence of superconductivity by electron transport or magnetic measurements, the finding calls for further investigations on whether the signatures of energy gap can be safely attributed to the emergence of superconductivity.

As for rare-earth based heavy-fermion compounds, there have been no report on a 2D superconductor with an atomic-scale thickness, but superlattices made of few-UC-thick layers of heavy-fermion superconductor CeCoIn$_5$ and normal metal YbCoIn$_5$ were successfully created [31, 195, 196]. 4$f$ electrons in CeCoIn$_5$ are localized at the Ce atomic sites at high temperatures, but as the temperature is decreased, they form a narrow conduction band together with $sp$ itinerant electrons as a consequence of the Kondo effect. The resulting electronic states are highly unusual because of strong electron correlations and a very large effective mass $m_{eff}$. In terms of superconductivity, this has a significant influence on $H_{c2}$, because $H_{c2}$ is usually determined by the orbital pair-breaking effect and its value, $H_{c2}^{orb}$, is enhanced due to the relation $H_{c2}^{orb} \propto m_{eff}^2$. Indeed, $\mu_0 H_{c2}^{orb}$ exceeds 10 T in a bulk CeCoIn$_5$, which is well above the Pauli limit $\mu_0 H_P = \sqrt{2}\Delta/g\mu_B$ ($\Delta$: superconducting energy gap, $g$: g-factor; see Sec.3.2). Therefore, $H_{c2}$ is dominated by the Pauli paramagnetic effect, not by the orbital effect as in the usual superconductors. This offers a unique opportunity for studying exotic phenomena such as the FFLO states and the helical states induced by the Rashba effect (see Sec. 3.2) [92, 100]. The crystal of CeCoIn$_5$ has the space inversion symmetry, but introduction of a superlattice with atomically thin layers can artificially break this symmetry, consequently inducing the Rashba effect in a controllable way [195, 196].

In Ref. [31], CeCoIn$_5$($n$)/YbCoIn$_5$($m$) superlattices ($n$, $m$: number of UC in each layer) were epitaxially grown by the MBE technique and their superconducting properties were revealed by electron transport measurements for the first time. As $n$ was decreased while $m = 5$ was fixed, $T_c$ was found to decrease from the bulk value of 2.3 K, but a clear superconducting transition at 1.0 K was still observed for $n = 3$ (see Fig. 23(a)). For $n = 2$ and 1, the zero resistance state was not observed but magnetotransport measurements indicated the onset of superconductivity. Although suppression of superconductivity may be expected because of the presence of the neighboring normal metal layers of YbCoIn$_5$, the observed decrease in $T_c$ should be mostly intrinsic. This is because the large Fermi velocity mismatch between the two layers suppresses the proximity effect and confines the Cooper pairs within the CeCoIn$_5$ layers. One of the intriguing finding here was that $H_{c2}$, measured in both out-of-plane and in-plane directions, did not decrease from the bulk values in proportion to the reduction in $T_c$. Since $H_{c2}$ can be identified with the Pauli limit $H_P$ here and $H_P$ is proportional to the superconducting energy gap $\Delta$, the ratio $\Delta/k_B T_c$ must be enhanced. The estimated ratio of $\Delta/k_B T_c >$ 10 is much larger than the BCS value of 3.54, suggesting the realization of an extremely strong-coupling superconductor.

Furthermore, anomalous behaviors were observed in the angular dependence of $H_{c2}(\theta)$, where $\theta$ is the angle of magnetic field relative to the in-plane direction [195]. When the conducting layers consisting of a quasi-2D superconductors are sufficiently decoupled, the angle dependence of $H_{c2}(\theta)$ obeys the 2D Tinkham model (see Eq. (9)) as far as the orbital pair-breaking effect is dominant. However, when $H_{c2}(\theta)$ is dominated by the Pauli paramagnetic pair-breaking effect, the angle dependence is expressed by the 3D anisotropic GL model regardless of the interlayer coupling strength (see Eq. (8)). In this case, the anisotropy of superconductivity reflects that of g-factor in the Zeeman term. In Ref. [195], for $n = 4,5$ ($n$: number of UC for the CeCoIn$_5$ layers), the angular dependence of $H_{c2}(\theta)$ was expressed by the 3D anisotropic GL model in a wide temperature range (see the lower panel of Fig. 23(b); the variation around $\theta = 0$ is smooth). This was ascribed to the strong Pauli paramagnetic effect in CeCoIn$_5$. For $n = 3$, however, the angular dependence of $H_{c2}(\theta)$ at $T = 0.8$ K ($T_c$

= 1.04 K) was described by the 2D Tinkham model (see the upper panel of Fig. 23(b); the variation around $\theta = 0$ is cusp-like). This means that the orbital pair breaking-effect is stronger than the Pauli effect in this regime, considering the fact that the CeCoIn$_5$ layers are well decoupled from each other. Therefore, it was concluded that decrease in the layer thickness makes the Pauli effect weaker compared to the orbital effect. This phenomenon can be explained by the fact that the thinning of the CeCoIn$_5$ layers leads to increasingly important roles of the CeCoIn$_5$-YbCoIn$_5$ interface and of the space inversion symmetry breaking. Namely, the resultant manifestation of the Rashba effect significantly weakens the Pauli effect, because the spins are already polarized due the Rashba effect. Nevertheless, the Pauli effect was found to dominate again at lower temperatures ($T << T_c$), which was interpreted as the effect of the FFLO phase appearing in CeCoIn$_5$. The influence of the Rashba effect on $H_{c2}(\theta)$ was confirmed in a subsequent experiment, where additional space inversion symmetry breaking was introduced by an asymmetric stacking sequence of CeCoIn$_5$($n$)/YbCoIn$_5$($m$)/CeCoIn$_5$($n$)/YbCoIn$_5$($m'$) ($m \neq m'$) [196].

## 4. Summary

In the present review, we have seen that the existence of 2D superconductors with atomic-scale thicknesses have already been established in a wide range of fields and that they are now under extensive investigations from different viewpoints. The relevant superconducting materials encompass almost all categories to think of, from simple elemental metals on semiconductor surfaces, graphene, transition-metal chalcogenides, and FeSe to more complex compounds and molecules such as cuprates, perovskite oxides, rare-earth based heavy-fermion systems, and organic conductors. This means that the degree of electron correlation, which is one of the key factors in the modern solid-state physics and materials science, ranges from the weak to the strong limit. In terms of experimental technique, the studies of 2D superconductors feature a large variety as well. For example, LT-STM/STS and ARPES are powerful tools to directly probe the superconducting energy gaps of atomically thin superconductors exposed to the surface. Their usage is not limited to the studies treated here, but they surely have great advantages because of their surface sensitivity. Highly advanced MBE and PLD techniques allow researchers to grow different kinds of heterostructures in a layer-by-layer fashion and to fabricate atomically well-defined interfaces and superlattices with ideal 2D characters. Two recent technical breakthroughs, the EDL-FET device and the mechanical exfoliation of atomic sheets, have also greatly activated and broadened the present topics. Undoubtedly, the state-of-the-art nanotechnology including the above examples has been essential for the studies on superconductors in the 2D limit and will surely continue to drive their developments in the coming future.

New phenomena and physics revealed in these systems are so diverse and significant that they are also worthwhile to repeat here. Contrary to the general belief that superconductivity is fragile and must be suppressed as the material thickness approaches the atomic-scale limit, many 2D systems have been found to exhibit robust superconductivity at low temperatures as far as the structural and compositional quality of the sample is sufficiently high. This promises practical applications of these materials and devices in future. In some cases, the observed $T_c$ is comparable to or even higher than that of the bulk counterpart, and particularly for FeSe monolayers grown on $SrTiO_3$ substrates, $T_c$ is surprisingly high. Its origin is now under hot debate, but it is clear that the interface (or charge transfer through the interface) rather than the atomic-scale thickness itself plays the essential role. This may be called an interface-induced high-$T_c$ superconductor, but the ultimately small layer thickness helps this effect manifest itself in a dramatic way. Aside from the $T_c$ issue, there are many interesting phenomena and effects that take place, or are expected to do so, because of the two-dimensionality and the presence of the interface. One of the examples is the formation of QWS, which leads to the $T_c$ oscillation as a function of layer thickness and the Bean-like hard superconductivity. The absence of the usual orbital pair-breaking effect under in-plane magnetic field should lead to dominance of the Pauli paramagnetic pair-breaking effect. The signatures of the Rashba effect have already been detected in some materials, including Pb atomic layers and heavy-fermion compounds. The reduction of the layer thickness down to the half-UC in transition-metal dichalcogenides unveils the intrinsic inversion symmetry breaking of the lattices, from which the Ising superconductivity arises. In addition to these exotic phenomena, traditional themes such as the S-I transition have seen substantial progress, mostly thanks to the invention of the EDL-FET device. In some cases, new phases such as the Bose metal was indicated beyond the conventional theoretical framework of the S-I transition. Competition and coexistence of superconductivity with other many-body electronic state such as magnetism and CDW also enrich the physics and phenomena found in the regime of the 2D limit.

Although the studies on 2D superconductors with atomic-scale thicknesses have already been developed to a very high level, further important progress is sure to come in the next decade. This will be brought not only by focusing on individual topics described in Sec. 3, but also by crossing their boundaries, i.e., by promoting the overlap among the categories shown in Fig. 1. Along the horizontal axis, different materials may be combined to fabricate 2D hybrid systems, for example, using organic molecules and metal atomic layers on semiconductor surfaces. In the vertical direction, different experimental techniques may be combined. For example, surface atom layers treated only in UHV so far may be investigated using an EDL-FET device after an appropriate protection process. The 2D superconductivity at the buried interface may be exposed to the surface for the purpose of LT-STM and ARPES studies. This certainly calls for an interdisciplinary approach involving material science, surface science, and device physics. In terms of new phenomena and physics, finding the topological superconductivity will be one direction to take, since combination of the conventional (BCS-type) superconductivity and the Rashba effect is one of the most promising methods to realize it [197]. The two dimensionality is also a key factor in this respect, in view of observing and manipulating the Majorana zero mode within the vortex or at the edge [198, 199]. This may be an alternative approach in place of topological insulator-superconductor heterostructures, for which unusual vortex core states were observed by LT-STM [200, 201].

**Acknowledgements**

The author thanks A. Tanaka, S. Yoshizawa, T. Yamaguchi, C. Liu, Y. Hasegawa, S. Hasegawa, X. Hu, T. Kawakami, and H. Weitering for fruitful discussions and their critical reading of the manuscript. This work was financially supported by JSPS KAKENHI Grants No. 25247053 /26610107 and by World Premier International Research Center (WPI) Initiative on Materials Nanoarchitectonics, MEXT, Japan.

# References

[1] Cardy J 1996 *Scaling and Renormalization in Statistical Physics* (Cambridge: Cambridge University Press)

[2] Rice T M 1965 Superconductivity in One and Two Dimensions *Phys. Rev.* **140** A1889-A91

[3] Hohenberg P C 1967 Existence of Long-Range Order in One and Two Dimensions *Phys. Rev.* **158** 383-6

[4] Likharev K K and Semenov V K 1991 RSFQ logic/memory family: a new Josephson-junction technology for sub-terahertz-clock-frequency digital systems *IEEE Trans. Appl. Supercond.* **1** 3-28

[5] Yazdani A 2006 Lean and mean superconductivity *Nat. Phys.* **2** 151

[6] Seidel P ed 2015 *Applied Superconductivity: Handbook on Devices and Applications, Vol.2, Chapt. 9 & 10* (New York: Wiley)

[7] Mermin N D and Wagner H 1966 Absence of Ferromagnetism or Antiferromagnetism in One- or Two-Dimensional Isotropic Heisenberg Models *Phys. Rev. Lett.* **17** 1133-6

[8] Coleman S 1973 There are no Goldstone bosons in two dimensions *Comm. Math. Phys.* **31** 259-64

[9] Kosterlitz J M and Thouless D J 1973 ORDERING, METASTABILITY AND PHASE-TRANSITIONS IN 2 DIMENSIONAL SYSTEMS *J. Phys. C* **6** 1181-203

[10] Berezinskii V L 1972 Destruction of long-range order in one-dimensional and 2-dimensional systems having a continuous symmetry group. II. Quantum systems *Sov. Phys. JETP* **34** 610–6

[11] Mooij J E 1984 *Percolation, Localization, and Superconductivity,* ed A M Goldman and S A Wolf (Heidelberg: Springer)

[12] Minnhagen P 1981 Kosterlitz- Thouless transition for a two-dimensional superconductor: Magnetic-field dependence from a Coulomb-gas analogy *Phys. Rev. B* **23** 5745-61

[13] Halperin B I and Nelson D R 1979 RESISTIVE TRANSITION IN SUPERCONDUCTING FILMS *J. Low Temp. Phys.* **36** 599-616

[14] Goldman A M and Markovic N 1998 Superconductor-insulator transitions in the two-dimensional limit *Phys. Today* **51** 39-44

[15] Anderson P W 1959 Theory of dirty superconductors *J. Phys. Chem. Solids* **11** 26-30

[16] Strongin M, Thompson R S, Kammerer O F and Crow J E 1970 Destruction of Superconductivity in Disordered Near-Monolayer Films *Phys. Rev. B* **1** 1078-91

[17] Özer M M, Thompson J R and Weitering H H 2006 Hard superconductivity of a soft metal in the quantum regime *Nat. Phys.* **2** 173-6

[18] Zhang T, Cheng P, Li W J, Sun Y J, Wang G, Zhu X G, He K, Wang L L, Ma X C, Chen X, Wang Y Y, Liu Y, Lin H Q, Jia J F and Xue Q K 2010 Superconductivity in one-atomic-layer metal films grown on Si(111) *Nat. Phys.* **6** 104-8

[19] Guo Y, Zhang Y F, Bao X Y, Han T Z, Tang Z, Zhang L X, Zhu W G, Wang E G, Niu Q, Qiu Z Q, Jia J F, Zhao Z X and Xue Q K 2004 Superconductivity modulated by quantum size effects *Science* **306** 1915-7

[20] Bollinger A T, Dubuis G, Yoon J, Pavuna D, Misewich J and Bozovic I 2011 Superconductor-insulator transition in La2-xSrxCuO4 at the pair quantum resistance *Nature* **472** 458-60

[21] Reyren N, Thiel S, Caviglia A D, Kourkoutis L F, Hammerl G, Richter C, Schneider C W, Kopp T, Rüetschi A-S, Jaccard D, Gabay M, Muller D A, Triscone J-M and Mannhart J 2007 Superconducting Interfaces Between Insulating Oxides *Science* **317** 1196-9

[22] Shimotani H, Asanuma H, Tsukazaki A, Ohtomo A, Kawasaki M and Iwasa Y 2007 Insulator-to-metal transition in ZnO by electric double layer gating *Appl. Phys. Lett.* **91** 082106

[23] Ueno K, Nakamura S, Shimotani H, Ohtomo A, Kimura N, Nojima T, Aoki H, Iwasa Y and Kawasaki M 2008 Electric-field-induced superconductivity in an insulator *Nat. Mater.* **7** 855-8

[24] Ye J T, Inoue S, Kobayashi K, Kasahara Y, Yuan H T, Shimotani H and Iwasa Y 2010 Liquid-gated interface superconductivity on an atomically flat film *Nat. Mater.* **9** 125-8

[25] Novoselov K S, Geim A K, Morozov S V, Jiang D, Zhang Y, Dubonos S V, Grigorieva I V and Firsov A A 2004 Electric field effect in atomically thin carbon films *Science* **306** 666-9

[26] Ugeda M M, Bradley A J, Zhang Y, Onishi S, Chen Y, Ruan W, Ojeda-Aristizabal C, Ryu H, Edmonds M T, Tsai H-Z, Riss A, Mo S-K, Lee D, Zettl A, Hussain Z, Shen Z-X and Crommie M F 2016 Characterization of collective ground states in single-layer NbSe2 *Nat. Phys.* **12** 92-7

[27] Xi X, Wang Z, Zhao W, Park J-H, Law K T, Berger H, Forro L, Shan J and Mak K F 2016 Ising pairing in superconducting NbSe2 atomic layers *Nat. Phys.* **12** 139-43

[28] Wang Q-Y, Li Z, Zhang W-H, Zhang Z-C, Zhang J-S, Li W, Ding H, Ou Y-B, Deng P, Chang K, Wen J, Song C-L, He K, Jia J-F, Ji S-H, Wang Y-Y, Wang L-L, Chen X, Ma X-C and Xue Q-K 2012 Interface-Induced High-Temperature Superconductivity in Single Unit-Cell FeSe Films on SrTiO3 *Chin. Phys. Lett.* **29** 037402

[29] Ge J-F, Liu Z-L, Liu C, Gao C-L, Qian D, Xue Q-K, Liu Y and Jia J-F 2015 Superconductivity above 100 K in single-layer FeSe films on doped SrTiO3 *Nat. Mater.* **14** 285-9

[30] Clark K, Hassanien A, Khan S, Braun K F, Tanaka H and Hla S W 2010 Superconductivity in just four pairs of (BETS)2GaCl4 molecules *Nat. Nanotechnol.* **5** 261-5

[31] Mizukami Y, Shishido H, Shibauchi T, Shimozawa M, Yasumoto S, Watanabe D, Yamashita M, Ikeda H, Terashima T, Kontani H and Matsuda Y 2011 Extremely strong-coupling superconductivity in artificial two-dimensional Kondo lattices *Nat. Phys.* **7** 849-53

[32] Tinkham M 1996 *Introduction to Superconductivity* (New York: McGraw-Hill)

[33] Katsumoto S 1995 SINGLE-ELECTRON TUNNELING AND PHASE-TRANSITIONS IN GRANULAR FILMS *J. Low Temp. Phys.* **98** 287-349

[34] Haviland D B, Liu Y and Goldman A M 1989 Onset of superconductivity in the two-dimensional

limit *Phys. Rev. Lett.* **62** 2180-3

[35] Yazdani A and Kapitulnik A 1995 Superconducting-Insulating Transition in Two-Dimensional α-MoGe Thin Films *Phys. Rev. Lett.* **74** 3037-40

[36] Hebard A F and Paalanen M A 1990 Magnetic-field-tuned superconductor-insulator transition in two-dimensional films *Phys. Rev. Lett.* **65** 927-30

[37] Aslamasov L G and Larkin A I 1968 The influence of fluctuation pairing of electrons on the conductivity of normal metal *Phys. Lett.* **26A** 238-9

[38] Thompson R S 1970 Microwave, Flux Flow, and Fluctuation Resistance of Dirty Type-II Superconductors *Phys. Rev. B* **1** 327-33

[39] Imry Y 1997 *Introduction to Mesoscopic Physics* (New York: Oxford University Press)

[40] Sondhi S L, Girvin S M, Carini J P and Shahar D 1997 Continuous quantum phase transitions *Rev. Mod. Phys.* **69** 315-33

[41] Fisher M P A 1990 Quantum phase transitions in disordered two-dimensional superconductors *Phys. Rev. Lett.* **65** 923-6

[42] Fisher M P A, Grinstein G and Girvin S M 1990 Presence of quantum diffusion in two dimensions: Universal resistance at the superconductor-insulator transition *Phys. Rev. Lett.* **64** 587-90

[43] Liu Y, Haviland D B, Nease B and Goldman A M 1993 Insulator-to-superconductor transition in ultrathin films *Phys. Rev. B* **47** 5931-46

[44] Smith R A, Reizer M Y and Wilkins J W 1995 Suppression of the order parameter in homogeneous disordered superconductors *Phys. Rev. B* **51** 6470-92

[45] Liu Y, McGreer K A, Nease B, Haviland D B, Martinez G, Halley J W and Goldman A M 1991 Scaling of the insulator-to-superconductor transition in ultrathin amorphous Bi films *Phys. Rev. Lett.* **67** 2068-71

[46] Parendo K A, Tan K H S B, Bhattacharya A, Eblen-Zayas M, Staley N E and Goldman A M 2005 Electrostatic Tuning of the Superconductor-Insulator Transition in Two Dimensions *Phys. Rev. Lett.* **94** 197004

[47] Epstein K, Goldman A and Kadin A 1981 Vortex-Antivortex Pair Dissociation in Two-Dimensional Superconductors *Phys. Rev. Lett.* **47** 534-7

[48] Hebard A and Fiory A 1983 Critical-Exponent Measurements of a Two-Dimensional Superconductor *Phys. Rev. Lett.* **50** 1603-6

[49] Wolf S, Gubser D, Fuller W, Garland J and Newrock R 1981 Two-Dimensional Phase Transition in Granular NbN Films *Phys. Rev. Lett.* **47** 1071-4

[50] Medvedyeva K, Kim B J and Minnhagen P 2000 Analysis of current-voltage characteristics of two-dimensional superconductors: Finite-size scaling behavior in the vicinity of the Kosterlitz-Thouless transition *Phys. Rev. B* **62** 14531-40

[51] Zhao W, Wang Q, Liu M, Zhang W, Wang Y, Chen M, Guo Y, He K, Chen X, Wang Y, Wang J,

Xie X, Niu Q, Wang L, Ma X, Jain J K, Chan M H W and Xue Q-K 2013 Evidence for Berezinskii–Kosterlitz–Thouless transition in atomically flat two-dimensional Pb superconducting films *Solid State Commun.* **165** 59-63

[52] Matetskiy A V, Ichinokura S, Bondarenko L V, Tupchaya A Y, Gruznev D V, Zotov A V, Saranin A A, Hobara R, Takayama A and Hasegawa S 2015 Two-Dimensional Superconductor with a Giant Rashba Effect: One-Atom-Layer Tl-Pb Compound on Si(111) *Phys. Rev. Lett.* **115** 147003

[53] Zhang W-H, Sun Y, Zhang J-S, Li F-S, Guo M-H, Zhao Y-F, Zhang H-M, Peng J-P, Xing Y, Wang H-C, Fujita T, Hirata A, Li Z, Ding H, Tang C-J, Wang M, Wang Q-Y, He K, Ji S-H, Chen X, Wang J-F, Xia Z-C, Li L, Wang Y-Y, Wang J, Wang L-L, Chen M-W, Xue Q-K and Ma X-C 2014 Direct Observation of High-Temperature Superconductivity in One-Unit-Cell FeSe Films *Chin. Phys. Lett.* **31** 017401

[54] Lu J M, Zheliuk O, Leermakers I, Yuan N F Q, Zeitler U, Law K T and Ye J T 2015 Evidence for two-dimensional Ising superconductivity in gated MoS2 *Science* **350** 1353-7

[55] Xu C, Wang L, Liu Z, Chen L, Guo J, Kang N, Ma X-L, Cheng H-M and Ren W 2015 Large-area high-quality 2D ultrathin Mo2C superconducting crystals *Nat. Mater.* **14** 1135-41

[56] Pfennigstorf O, Petkova A, Guenter H L and Henzler M 2002 Conduction mechanism in ultrathin metallic films *Phys. Rev. B* **65** 045412

[57] Jałochowski M and Bauer E 1988 Reflection high‐energy electron diffraction intensity oscillations during the growth of Pb on Si(111) *J. Appl. Phys.* **63** 4501-4

[58] Paggel J J, Miller T and Chiang T C 1999 Quantum-well states as Fabry-Perot modes in a thin-film electron interferometer *Science* **283** 1709-11

[59] Carbotte J P and Marsiglio F 2003 *The Physics of Superconductivity Vol. I Convenrtional and High-Tc superconductors,* ed K H Bennemann and J B Ketterson (Heidelberg: Springer)

[60] Su W B, Chang S H, Jian W B, Chang C S, Chen L J and Tsong T T 2001 Correlation between Quantized Electronic States and Oscillatory Thickness Relaxations of 2D Pb Islands on Si(111)-(7×7) Surfaces *Phys. Rev. Lett.* **86** 5116-9

[61] Özer M M, Thompson J R and Weitering H H 2006 Robust superconductivity in quantum-confined Pb: Equilibrium and irreversible superconductive properties *Phys. Rev. B* **74**

[62] Özer M M, Jia Y, Zhang Z, Thompson J R and Weitering H H 2007 Tuning the quantum stability and superconductivity of ultrathin metal alloys *Science* **316** 1594-7

[63] Hess H F, Robinson R B, Dynes R C, Valles J M, Jr. and Waszczak J V 1989 Scanning-Tunneling-Microscope Observation of the Abrikosov Flux Lattice and the Density of States near and inside a Fluxoid *Phys. Rev. Lett.* **62** 214-6

[64] Roditchev D, Brun C, Serrier-Garcia L, Cuevas J C, Bessa V H L, Milosevic M V, Debontridder F, Stolyarov V and Cren T 2015 Direct observation of Josephson vortex cores *Nat. Phys.* **11** 332-7

[65] Tominaga T, Sakamoto T, Kim H, Nishio T, Eguchi T and Hasegawa Y 2013 Trapping and squeezing of vortices in voids directly observed by scanning tunneling microscopy and spectroscopy *Phys. Rev. B* **87** 195434

[66] Nishio T, An T, Nomura A, Miyachi K, Eguchi T, Sakata H, Lin S, Hayashi N, Nakai N, Machida M and Hasegawa Y 2008 Superconducting Pb Island Nanostructures Studied by Scanning Tunneling Microscopy and Spectroscopy *Phys. Rev. Lett.* **101** 167001

[67] Bean C P and Livingston J D 1964 Surface Barrier in Type-II Superconductors *Phys. Rev. Lett.* **12** 14-6

[68] Cren T, Serrier-Garcia L, Debontridder F and Roditchev D 2011 Vortex Fusion and Giant Vortex States in Confined Superconducting Condensates *Phys. Rev. Lett.* **107** 097202

[69] de Gennes P G 1989 *Superconductivity of Metals and Alloys* ( : Addion-Wesley)

[70] Kim J, Chua V, Fiete G A, Nam H, MacDonald A H and Shih C K 2012 Visualization of geometric influences on proximity effects in heterogeneous superconductor thin films *Nat. Phys.* **8** 463-8

[71] Serrier-Garcia L, Cuevas J C, Cren T, Brun C, Cherkez V, Debontridder F, Fokin D, Bergeret F S and Roditchev D 2013 Scanning Tunneling Spectroscopy Study of the Proximity Effect in a Disordered Two-Dimensional Metal *Phys. Rev. Lett.* **110** 157003

[72] Andreev A F 1964 Thermal conductivity of the intermediate state of superconductors *Sov. Phys. JETP* **19** 1228.

[73] Blonder G, Tinkham M and Klapwijk T 1982 Transition from metallic to tunneling regimes in superconducting microconstrictions: Excess current, charge imbalance, and supercurrent conversion *Phys. Rev. B* **25** 4515-32

[74] Kim H, Lin S-Z, Graf M J, Kato T and Hasegawa Y Enhancement and termination of the superconducting proximity effect due to atomic-scale defects visualized by scanning tunneling microscopy *arXiv:1401.2602*

[75] Likharev K 1979 Superconducting weak links *Rev. Mod. Phys.* **51** 101-59

[76] Simonin J 1986 Surface term in the superconductive Ginzburg-Landau free energy: Application to thin films *Phys. Rev. B* **33** 7830-2

[77] Eom D, Qin S, Chou M Y and Shih C 2006 Persistent Superconductivity in Ultrathin Pb Films: A Scanning Tunneling Spectroscopy Study *Phys. Rev. Lett.* **96** 027005

[78] Qin S Y, Kim J, Niu Q and Shih C K 2009 Superconductivity at the Two-Dimensional Limit *Science* **324** 1314-7

[79] Brun C, Hong I P, Patthey F, Sklyadneva I, Heid R, Echenique P, Bohnen K, Chulkov E and Schneider W-D 2009 Reduction of the Superconducting Gap of Ultrathin Pb Islands Grown on Si(111) *Phys. Rev. Lett.* **102** 207002

[80] Zangwill A 1988 *Physics at Surfaces* ( : Cambridge Univsersity Press)

[81] Lifshits V G, Saranin A A and Zotov A V 1994 *Surface Phases on Silicon: Preparation, Structures,*

*and Properties* (Chichester: Wiley)

[82] Rotenberg E, Koh H, Rossnagel K, Yeom H, Schäfer J, Krenzer B, Rocha M and Kevan S 2003 Indium $\sqrt{7}\times\sqrt{3}$ on Si(111): A Nearly Free Electron Metal in Two Dimensions *Phys. Rev. Lett.* **91** 246404

[83] Uchihashi T, Mishra P, Aono M and Nakayama T 2011 Macroscopic Superconducting Current through a Silicon Surface Reconstruction with Indium Adatoms: Si(111)-($\sqrt{7}\times\sqrt{3}$)-In *Phys. Rev. Lett.* **107**

[84] Yamada M, Hirahara T and Hasegawa S 2013 Magnetoresistance Measurements of a Superconducting Surface State of In-Induced and Pb-Induced Structures on Si(111) *Phys. Rev. Lett.* **110** 237001

[85] Kraft J, Surnev S L and Netzer F P 1995 The structure of the indium-Si(111) ($7\times\sqrt{3}$) monolayer surface *Surf. Sci.* **340** 36-48

[86] Uchihashi T, Mishra P and Nakayama T 2013 Resistive phase transition of the superconducting Si(111)-($\sqrt{7}\times\sqrt{3}$)-In surface *Nanoscale Res. Lett.* **8** 167

[87] Ambegaokar V and Baratoff A 1963 Tunneling Between Superconductors *Phys. Rev. Lett.* **10** 486-9

[88] Yoshizawa S, Kim H, Kawakami T, Nagai Y, Nakayama T, Hu X, Hasegawa Y and Uchihashi T 2014 Imaging Josephson Vortices on the Surface Superconductor Si(111)-($\sqrt{7}\times\sqrt{3}$)-In using a Scanning Tunneling Microscope *Phys. Rev. Lett.* **113** 247004

[89] Brun C, Cren T, Cherkez V, Debontridder F, Pons S, Fokin D, Tringides M C, Bozhko S, Ioffe L B, Altshuler B L and Roditchev D 2014 Remarkable effects of disorder on superconductivity of single atomic layers of lead on silicon *Nat. Phys.* **10** 444-50

[90] Fulde P and Ferrell R A 1964 Superconductivity in a Strong Spin-Exchange Field *Phys. Rev.* **135** A550-A63

[91] Larkin A I and Ovchinnikov Y, N. 1965 INHOMOGENEOUS STATE OF SUPERCONDUCTORS *Sov. Phys. JETP* **20** 762

[92] Matsuda Y and Shimahara H 2007 Fulde-Ferrell-Larkin-Ovchinnikov State in Heavy Fermion Superconductors *J. Phys. Soc. Jpn.* **76** 051005

[93] Bychkov Y A and Rashba E I 1984 PROPERTIES OF A 2D ELECTRON-GAS WITH LIFTED SPECTRAL DEGENERACY *JETP Lett.* **39** 78-81

[94] Sakamoto K, Kim T-H, Kuzumaki T, Müller B, Yamamoto Y, Ohtaka M, Osiecki J R, Miyamoto K, Takeichi Y, Harasawa A, Stolwijk S D, Schmidt A B, Fujii J, Uhrberg R I G, Donath M, Yeom H W and Oda T 2013 Valley spin polarization by using the extraordinary Rashba effect on silicon *Nat. Commun.* **4** 2073

[95] Yaji K, Ohtsubo Y, Hatta S, Okuyama H, Miyamoto K, Okuda T, Kimura A, Namatame H, Taniguchi M and Aruga T 2010 Large Rashba spin splitting of a metallic surface-state band on a

semiconductor surface *Nat. Commun.* **1** 17

[96] Fujimoto S 2007 Electron Correlation and Pairing States in Superconductors without Inversion Symmetry *J. Phys. Soc. Jpn.* **76** 051008

[97] Barzykin V and Gor'kov L P 2002 Inhomogeneous Stripe Phase Revisited for Surface Superconductivity *Phys. Rev. Lett.* **89** 227002

[98] Kaur R P, Agterberg D F and Sigrist M 2005 Helical Vortex Phase in the Noncentrosymmetric CePt3Si *Phys. Rev. Lett.* **94** 137002

[99] Agterberg D F and Kaur R P 2007 Magnetic-field-induced helical and stripe phases in Rashba superconductors *Phys. Rev. B* **75** 064511

[100] Bauer E and Sigrist M eds 2012 *Non-centrosymmetric Superconductors* (Heidelberg: Springer)

[101] Sekihara T, Masutomi R and Okamoto T 2013 Two-Dimensional Superconducting State of Monolayer Pb Films Grown on GaAs(110) in a Strong Parallel Magnetic Field *Phys. Rev. Lett.* **111** 057005

[102] Zhang H-M, Sun Y, Li W, Peng J-P, Song C-L, Xing Y, Zhang Q, Guan J, Li Z, Zhao Y, Ji S, Wang L, He K, Chen X, Gu L, Ling L, Tian M, Li L, Xie X C, Liu J, Yang H, Xue Q-K, Wang J and Ma X 2015 Detection of a Superconducting Phase in a Two-Atom Layer of Hexagonal Ga Film Grown on Semiconducting GaN(0001) *Phys. Rev. Lett.* **114** 107003

[103] Otte H R 2003 *The Physics of Superconductors Vol. I Conventional and High-Tc Superconductors,* ed K H Bennemann and J B Kettemann: Springer-Verlag)

[104] Triscone J M, Fischer O, Brunner O, Antognazza L, Kent A D and Karkut M G 1990 YBa2Cu3O7/PrBa2Cu3O7 superlattices: Properties of Ultrathin superconducting layers separated by insulating layers *Phys. Rev. Lett.* **64** 804-7

[105] Lowndes D H, Norton D P and Budai J D 1990 Superconductivity in nonsymmetric epitaxial YBa2Cu3O7-x/PrBa2Cu3O7-x superlattices: The superconducting behavior of Cu-O bilayers *Phys. Rev. Lett.* **65** 1160-3

[106] Li Q, Xi X X, Wu X D, Inam A, Vadlamannati S, McLean W L, Venkatesan T, Ramesh R, Hwang D M, Martinez J A and Nazar L 1990 Interlayer coupling effect in high-Tc superconductors probed by YBa2Cu3O7-x/PrBa2Cu3O7-x superlattices *Phys. Rev. Lett.* **64** 3086-9

[107] Terashima T, Shimura K, Bando Y, Matsuda Y, Fujiyama A and Komiyama S 1991 Superconductivity of one-unit-cell thick YBa2Cu3O7 thin film *Phys. Rev. Lett.* **67** 1362-5

[108] Wheatley J M, Hsu T C and Anderson P W 1988 Interlayer effects in high-Tc superconductors *Nature* **333** 121-

[109] Gozar A, Logvenov G, Kourkoutis L F, Bollinger A T, Giannuzzi L A, Muller D A and Bozovic I 2008 High-temperature interface superconductivity between metallic and insulating copper oxides *Nature* **455** 782-5

[110] Logvenov G, Gozar A and Bozovic I 2009 High-Temperature Superconductivity in a Single

Copper-Oxygen Plane *Science* **326** 699-702

[111] Smadici S, Lee J C T, Wang S, Abbamonte P, Logvenov G, Gozar A, Cavellin C D and Bozovic I 2009 Superconducting Transition at 38 K in Insulating-Overdoped La2CuO4-La1.64Sr0.36CuO4 Superlattices: Evidence for Interface Electronic Redistribution from Resonant Soft X-Ray Scattering *Phys. Rev. Lett.* **102** 107004

[112] Caviglia A D, Gariglio S, Reyren N, Jaccard D, Schneider T, Gabay M, Thiel S, Hammerl G, Mannhart J and Triscone J M 2008 Electric field control of the LaAlO3/SrTiO3 interface ground state *Nature* **456** 624-7

[113] Leng X, Garcia-Barriocanal J, Bose S, Lee Y and Goldman A M 2011 Electrostatic Control of the Evolution from a Superconducting Phase to an Insulating Phase in Ultrathin YBa2Cu3O7-x Films *Phys. Rev. Lett.* **107** 027001

[114] Zeng S W, Huang Z, Lv W M, Bao N N, Gopinadhan K, Jian L K, Herng T S, Liu Z Q, Zhao Y L, Li C J, Harsan Ma H J, Yang P, Ding J, Venkatesan T and Ariando 2015 Two-dimensional superconductor-insulator quantum phase transitions in an electron-doped cuprate *Phys. Rev. B* **92** 020503

[115] Hetel I, Lemberger T R and Randeria M 2007 Quantum critical behaviour in the superfluid density of strongly underdoped ultrathin copper oxide films *Nat. Phys.* **3** 700-2

[116] Novoselov K S, Jiang D, Schedin F, Booth T J, Khotkevich V V, Morozov S V and Geim A K 2005 Two-dimensional atomic crystals *PNAS* **102** 10451-3

[117] Jiang D, Hu T, You L, Li Q, Li A, Wang H, Mu G, Chen Z, Zhang H, Yu G, Zhu J, Sun Q, Lin C, Xiao H, Xie X and Jiang M 2014 High-Tc superconductivity in ultrathin Bi2Sr2CaCu2O8+x down to half-unit-cell thickness by protection with graphene *Nat. Commun.* **5**

[118] Mannhart J and Schlom D G 2010 Oxide Interfaces-An Opportunity for Electronics *Science* **327** 1607-11

[119] Ohtomo A and Hwang H Y 2004 A high-mobility electron gas at the LaAlO3/SrTiO3 heterointerface *Nature* **427** 423-6

[120] Thiel S, Hammerl G, Schmehl A, Schneider C W and Mannhart J 2006 Tunable Quasi-Two-Dimensional Electron Gases in Oxide Heterostructures *Science* **313** 1942-5

[121] Nakagawa N, Hwang H Y and Muller D A 2006 Why some interfaces cannot be sharp *Nat. Mater.* **5** 204-9

[122] Basletic M, Maurice J L, Carretero C, Herranz G, Copie O, Bibes M, Jacquet E, Bouzehouane K, Fusil S and Barthelemy A 2008 Mapping the spatial distribution of charge carriers in LaAlO3/SrTiO3 heterostructures *Nat. Mater.* **7** 621-5

[123] Koonce C S, Cohen M L, Schooley J F, Hosler W R and Pfeiffer E R 1967 Superconducting Transition Temperatures of Semiconducting SrTiO3 *Phys. Rev.* **163** 380-90

[124] Gabay M and Kapitulnik A 1993 Vortex-antivortex crystallization in thin superconducting and

superfluid films *Phys. Rev. Lett.* **71** 2138-41

[125] Zhang S-C 1993 Vortex-antivortex lattice in superfluid films *Phys. Rev. Lett.* **71** 2142-5

[126] Reyren N, Gariglio S, Caviglia A D, Jaccard D, Schneider T and Triscone J-M 2009 Anisotropy of the superconducting transport properties of the LaAlO3/SrTiO3 interface *Appl. Phys. Lett.* **94** 112506

[127] Sing M, Berner G, Goß K, Müller A, Ruff A, Wetscherek A, Thiel S, Mannhart J, Pauli S A, Schneider C W, Willmott P R, Gorgoi M, Schäfers F and Claessen R 2009 Profiling the Interface Electron Gas of LaAlO3/SrTiO3 Heterostructures with Hard X-Ray Photoelectron Spectroscopy *Phys. Rev. Lett.* **102** 176805

[128] Pentcheva R and Pickett W E 2006 Charge localization or itineracy at LaAlO3/SrTiO3 interfaces: Hole polarons, oxygen vacancies, and mobile electrons *Phys. Rev. B* **74** 035112

[129] Brinkman A, Huijben M, van Zalk M, Huijben J, Zeitler U, Maan J C, van der Wiel W G, Rijnders G, Blank D H A and Hilgenkamp H 2007 Magnetic effects at the interface between non-magnetic oxides *Nat. Mater.* **6** 493-6

[130] Dikin D A, Mehta M, Bark C W, Folkman C M, Eom C B and Chandrasekhar V 2011 Coexistence of Superconductivity and Ferromagnetism in Two Dimensions *Phys. Rev. Lett.* **107** 056802

[131] Li L, Richter C, Mannhart J and Ashoori R C 2011 Coexistence of magnetic order and two-dimensional superconductivity at LaAlO3/SrTiO3 interfaces *Nat. Phys.* **7** 762-6

[132] Bert J A, Kalisky B, Bell C, Kim M, Hikita Y, Hwang H Y and Moler K A 2011 Direct imaging of the coexistence of ferromagnetism and superconductivity at the LaAlO3/SrTiO3 interface *Nat. Phys.* **7** 767-71

[133] Ben Shalom M, Sachs M, Rakhmilevitch D, Palevski A and Dagan Y 2010 Tuning Spin-Orbit Coupling and Superconductivity at the SrTiO3/LaAlO3 Interface: A Magnetotransport Study *Phys. Rev. Lett.* **104** 126802

[134] Ahn C H, Triscone J M and Mannhart J 2003 Electric field effect in correlated oxide systems *Nature* **424** 1015-8

[135] Dhoot A S, Yuen J D, Heeney M, McCulloch I, Moses D and Heeger A J 2006 Beyond the metal-insulator transition in polymer electrolyte gated polymer field-effect transistors *PNAS* **103** 11834-7

[136] Panzer M J and Frisbie C D 2006 High Carrier Density and Metallic Conductivity in Poly(3-hexylthiophene) Achieved by Electrostatic Charge Injection *Adv. Func. Mater.* **16** 1051-6

[137] Misra R, McCarthy M and Hebard A F 2007 Electric field gating with ionic liquids *Appl. Phys. Lett.* **90** 052905

[138] Yuan H, Shimotani H, Tsukazaki A, Ohtomo A, Kawasaki M and Iwasa Y 2009 High-Density Carrier Accumulation in ZnO Field-Effect Transistors Gated by Electric Double Layers of Ionic Liquids *Adv. Func. Mater.* **19** 1046-53

[139] Ueno K, Nakamura S, Shimotani H, Yuan H T, Kimura N, Nojima T, Aoki H, Iwasa Y and Kawasaki M 2011 Discovery of superconductivity in KTaO3 by electrostatic carrier doping *Nat. Nanotechnol.* **6** 408-12

[140] Lee Y, Clement C, Hellerstedt J, Kinney J, Kinnischtzke L, Leng X, Snyder S D and Goldman A M 2011 Phase Diagram of Electrostatically Doped SrTiO3 *Phys. Rev. Lett.* **106** 136809

[141] Li L J, O'Farrell E C T, Loh K P, Eda G, Özyilmaz B and Castro Neto A H 2016 Controlling many-body states by the electric-field effect in a two-dimensional material *Nature* **529** 185-9

[142] Saito Y, Kasahara Y, Ye J, Iwasa Y and Nojima T 2015 Metallic ground state in an ion-gated two-dimensional superconductor *Science* **350** 409-13

[143] Tinkham M 1963 Effect of Fluxoid Quantization on Transitions of Superconducting Films *Phys. Rev.* **129** 2413-22

[144] Shimshoni E, Auerbach A and Kapitulnik A 1998 Transport through Quantum Melts *Phys. Rev. Lett.* **80** 3352-5

[145] Taniguchi K, Matsumoto A, Shimotani H and Takagi H 2012 Electric-field-induced superconductivity at 9.4 K in a layered transition metal disulphide MoS2 *Appl. Phys. Lett.* **101** 042603

[146] Ye J T, Zhang Y J, Akashi R, Bahramy M S, Arita R and Iwasa Y 2012 Superconducting Dome in a Gate-Tuned Band Insulator *Science* **338** 1193-6

[147] Saito Y, Nakamura Y, Bahramy M S, Kohama Y, Ye J, Kasahara Y, Nakagawa Y, Onga M, Tokunaga M, Nojima T, Yanase Y and Iwasa Y 2016 Superconductivity protected by spin-valley locking in ion-gated MoS2 *Nat. Phys.* **12** 144-9

[148] Molina-Sánchez A, Sangalli D, Hummer K, Marini A and Wirtz L 2013 Effect of spin-orbit interaction on the optical spectra of single-layer, double-layer, and bulk MoS2 *Phys. Rev. B* **88** 045412

[149] Xiao D, Liu G-B, Feng W, Xu X and Yao W 2012 Coupled Spin and Valley Physics in Monolayers of MoS2 and Other Group-VI Dichalcogenides *Phys. Rev. Lett.* **108** 196802

[150] Novoselov K S, Geim A K, Morozov S V, Jiang D, Katsnelson M I, Grigorieva I V, Dubonos S V and Firsov A A 2005 Two-dimensional gas of massless Dirac fermions in graphene *Nature* **438** 197-200

[151] Zhang Y, Tan Y-W, Stormer H L and Kim P 2005 Experimental observation of the quantum Hall effect and Berry's phase in graphene *Nature* **438** 201-4

[152] Heersche H B, Jarillo-Herrero P, Oostinga J B, Vandersypen L M K and Morpurgo A F 2007 Bipolar supercurrent in graphene *Nature* **446** 56-9

[153] Hannay N B, Geballe T H, Matthias B T, Andres K, Schmidt P and MacNair D 1965 Superconductivity in Graphitic Compounds *Phys. Rev. Lett.* **14** 225-6

[154] Emery N, Hérold C, d'Astuto M, Garcia V, Bellin C, Marêché J F, Lagrange P and Loupias G 2005

Superconductivity of Bulk CaC6 *Phys. Rev. Lett.* **95** 087003

[155] Weller T E, Ellerby M, Saxena S S, Smith R P and Skipper N T 2005 Superconductivity in the intercalated graphite compounds C6Yb and C6Ca *Nat. Phys.* **1** 39-41

[156] Xue M, Chen G, Yang H, Zhu Y, Wang D, He J and Cao T 2012 Superconductivity in Potassium-Doped Few-Layer Graphene *Journal of the American Chemical Society* **134** 6536-9

[157] Tiwari A P, Shin S, Hwang E, Jung S-G, Park T and Lee H Superconductivity at 7.4 K in Few Layer Graphene by Li-intercalation *arXiv:1508.06360*

[158] Li K, Feng X, Zhang W, Ou Y, Chen L, He K, Wang L-L, Guo L, Liu G, Xue Q-K and Ma X 2013 Superconductivity in Ca-intercalated epitaxial graphene on silicon carbide *Appl. Phys. Lett.* **103** 062601

[159] Kanetani K, Sugawara K, Sato T, Shimizu R, Iwaya K, Hitosugi T and Takahashi T 2012 Ca intercalated bilayer graphene as a thinnest limit of superconducting C6Ca *PNAS* **109** 19610-3

[160] Ichinokura S, Sugawara K, Takayama A, Takahashi T and Hasegawa S 2016 Superconducting Calcium-Intercalated Bilayer Graphene *Acs Nano* **10** 2761-5

[161] Ludbrook B M, Levy G, Nigge P, Zonno M, Schneider M, Dvorak D J, Veenstra C N, Zhdanovich S, Wong D, Dosanjh P, Straßer C, Stöhr A, Forti S, Ast C R, Starke U and Damascelli A 2015 Evidence for superconductivity in Li-decorated monolayer graphene *PNAS* **112** 11795-9

[162] Profeta G, Calandra M and Mauri F 2012 Phonon-mediated superconductivity in graphene by lithium deposition *Nat. Phys.* **8** 131-4

[163] Chapman J, Su Y, Howard C A, Kundys D, Grigorenko A, Guinea F, Geim A K, Grigorieva I V and Nair R R Superconductivity in Ca-doped graphene *arXiv: 1508.06931*

[164] Xu X, Yao W, Xiao D and Heinz T F 2014 Spin and pseudospins in layered transition metal dichalcogenides *Nat. Phys.* **10** 343-50

[165] Frindt R F 1972 Superconductivity in Ultrathin $NbSe_{2}$ Layers *Phys. Rev. Lett.* **28** 299-301

[166] Mohammed S E-B, Daniel W, Saverio R, Geetha B, Don Mck P and Simon J B 2013 Superconductivity in two-dimensional NbSe2 field effect transistors *Superconductor Science and Technology* **26** 125020

[167] Cao Y, Mishchenko A, Yu G L, Khestanova E, Rooney A P, Prestat E, Kretinin A V, Blake P, Shalom M B, Woods C, Chapman J, Balakrishnan G, Grigorieva I V, Novoselov K S, Piot B A, Potemski M, Watanabe K, Taniguchi T, Haigh S J, Geim A K and Gorbachev R V 2015 Quality Heterostructures from Two-Dimensional Crystals Unstable in Air by Their Assembly in Inert Atmosphere *Nano Lett.* **15** 4914-21

[168] Xi X, Zhao L, Wang Z, Berger H, Forró L, Shan J and Mak K F 2015 Strongly enhanced charge-density-wave order in monolayer NbSe2 *Nat. Nanotechnol.* **10** 765-9

[169] Tsen A W, Hunt B, Kim Y D, Yuan Z J, Jia S, Cava R J, Hone J, Kim P, Dean C R and Pasupathy

[169] A N 2016 Nature of the quantum metal in a two-dimensional crystalline superconductor *Nat. Phys.* **12** 208-12

[170] Das D and Doniach S 1999 Existence of a Bose metal at T=0 *Phys. Rev. B* **60** 1261-75

[171] Dalidovich D and Phillips P 2002 Phase Glass is a Bose Metal: A New Conducting State in Two Dimensions *Phys. Rev. Lett.* **89** 027001

[172] Hsu F-C, Luo J-Y, Yeh K-W, Chen T-K, Huang T-W, Wu P M, Lee Y-C, Huang Y-L, Chu Y-Y, Yan D-C and Wu M-K 2008 Superconductivity in the PbO-type structure α-FeSe *PNAS* **105** 14262-4

[173] Medvedev S, McQueen T M, Troyan I A, Palasyuk T, Eremets M I, Cava R J, Naghavi S, Casper F, Ksenofontov V, Wortmann G and Felser C 2009 Electronic and magnetic phase diagram of β-Fe1.01Se with superconductivity at 36.7 K under pressure *Nat. Mater.* **8** 630-3

[174] Mizuguchi Y and Takano Y 2010 Review of Fe Chalcogenides as the Simplest Fe-Based Superconductor *J. Phys. Soc. Jpn.* **79** 102001

[175] Wu G, Xie Y L, Chen H, Zhong M, Liu R H, Shi B C, Li Q J, Wang X F, Wu T, Yan Y J, Ying J J and Chen X H 2009 Superconductivity at 56 K in samarium-doped SrFeAsF *J. of Phys.: Condens. Matter* **21** 142203

[176] Song C-L, Wang Y-L, Jiang Y-P, Li Z, Wang L, He K, Chen X, Ma X-C and Xue Q-K 2011 Molecular-beam epitaxy and robust superconductivity of stoichiometric FeSe crystalline films on bilayer graphene *Phys. Rev. B* **84** 020503

[177] Sun Y, Zhang W, Xing Y, Li F, Zhao Y, Xia Z, Wang L, Ma X, Xue Q-K and Wang J 2014 High temperature superconducting FeSe films on SrTiO3 substrates *Scientific Reports* **4** 6040

[178] Miyata Y, Nakayama K, Sugawara K, Sato T and Takahashi T 2015 High-temperature superconductivity in potassium-coated multilayer FeSe thin films *Nat. Mater.* **14** 775-9

[179] He S, He J, Zhang W, Zhao L, Liu D, Liu X, Mou D, Ou Y-B, Wang Q-Y, Li Z, Wang L, Peng Y, Liu Y, Chen C, Yu L, Liu G, Dong X, Zhang J, Chen C, Xu Z, Chen X, Ma X, Xue Q and Zhou X J 2013 Phase diagram and electronic indication of high-temperature superconductivity at 65 K in single-layer FeSe films *Nat. Mater.* **12** 605-10

[180] Tan S, Zhang Y, Xia M, Ye Z, Chen F, Xie X, Peng R, Xu D, Fan Q, Xu H, Jiang J, Zhang T, Lai X, Xiang T, Hu J, Xie B and Feng D 2013 Interface-induced superconductivity and strain-dependent spin density waves in FeSe/SrTiO3 thin films *Nat. Mater.* **12** 634-40

[181] Liu D, Zhang W, Mou D, He J, Ou Y-B, Wang Q-Y, Li Z, Wang L, Zhao L, He S, Peng Y, Liu X, Chen C, Yu L, Liu G, Dong X, Zhang J, Chen C, Xu Z, Hu J, Chen X, Ma X, Xue Q and Zhou X J 2012 Electronic origin of high-temperature superconductivity in single-layer FeSe superconductor *Nat. Commun.* **3** 931

[182] Lee J J, Schmitt F T, Moore R G, Johnston S, Cui Y T, Li W, Yi M, Liu Z K, Hashimoto M, Zhang Y, Lu D H, Devereaux T P, Lee D H and Shen Z X 2014 Interfacial mode coupling as the origin of the enhancement of Tc in FeSe films on SrTiO3 *Nature* **515** 245-8

[183] Song C-L, Zhang H-M, Zhong Y, Hu X-P, Ji S-H, Wang L, He K, Ma X-C and Xue Q-K 2016 Observation of Double-Dome Superconductivity in Potassium-Doped FeSe Thin Films *Phys. Rev. Lett.* **116** 157001

[184] Shiogai J, Ito Y, Mitsuhashi T, Nojima T and Tsukazaki A 2016 Electric-field-induced superconductivity in electrochemically etched ultrathin FeSe films on SrTiO3 and MgO *Nat. Phys.* **12** 42-6

[185] Tang C, Liu C, Zhou G, Li F, Ding H, Li Z, Zhang D, Li Z, Song C, Ji S, He K, Wang L, Ma X and Xue Q-K 2016 Interface-enhanced electron-phonon coupling and high-temperature superconductivity in potassium-coated ultrathin FeSe films on SrTiO3 *Phys. Rev. B* **93** 020507

[186] Zhang W H, Liu X, Wen C H P, Peng R, Tan S Y, Xie B P, Zhang T and Feng D L 2016 Effects of Surface Electron Doping and Substrate on the Superconductivity of Epitaxial FeSe Films *Nano Lett.* **16** 1969-73

[187] Tang C, Zhang D, Zang Y, Liu C, Zhou G, Li Z, Zheng C, Hu X, Song C, Ji S, He K, Chen X, Wang L, Ma X and Xue Q-K 2015 Superconductivity dichotomy in K-coated single and double unit cell FeSe films on SrTiO3 *Phys. Rev. B* **92** 180507

[188] Pfleiderer C 2009 Superconducting phases of f-electron compounds *Rev. Mod. Phys.* **81** 1551-624

[189] Ishiguro T, Yamaji K and Saito G 1998 *Organic Superconductors 2nd ed.* (Heidelberg: Springer)

[190] Kushch N D, Dyachenko O A, Lyubovskii R B, Pesotskii S I, Kartsovnik M V, Kovalev A E, Cassoux P and Kobayashi H 1997 New BETS salts based on magnetic (CuCl3, FeCl4) and non-magnetic (GaCl4) anions *Adv. Mater. Opt. Electr.* **7** 57-60

[191] Uji S, Terashima T, Terakura C, Yakabe T, Terai Y, Yasuzuka S, Imanaka Y, Tokumoto M, Kobayashi A, Sakai F, Tanaka H, Kobayashi H, Balicas L and S. brooks J 2003 Global Phase Diagram of the Magnetic Field-Induced Organic Superconductors λ-(BETS)2FexGa1-xCl4 *J. Phys. Soc. Jpn.* **72** 369-73

[192] Lang M and Müller J 2004 *The Physics of Superconductors Vol. II Superconductivity in Nanostructures, High-Tc and Novel Superconductors, Organic Superconductors* ed K H Bennemann and J B Ketterson (Heidelberg: Springer)

[193] Hassanien A, Zhou B, Tanaka H, Miyazaki A, Tokumoto M, Kobayashi A, Zupanič E and Muševič I 2015 Epitaxial growth of insulating and superconducting monolayers of (BETS)2GaCl4 on Ag(111) *Phys. Status Solidi B* **252** 2574-9

[194] Uji S, Shinagawa H, Terashima T, Yakabe T, Terai Y, Tokumoto M, Kobayashi A, Tanaka H and Kobayashi H 2001 Magnetic-field-induced superconductivity in a two-dimensional organic conductor *Nature* **410** 908-10

[195] Goh S, Mizukami Y, Shishido H, Watanabe D, Yasumoto S, Shimozawa M, Yamashita M, Terashima T, Yanase Y, Shibauchi T, Buzdin A and Matsuda Y 2012 Anomalous Upper Critical Field in CeCoIn5/YbCoIn5 Superlattices with a Rashba-Type Heavy Fermion Interface *Phys. Rev.*

*Lett.* **109** 157006

[196] Shimozawa M, Goh S K, Endo R, Kobayashi R, Watashige T, Mizukami Y, Ikeda H, Shishido H, Yanase Y, Terashima T, Shibauchi T and Matsuda Y 2014 Controllable Rashba Spin-Orbit Interaction in Artificially Engineered Superlattices Involving the Heavy-Fermion Superconductor CeCoIn5 *Phys. Rev. Lett.* **112** 156404

[197] Sau J D, Lutchyn R M, Tewari S and Das Sarma S 2010 Generic New Platform for Topological Quantum Computation Using Semiconductor Heterostructures *Phys. Rev. Lett.* **104** 040502

[198] Liang Q-F, Wang Z and Hu X 2012 Manipulation of Majorana fermions by point-like gate voltage in the Vortex state of a topological superconductor *Europhys. Lett.* **99** 50004

[199] Ivanov D A 2001 Non-Abelian Statistics of Half-Quantum Vortices in p-Wave Superconductors *Phys. Rev. Lett.* **86** 268-71

[200] Wang M X, Liu C H, Xu J P, Yang F, Miao L, Yao M Y, Gao C L, Shen C Y, Ma X C, Chen X, Xu Z A, Liu Y, Zhang S C, Qian D, Jia J F and Xue Q K 2012 The Coexistence of Superconductivity and Topological Order in the Bi2Se3 Thin Films *Science* **336** 52-5

[201] Xu J-P, Wang M-X, Liu Z L, Ge J-F, Yang X, Liu C, Xu Z A, Guan D, Gao C L, Qian D, Liu Y, Wang Q-H, Zhang F-C, Xue Q-K and Jia J-F 2015 Experimental Detection of a Majorana Mode in the core of a Magnetic Vortex inside a Topological Insulator-Superconductor Bi2Te3/NbSe2 Heterostructure *Phys. Rev. Lett.* **114** 017001

**Figure and table captions**

**Figure 1.** Schematic chart categorizing materials dealt in the present paper. Note that the positioning of each topic is only qualitative and subjective.

**Figure 2.** Disorder-induced S-I transition of quench-condensed Bi films. The evolution of the temperature dependence of the sheet resistance $R(T)$ is displayed. The film thickness ranges from 4.36 to 74.27 Å. Reprinted figure with permission from Haviland D B et al. 1989 *Phys. Rev. Lett.* **62** 2180 [34]. Copyright 1989 by the American Physical Society. http://dx.doi.org/10.1103/PhysRevLett.62.2180.

**Figure 3.** Experimental identification of the KTB transition. (a) Plot for $T > T_c$ of the logarithm of the resistance as a function of $(T-T_c)^{-1/2}$ where $T_c = 1.903$ K. (b) Plot on logarithmic axes of the *I-V* characteristics taken at thirteen successively lower temperatures ranging from 1.939 K to 1.460 K. Reprinted figure with permission from Hebard A and Fiory A 1983 *Phys. Rev. Lett.* **50** 1603 [48]. Copyright 1983 by the American Physical Society. http://dx.doi.org/10.1103/PhysRevLett.50.1603.

**Figure 4.** Quantum oscillation of $T_c$ in ultrathin Pb films. (a) Observation of quantum well states in Pb films through *in situ* photoemission spectroscopy at 75 K. (b) $T_c$ (black solid dots) and the density of states $N(E_F) \propto -\sigma(dH_{c2}/dT)_{T_c}$ (red stars) as a function of Pb film thickness, demonstrating an oscillatory behavior in both $T_c$ and $N(E_F)$. From Guo Y et al, 2004 *Science* **306** 1915 [19]. Reprinted with permission from AAAS.

**Figure 5.** STM images and the corresponding d.c. magnetization loops of ultrathin Pb films. (a)(b) Quantum growth defects in 7-ML Pb film consist of either two-atom layer deep voids (a) or two-atom layer tall mesas (b), in films with a small shortage or excess of Pb, respectively. The image size is 700×700 nm$^2$ for both (a) and (b). (c)(d) The corresponding d.c. magnetic response of these films. Quantum voids produce 'hard' hysteresis loops (c), whereas quantum mesas produce 'soft' hysteresis loops (d). Reprinted by permission from Macmillan Publishers Ltd: Özer M M et al., 2006 *Nat. Phys.* **2** 173 [17], copyright 2006. http://dx.doi.org/10.1038/nphys244.

**Figure 6.** STM/STS observation of vortices in superconducting Pb islands and the proximity regions. (a) Topographic STM image of Pb islands surrounded by an atomically thin two-dimensional Pb wetting layer. The image size is 1000×1000 nm$^2$. (b) Colour-coded *dI/dV* mapping at zero-bias voltage of the same sample area acquired at $\mu_0 H_\perp = 60$ mT. (c) Characteristic local *dI/dV* spectra measured at the locations indicated in (a). (d) Zoom of the J1 SNS proximity junction together with corresponding

tunneling spectra at three locations at $\mu_0 H_\perp$ = 60 mT. Reprinted by permission from Macmillan Publishers Ltd: Roditchev D, et al 2015 *Nat. Phys.* **11** 332 [64], copyright 2015. http://dx.doi.org/10.1038/nphys3240.

**Figure 7.** Atomic structural models (a)-(c) and STM images (d)-(f) of metal-induced silicon surface reconstructions: Si(111)-SIC-Pb (a)(d), Si(111)-(√7×√3)-Pb (b)(e), and Si(111)-(√7×√3)-In (c)(f). Note that the Si(111)-(√7×√3)-In surface consists of 2 ML of indium according to a recent atomic structure model [Park J and Kang M 2012 *Phys. Rev. Lett.* **109** 166102]. Reprinted by permission from Macmillan Publishers Ltd: Zhang T et al., 2010 *Nat. Phys.* **6** 104 [18], copyright 2010. http://dx.doi.org/doi:10.1038/nphys1499.

**Figure 8.** Superconductivity of Si(111)-(√7×√3)-In. (a) $dI/dV$ spectra measured on the (√7×√3)-In phase as a function of temperature using a superconducting Nb tip. (b) Superconducting energy gap Δ as a function of temperature (open circles) and the fitting by the BCS gap function (solid curve). (c) Temperature dependence of zero bias resistances measured through electron transport measurement. (d) Temperature dependences of critical current $I_c$ (green squares) and retrapping current $I_r$ (pink squares). The blue solid and red dotted lines show theoretical fits. (a)(b) Reprinted by permission from Macmillan Publishers Ltd: Zhang T et al., 2010 *Nat. Phys.* **6** 104 [18], copyright 2010. http://dx.doi.org/doi:10.1038/nphys1499. (c)(d) Reprinted figure from Uchihashi T et al., 2011 *Phys. Rev. Lett.* **107** 207001 [83]. Copyright 2011 by the American Physical Society. http://dx.doi.org/10.1103/PhysRevLett.107.207001.

**Figure 9.** Schematic illustrations of electrons and superconductivity in a 2D system. (a) Effects of the out-of-plane ($H_\perp$) and in-plane ($H_{//}$) magnetic fields on the electron orbital motion of a Cooper pair. (b) Spin splitting and spin-momentum locking due to the Rashba effect. (c) Shifts of the Rashba-split Fermi surfaces caused by application of $H_{//}$. The arrows on the circles indicate the directions and the sizes of spin magnetizations. (d) Cooper pair formation on each Fermi circle depicted in (c) and the unidirectionally induced supercurrent $J$.

**Figure 10.** Rashba effects detected for atomic-layer superconductors on semiconductor surfaces. (a) Experimental (upper panel) and calculated (lower panel) Fermi contours of the 2D Tl-Pb compound on Si(111) shown in the √3×√3 surface Brillouin zone. (b) $H_{//}$ dependence of $\Delta T_c$ ($\equiv T_c - T_{c0}$) for Pb films on a GaAs substrate with $n$ = 9.4 and 11.3 nm$^{-2}$. The dashed lines are the best parabolic fits. (a) Reprinted figure with permission from Matetskiy A V et al., 2015 *Phys. Rev. Lett.* **115** 147003 [52]. Copyright 2015 by the American Physical Society. http://dx.doi.org/10.1103/PhysRevLett.115.147003. (b) Reprinted figure with permission from Sekihara T et al., 2013 *Phys. Rev. Lett.* **111** 057005 [101]. Copyright 2013 by the American Physical Society. http://dx.doi.org/10.1103/PhysRevLett.111.057005.

**Figure 11.** Superconductivity of the $La_{2-x}Sr_xCuO_4/La_2CuO_4$ interface. (a) Schematic illustration of 6-UC-thick $La_{2-x}Sr_xCuO_4/La_2CuO_4$ with δ-doping of Zn atoms at the $N = 2$ $CuO_2$ plane. (b) Effect of δ-doping with Zn on superconductivity in $La_{2-x}Sr_xCuO_4/La_2CuO_4$ bilayer films grown by MBE. From Logvenov G et al., 2009 *Science* **326** 699 [110]. Reprinted with permission from AAAS.

**Figure 12.** Electric-field induced S-I transition of 1-UC-thick $La_{2-x}Sr_xCuO_4$. (a) Temperature dependence of normalized resistance $r = R_\square(x,T)/R_Q$ of 1-UC-thick $La_{2-x}Sr_xCuO_4$ which is an initially heavily underdoped and insulating film. (b) Scaling of the data with respect to a single variable, with $zv = 1.5$. The discrete groups of points of collapse accurately onto a two-valued function. Reprinted by permission from Macmillan Publishers Ltd: Bollinger A T et al., 2011 *Nature* **472** 458 [20], copyright 2011. http://dx.doi.org/doi:10.1038/nature09998.

**Figure 13.** Superconductivity of the $LaAlO_3/SrTiO_3$ interface. (a) Schematic illustration of the $LaAlO_3/SrTiO_3$ interface. (b) Scanning transmission electron microscopy (STEM) image of a $LaAlO_3/SrTiO_3$ heterostructure. A 15-UC-thick $LaAlO_3$ film was grown on $SrTiO_3$ showing a coherent interface. (c) Dependence of the sheet resistance on $T$ of the 8-UC and 15-UC samples. (d) Temperature dependence of the out-of-plane upper critical field $H_{c2\perp}$ of the two samples. (e) $V$-$I$ curves on a logarithmic scale at different temperatures. The two long black lines correspond to $V = RI$ and $V \sim I^3$ dependencies. (a) Reprinted by permission from Macmillan Publishers Ltd: Ohtomo A and Hwang H Y 2004 *Nature* **427** 423 [119], copyright 2004. http://dx.doi.org/10.1038/nature02308. (b)-(e) From Reyren N et al., 2007 *Science* **317** 1196 [21]. Reprinted with permission from AAAS.

**Figure 14.** Coexistence of superconductivity and magnetism in the $LaAlO_3/SrTiO_3$ interface. (a) Magnetometry image mapping the ferromagnetic order. (b) Susceptometry image mapping the superfluid density at 40 mK. (c) The temperature dependence of the susceptibility taken at the two positions indicated in (b). Reprinted by permission from Macmillan Publishers Ltd: Bert J A et al. 2011 *Nat. Phys.* **7** 767 [132], copyright 2011. http://dx.doi.org/10.1038/nphys2079.

**Figure 15.** Electric-field-induced Superconductivity of ZrNCl. (a) Schematic of charge accumulation by an EDL formed at an interface between ionic liquid and solid. (b) Ball-and-stick model of a ZrNCl single crystal. (c) Channel sheet conductance $\sigma_s$, sheet carrier density $n_{2D}$ and Hall mobility $\mu_H$ of the ZrNCl EDL transistor modulated by the gate voltage $V_G$ from 0 to 4.5 V at 220 K. (d) Temperature

dependence of sheet resistance $R_s$ showing superconductivity transitions at different gate voltages, $V_G$. Reprinted by permission from Macmillan Publishers Ltd: Ye J T et al., 2010 *Nat. Mater.* **9** 125 [24], copyright 2010. http://dx.doi.org/10.1038/nmat2587.

**Figure 16.** Superconductivity of ZrNCl in the 2D limit. (a) Angular dependence of the upper critical fields $\mu_0 H_{c2}(\theta)$. Note that the definition of the magnetic field angle $\theta$ is different from that in the text. (b) Arrhenius plot of the sheet resistance of a ZrNCl-EDLT at $V_G$ = 6.5 V for different magnetic fields perpendicular to the surface of ZrNCl. (c) Vortex phase diagram of the ZrNCl-EDLT. From Saito Y et al., 2015 *Science* **350** 409 [142]. Reprinted with permission from AAAS.

**Figure 17.** Spin-valley locking and Ising superconductivity. (a) Spin-split conduction-band electron pockets near the *K* and *K'* points in the hexagonal Brillouin zone of monolayer $MoS_2$. (b) Side view (left) and top view (right) of the four outermost layers in a multilayered $MoS_2$ flake. (c) Temperature dependence of $B_{c2}$ (= $\mu_0 H_{c2}$) for superconducting states with different $T_c$. (d) $B_{c2}$ normalized by the Pauli limit $B_p$ (= $\mu_0 H_p$) as a function of reduced temperature $T/T_c$, including superconducting states from alkali-doped bulk phases and gated-induced phases. From Lu J M et al., 2015 *Science* **350** 1353 [54]. Reprinted with permission from AAAS.

**Figure 18.** Superconductivity of Ca-intercalated bilayer graphene. (a) Schematic view of crystal structure of $C_6CaC_6$ on SiC. (b) Band dispersions at K point of pristine bilayer graphene obtained by ARPES. (c) RHEED pattern showing √3×√3 R30° spots and streaks. (d) Schematic of four-point-probe measurement setup. (e) Comparison of temperature dependence of sheet resistance $R_{sheet}$ between $C_6LiC_6$ and $C_6CaC_6$. Reprinted from Ichinokura S et al., 2016 *Acs Nano* **10** 2761 [160] under an ACS AuthorChoice License. http://dx.doi.org/10.1021/acsnano.5b07848.

**Figure 19.** Superconductivity of 1-UC-thick $NbSe_2$ layer. (a) Schematic of the heterostructure assembly process. BN/graphite (G) on a polymer stamp (PDMS) is used to electrically contact and encapsulate $NbSe_2$ in an inert atmosphere. (b) Optical images of the heterostructure before device fabrication. The scale bar is 5 μm. (c) Arrhenius plot of resistance for several magnetic fields. (d) Full *H*–*T* phase diagram of the bilayer $NbSe_2$ device. Reprinted by permission from Macmillan Publishers Ltd: Tsen A W et al., 2016 *Nat. Phys.* **12** 208 [169], copyright 2106. http://dx.doi.org/10.1038/nphys3579.

**Figure 20.** Superconductivity of 1-UC-thick FeSe on $SrTiO_3$(001). (a) Schematic crystal structure of bulk FeSe. (b)-(e) Experimental results on 1-UC-thick FeSe film on $SrTiO_3$(001). (b) Atomically

resolved STM topography of the surface. (c) Tunneling spectrum taken at 4.2 K revealing the appearance of superconducting energy gap. (d) Zero-bias $dI/dV$ mapping of the vortex state at 4.2 K. The image size is 10.6 nm×10.6 nm. (e) Temperature dependence of the local resistance measured using a micro four-point-probe technique. (a) Reprinted from Hsu F-C, et al., 2008 *PNAS* **105** 14262 [172]. Copyright 2008 National Academy of Sciences. (b)-(d) Reprinted by permission from Wang Q-Y et al., 2012 *Chin. Phys. Lett.* **29** 037402 [28]. (e) Reprinted by permission from Macmillan Publishers Ltd: Ge J-F et al., 2015 *Nat. Mater.* **14** 285 [29], copyright 2015. http://dx.doi.org/10.1038/nmat4153.

**Figure 21.** Surface K-doping of FeSe mono- and multi-layer. (a)-(c) Schematic views of a 1 ML FeSe film (a), 3 ML FeSe film (b), and K-doped 3 ML FeSe film (c) grown on a SrTiO3 substrate. (e)-(f) ARPES intensity mapping at $T = 30$ K plotted as a function of a two-dimensional wavevector, corresponding to (a)-(c). (g) Electronic phase diagrams for different film thickness. Experimentally determined superconducting-gap sizes for 1, 3 and 20 ML films were plotted as a function of electron doping (red filled circles). Reprinted by permission from Macmillan Publishers Ltd: Reprinted from Miyata Y et al., 2015 *Nat. Mater.* **14** 775 [178], copyright 2015. http://dx.doi.org/10.1038/nmat4302.

**Figure 22.** Superconductivity of monolayer organic conductor λ-(BETS)$_2$GaCl$_4$. (a) STM image of a λ-(BETS)$_2$GaCl$_4$ molecular layer, revealing a double stacked BETS row, together with an unfinished packing of the first layer of BETS and GaCl$_4$ at the edge. (b) Model illustrating the molecular packing on Ag(111). (c) $dI/dV$ curve showing a superconducting gap with a corresponding fit. Reprinted by permission from Macmillan Publishers Ltd: Clark K et al., S W 2010 *Nat. Nanotechnol.* **5** 261 [30], copyright 2010. http://dx.doi.org/10.1038/nnano.2010.41.

**Figure 23.** Superconductivity of CeCoIn$_5$($n$)/YbCoIn$_5$($m$) superlattices. (a) Temperature dependence of electrical resistivity $\rho(T)$ for $n = 1, 2, 3, 5, 7$ and 9, compared with those of 300-nm-thick CeCoIn$_5$ and YbCoIn$_5$ single-crystalline thin films. (b) Normalized $H_{c2}(\theta)$ of the $n = 3$ superlattice at 0.16 K and 0.8 K and of the $n = 4$ superlattice at 0.25 K and 0.9 K. (a) Reprinted by permission from Macmillan Publishers Ltd: Mizukami Y et al., 2011 *Nat. Phys.* **7** 849 [31], copyright 2011. http://dx.doi.org/10.1038/nphys2112. (b) Reprinted figure with permission from Goh S et al., 2012 *Phys. Rev. Lett.* **109** 157006 [195]. Copyright 2012 by the American Physical Society. http://dx.doi.org/10.1103/PhysRevLett.109.157006.

**Table 1.** List of the categories and attributes of representative 2D superconductors with material type, form, thickness ($d$), superconducting transition temperature ($T_c$), $T_c$ of the bulk counterpart, GL coherence length at zero temperature ($\xi_{GL}(0)$), out-of-plane and in-plane upper critical magnetic fields

($\mu_0H_{c2\perp}(0)$, $\mu_0H_{c2//}(0)$). Thickness $d$ is defined as a distance between parallel atomic planes that the relevant 2D material would take in a bulk form. Regarding $T_c$, $T_c^{mid}$ was determined from the middle point of the resistive phase transition while $T_c^{gap}$ the opening of the superconducting energy gap. In some clear cases, $\mu_0H_{c2\perp}(0)$ values were calculated from $\xi_{GL}(0)$ using the relation $\mu_0H_{c2}(0) = \Phi_0/2\pi\xi_{GL}(0)^2$ by the author ($\Phi_0 = h/2e$: magnetic flux quantum). Note that bulk $\xi_{GL}(0)$ values are shown for some materials.

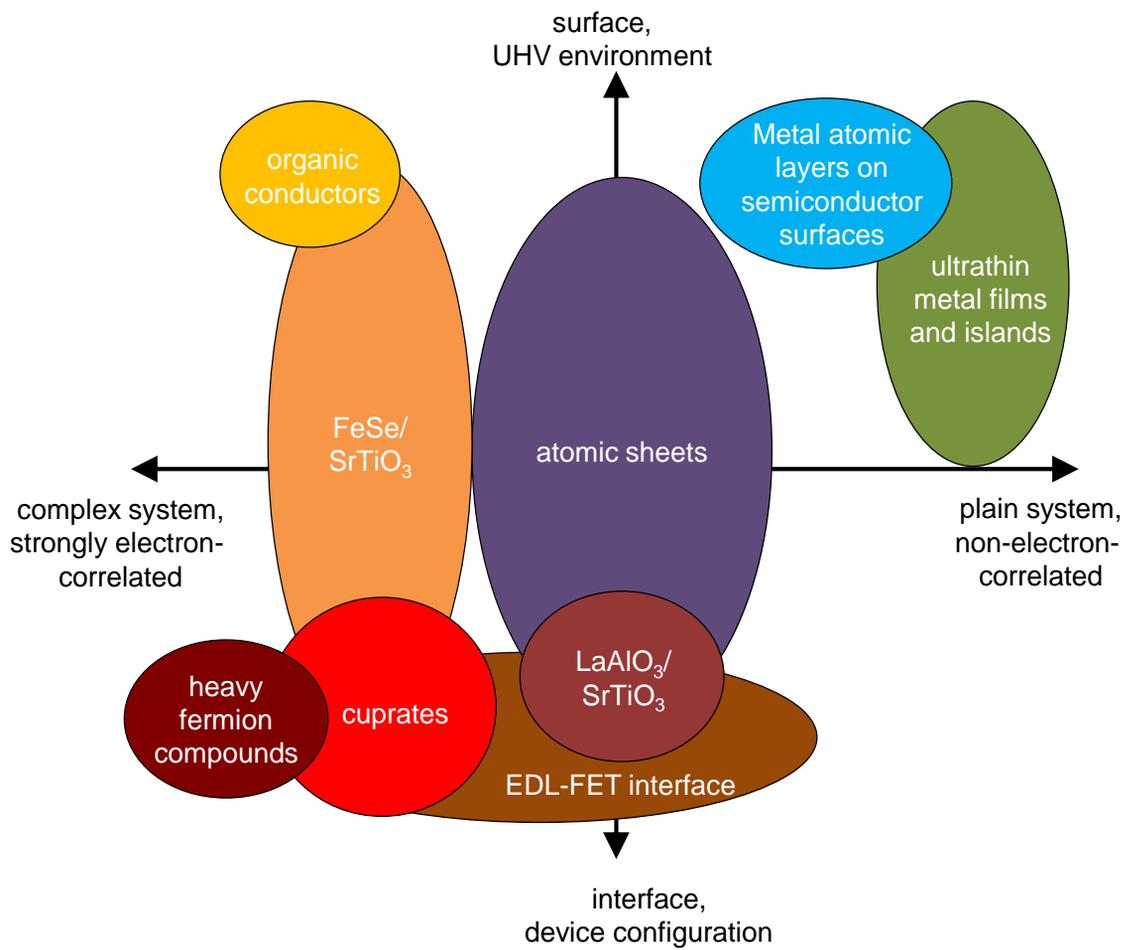

Figure 1

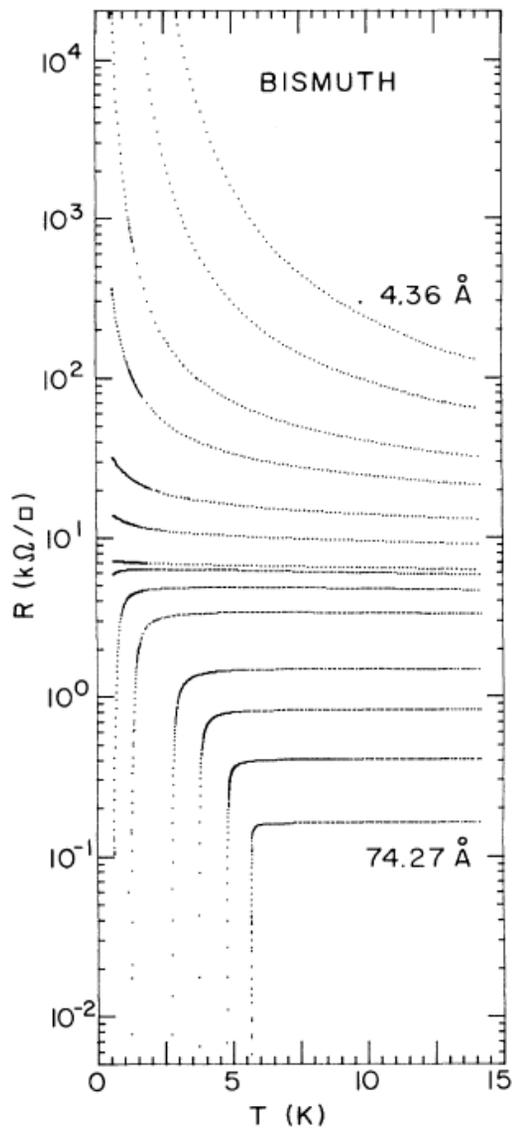

Figure 2

(a) 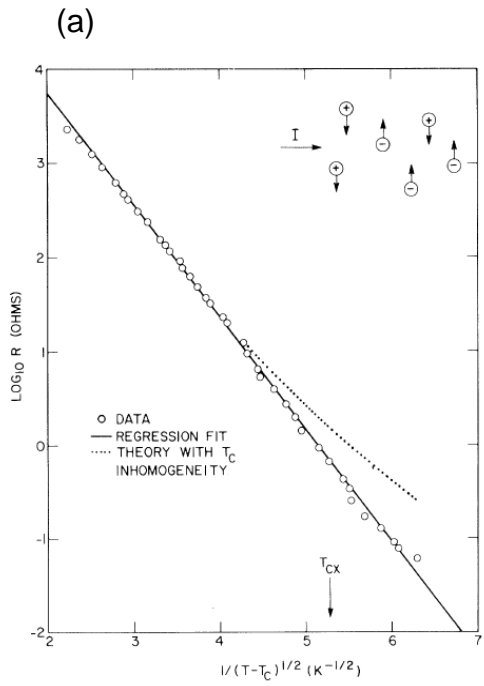
(b) 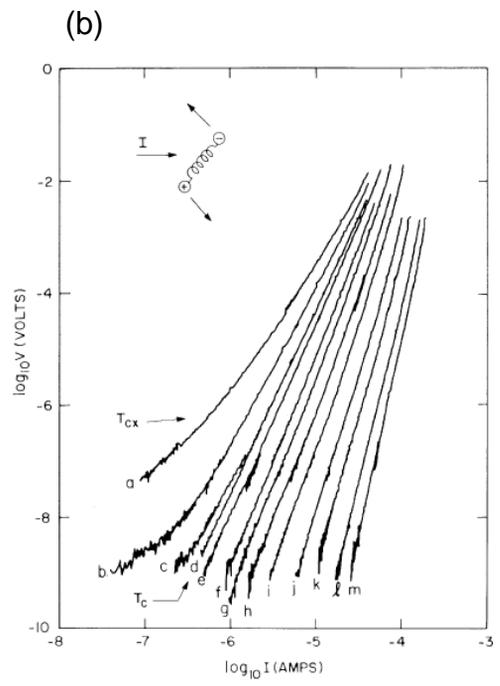

Figure 3

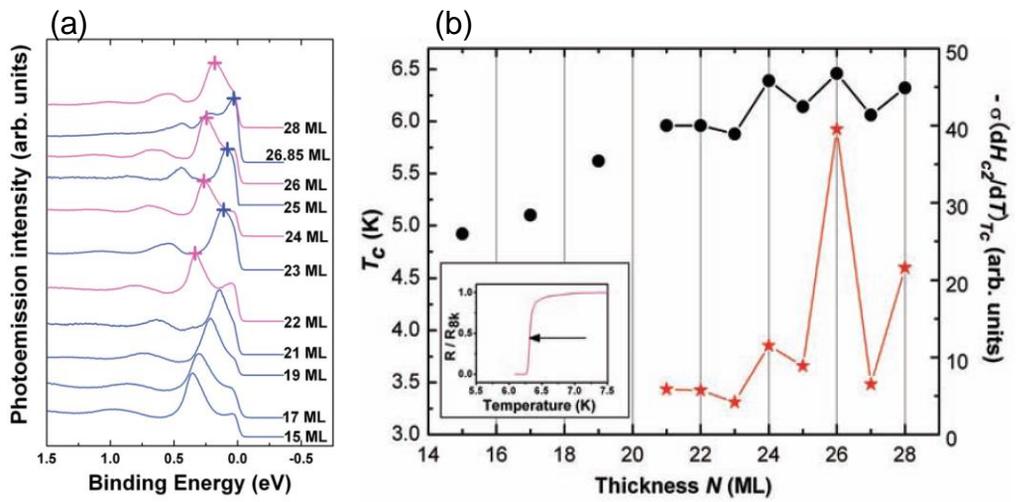

Figure 4

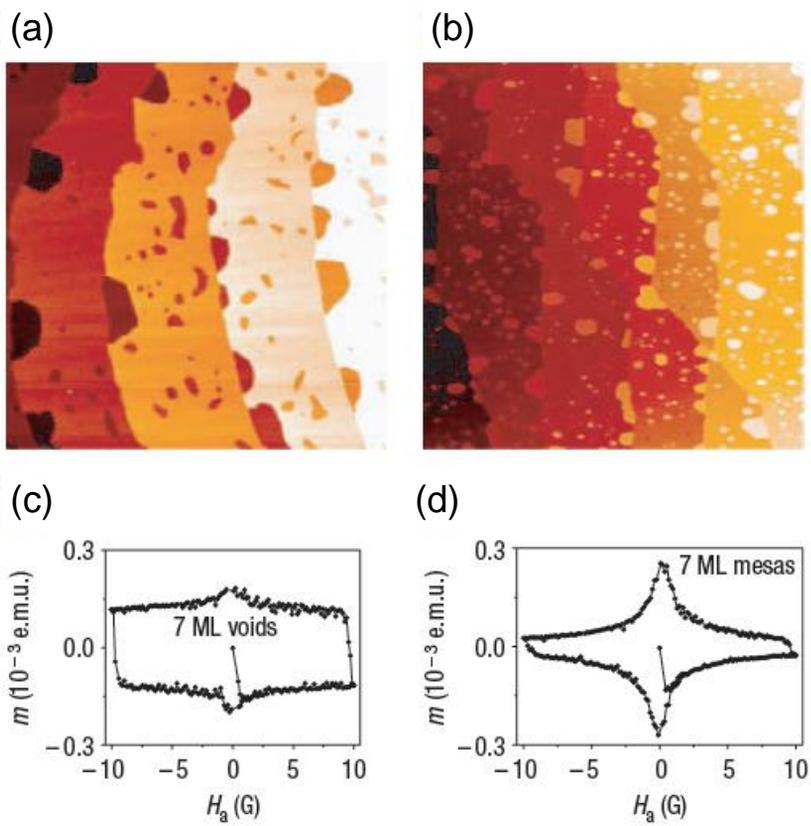

Figure 5

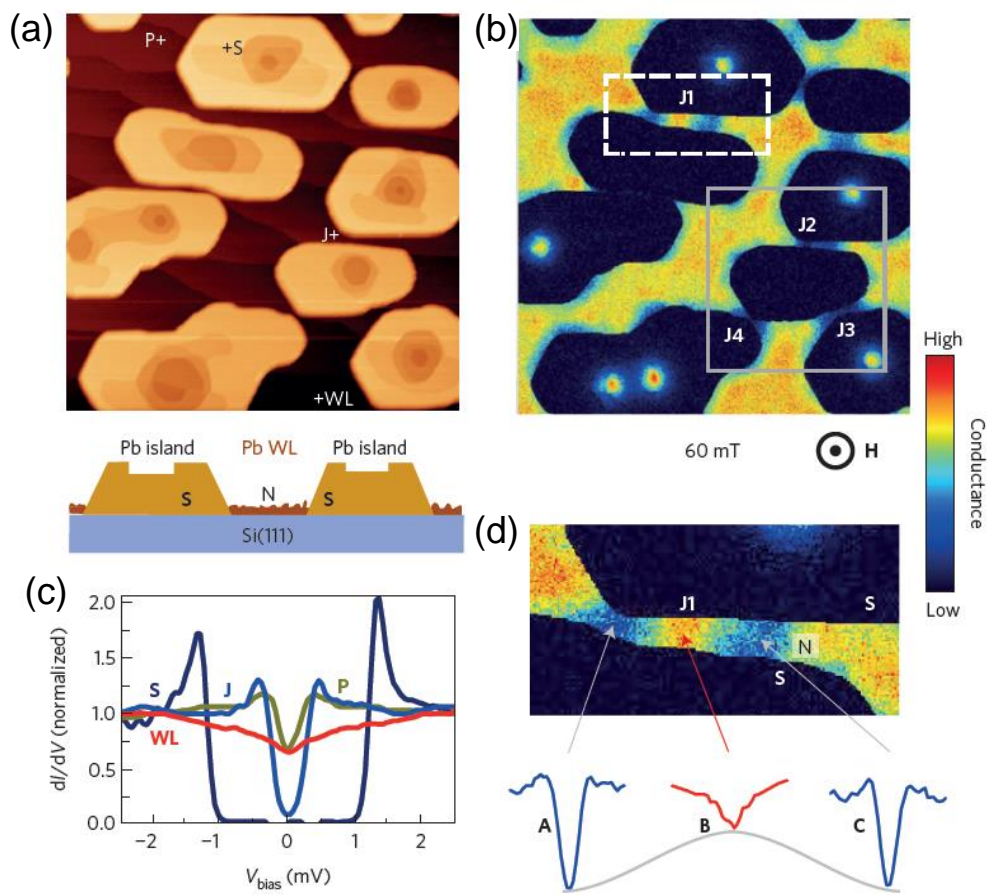

Figure 6

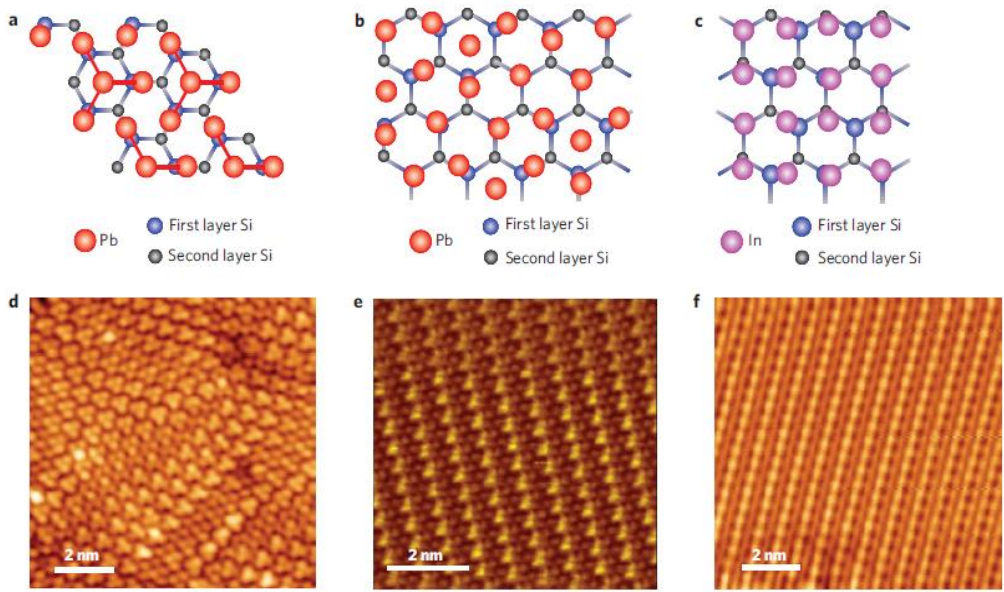

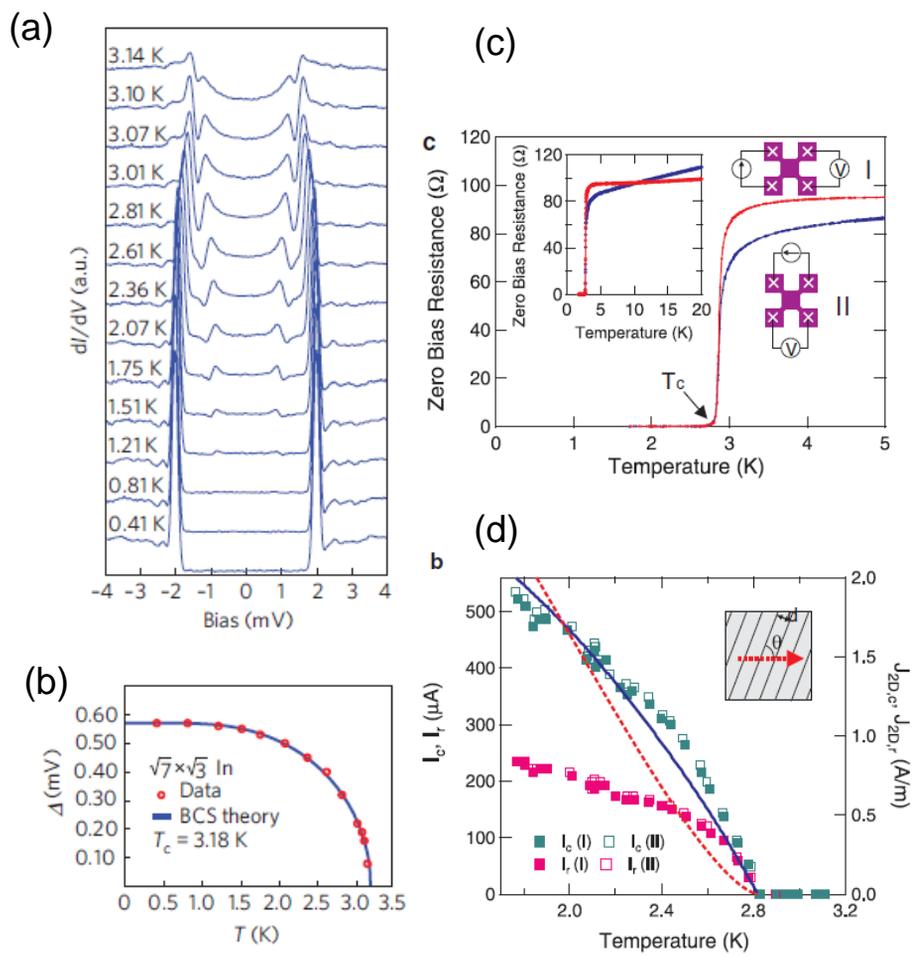

Figure 8

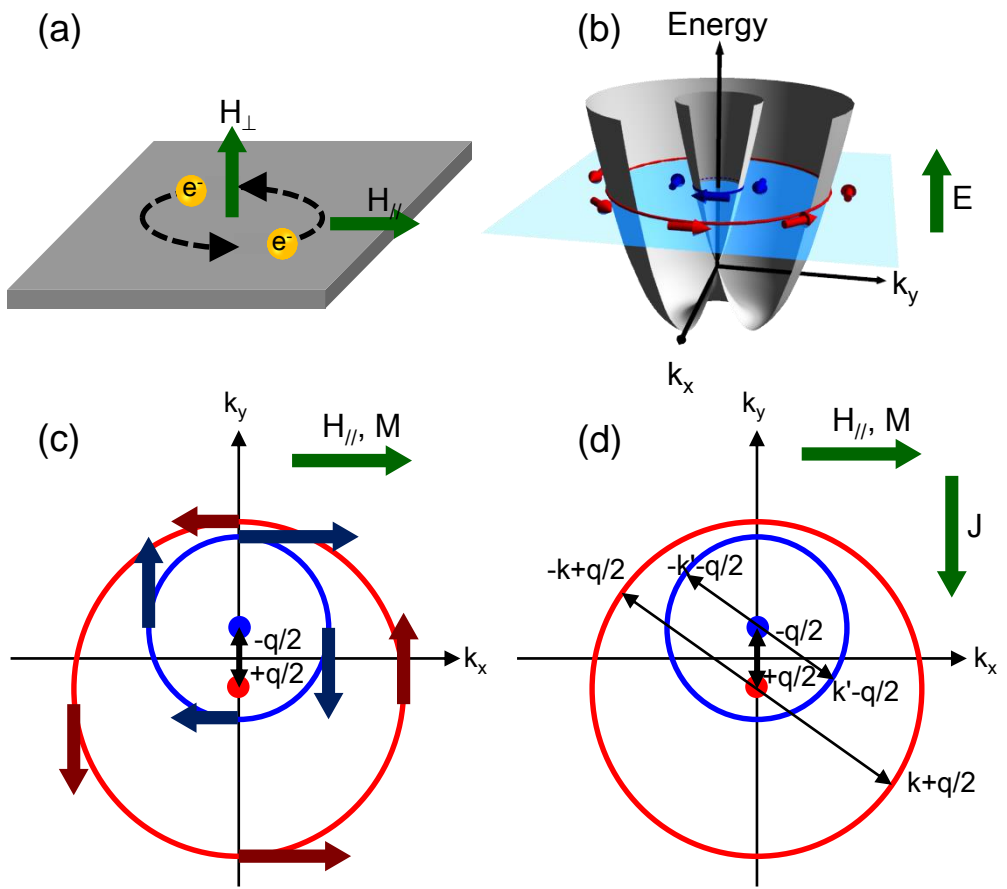

Figure 9

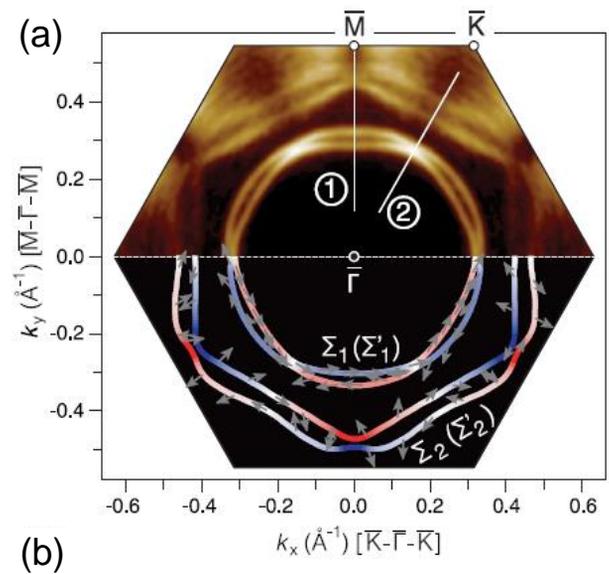

Figure 10

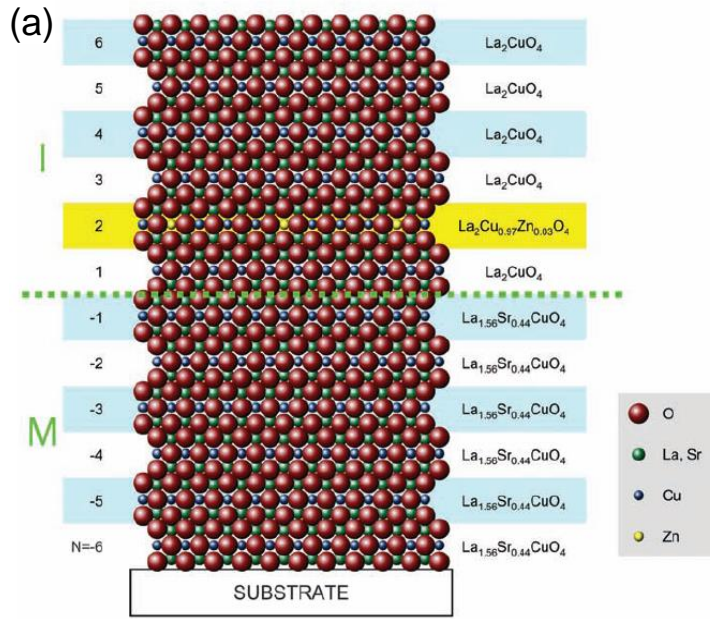

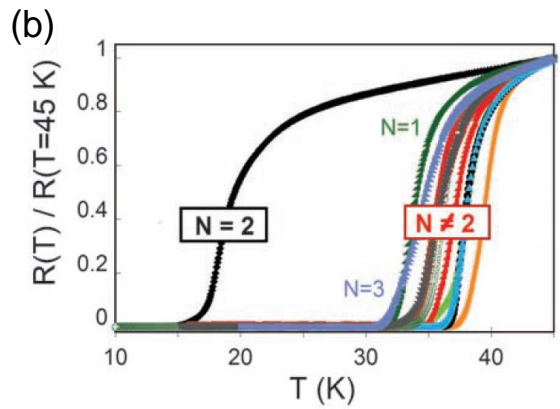

Figure 11

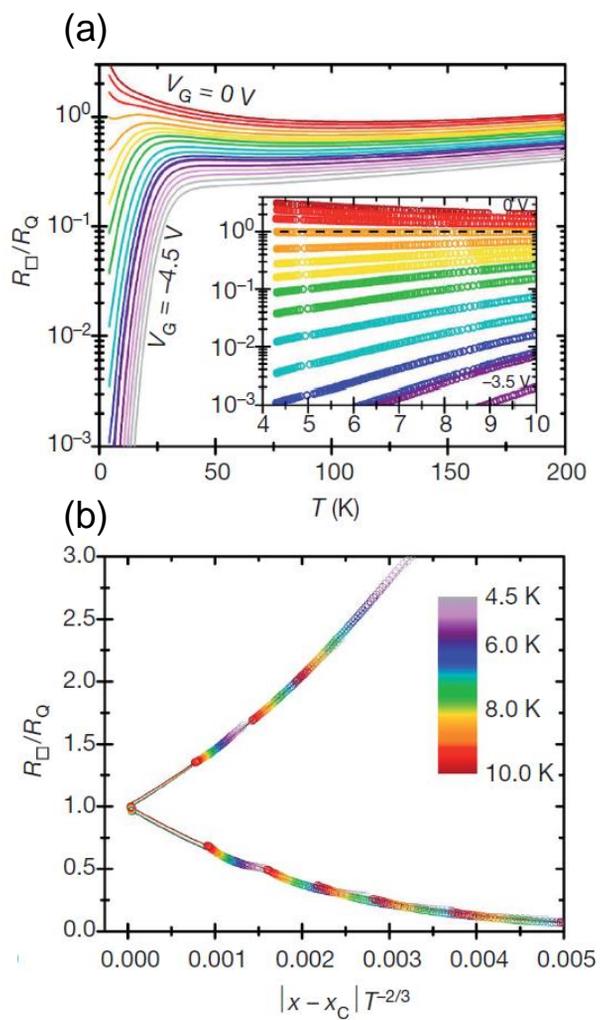

Figure 12

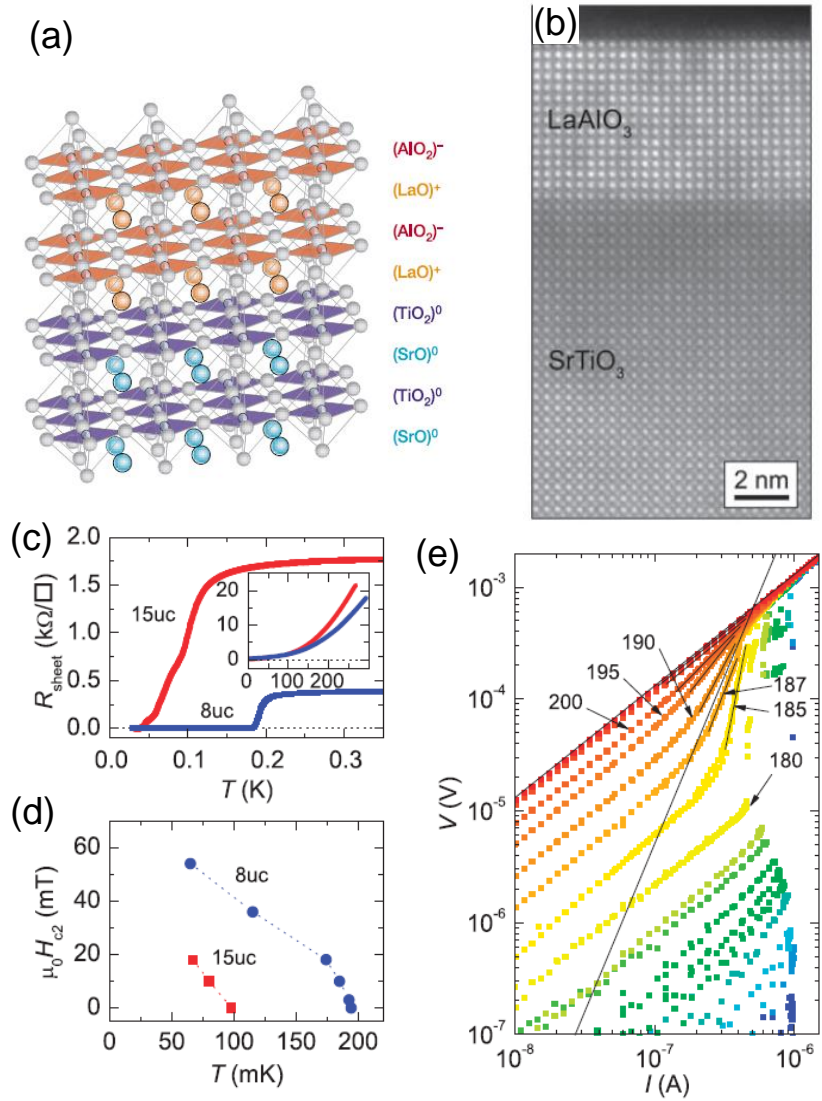

Figure 13

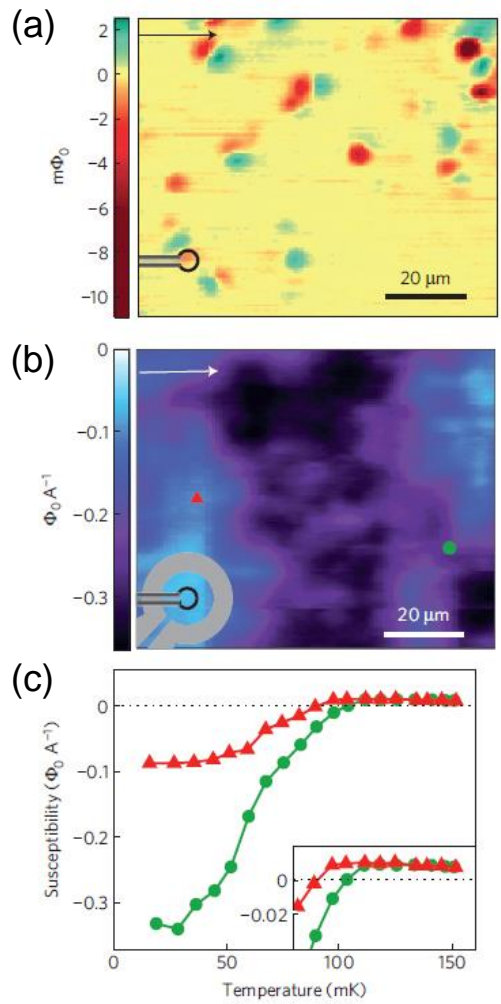

Figure 14

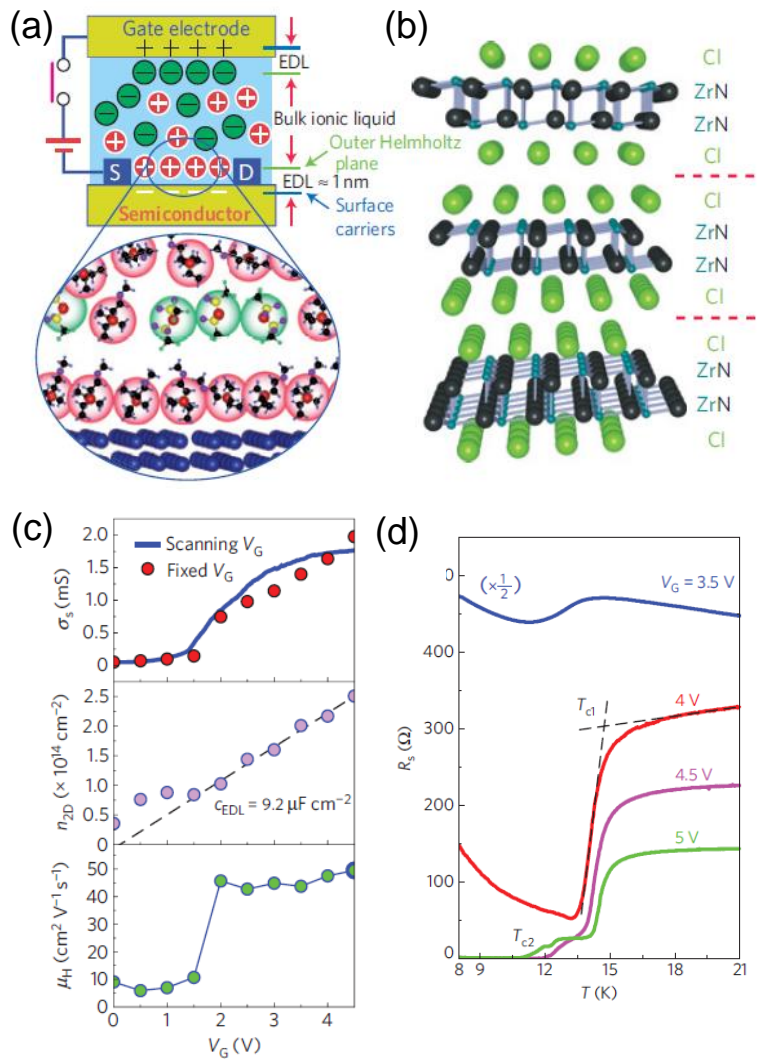

Figure 15

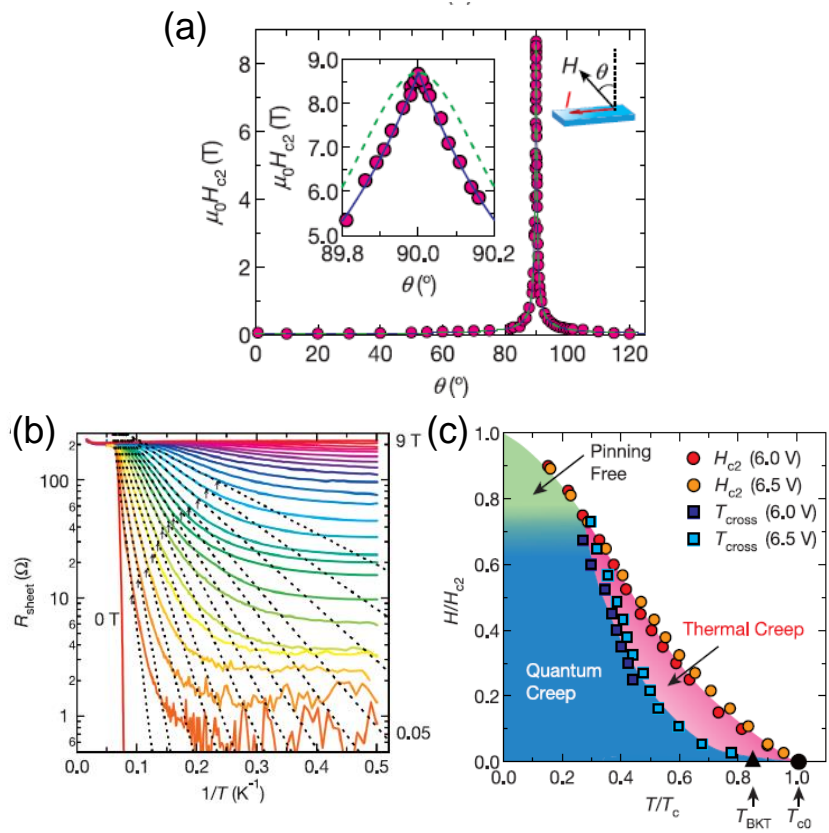

Figure 16

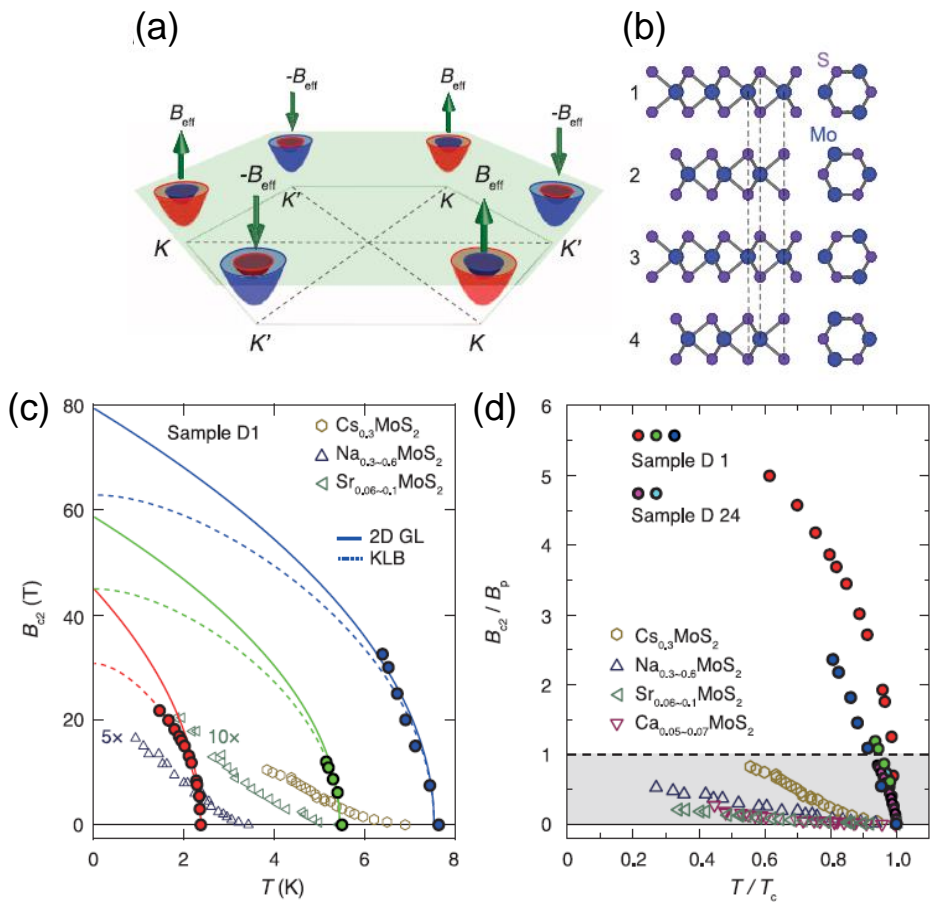

Figure 17

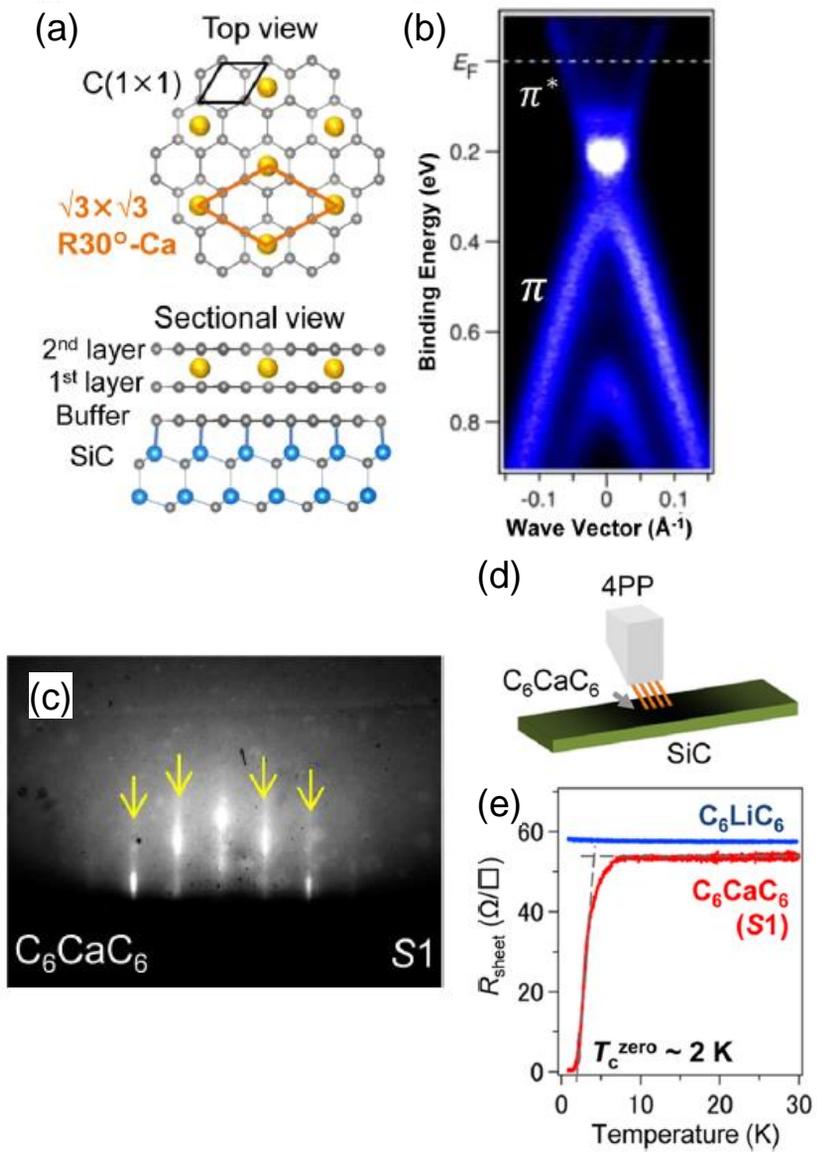

Figure 18

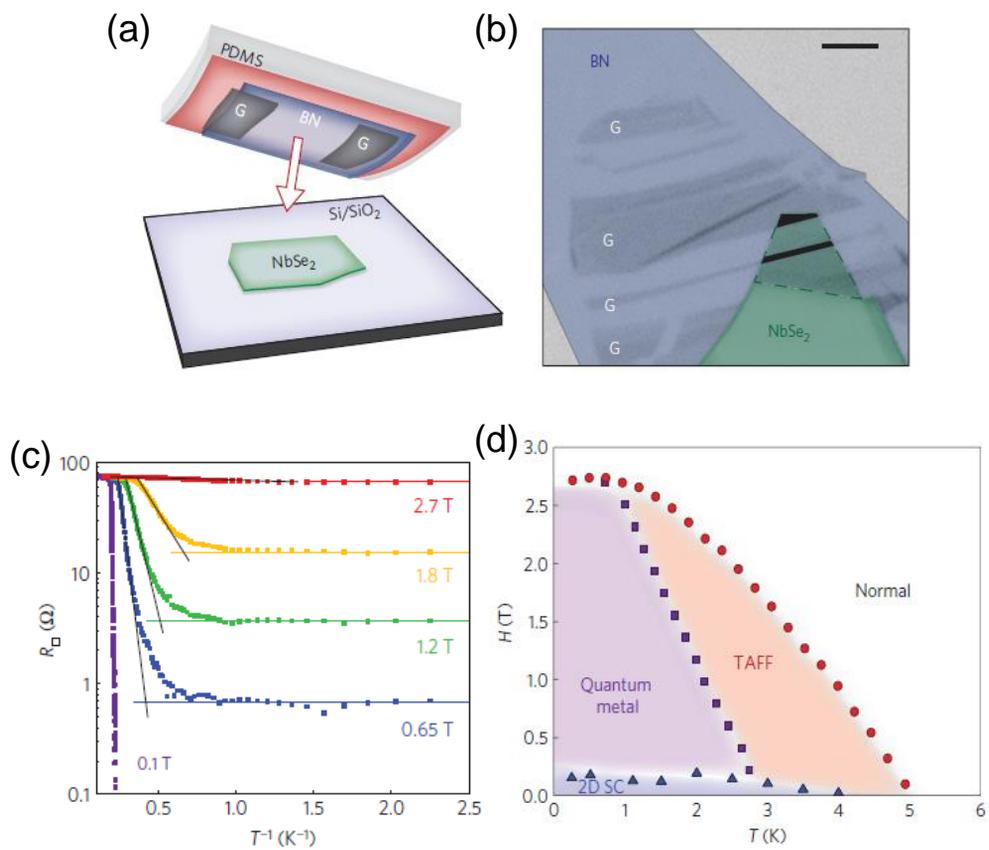

Figure 19

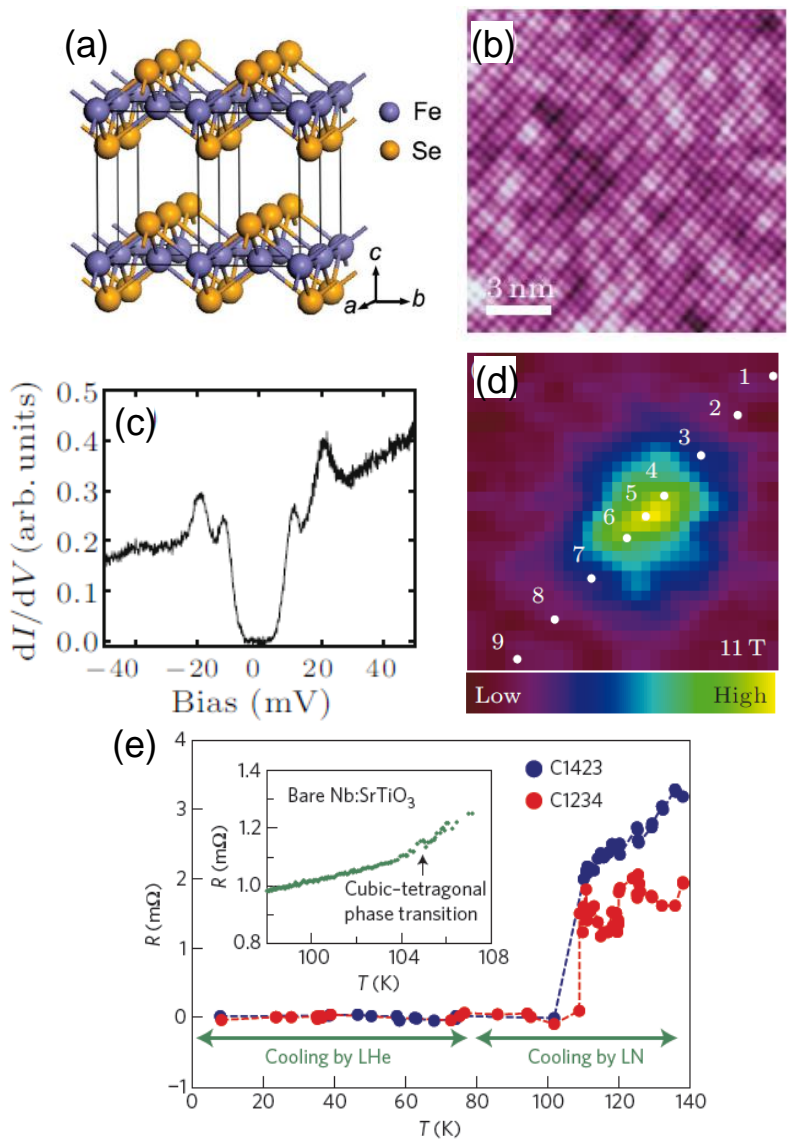

Figure 20

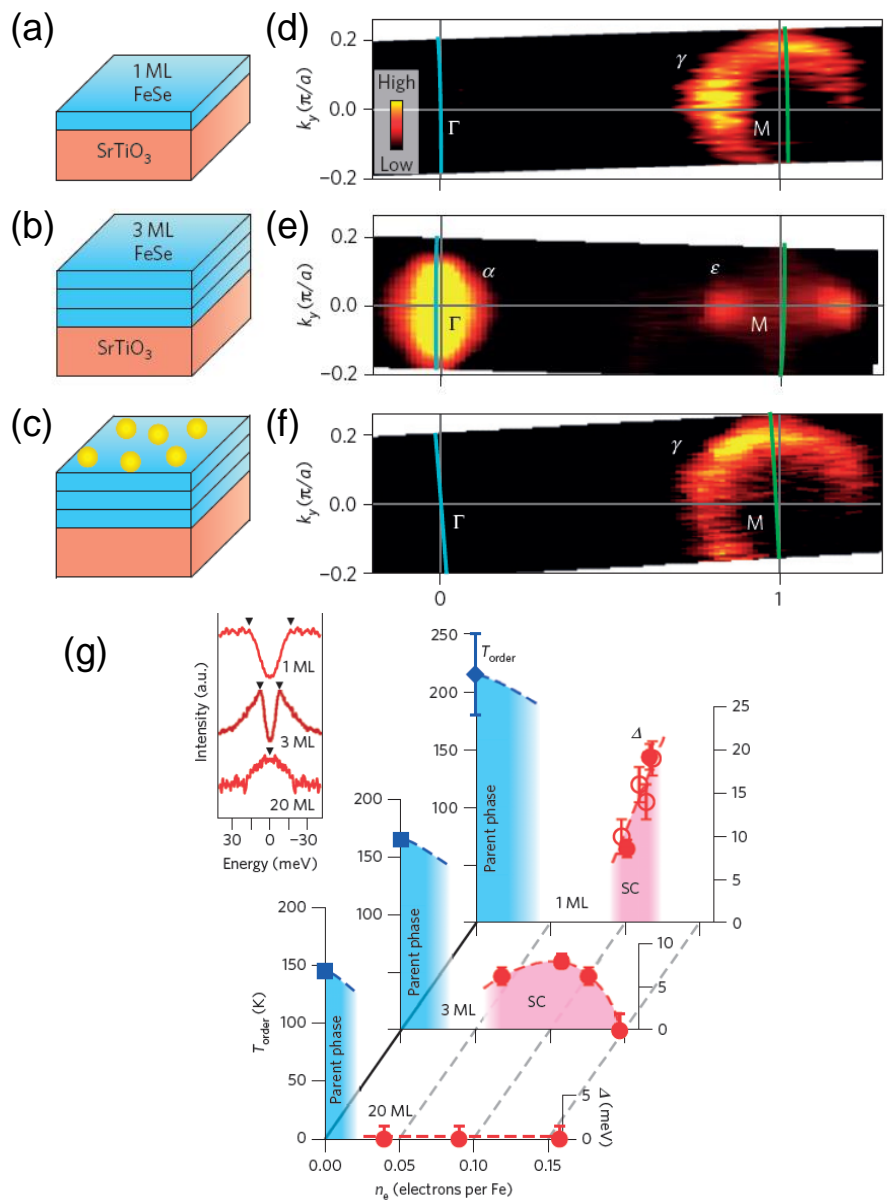

Figure 21

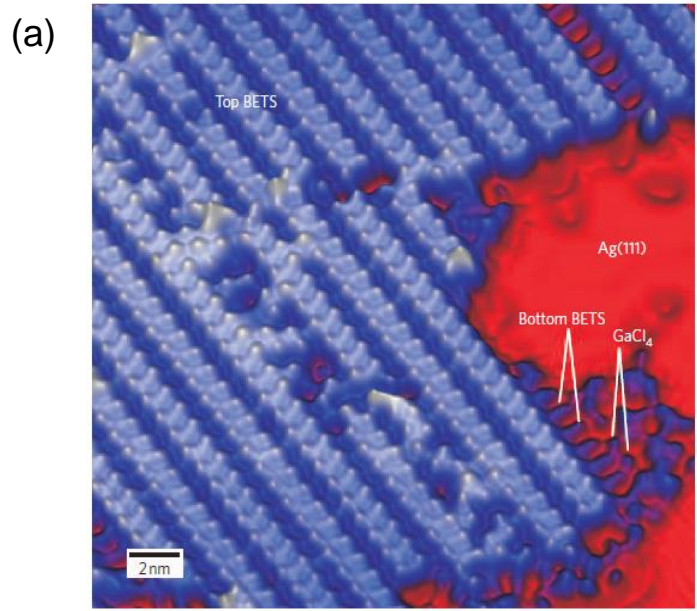
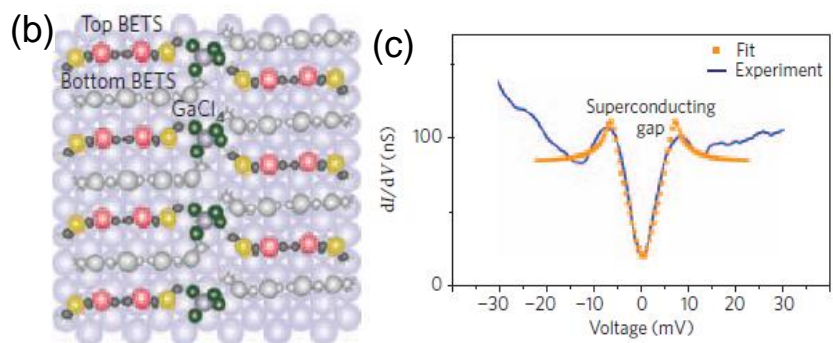

Figure 22

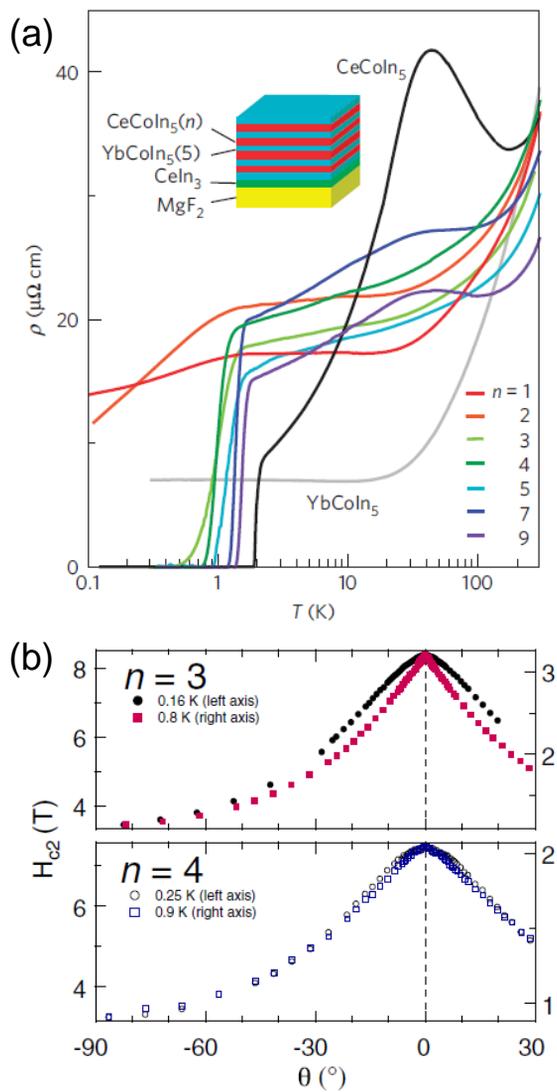

Figure 23

| 2D superconductor | material type | form | $d$ | $T_c$ | bulk $T_c$ | $\xi_{GL}(0)$ | $\mu_0 H_{c2\perp}(0)$ | $\mu_0 H_{c2//}(0)$ | Ref. |
|---|---|---|---|---|---|---|---|---|---|
| Pb/Si(111) | elemental metal | surface ultrathin film | 2 ML (+ 1 ML buffer) (0.57 nm) | 4.9 K ($T_c^{gap}$) | 7.2 K | 23 nm (9 ML) | 0.62 T (9 ML) | — | [78] [17] |
| Si(111)-($\sqrt{7}\times\sqrt{3}$)-In (Rect-phase) | elemental metal | surface atomic layer | 2 ML (0.495 nm) | 3.18 K ($T_c^{gap}$) | 3.4 K | 25 nm 47 nm | 0.49 T | — | [18] [84] [88] |
| Pb/GaAs(110) | elemental metal | surface atomic layer | 1 ML (0.29 nm) | 0.9 K ($T_c^{mid}$) | 7.2 K | 23 nm | 0.6 T | >> 15 T | [101] |
| La$_{2-x}$Sr$_x$CuO$_4$/ La$_2$CuO$_4$ | cuprate | heteroepitaxial interface | 1 UC (1.3 nm) | 40 K ($T_c^{mid}$, max) | 40 K | ~2 nm ($\xi_{GL\perp}(0)$, bulk) | — | — | [20] |
| LaAlO$_3$/SrTiO$_3$ | perovskite oxide | heteroepitaxial interface | 4~11 nm | 0.22 K | 0.3 K (max) | 70 nm | 0.1 T | 2 T | [21] [126] [127] |
| ZrNCl | nitride compound | FET interface | 0.65 UC (1.8 nm) | 14.8 K ($T_c^{mid}$, max) | 15.2 K (max) | 12.8 nm | 1.7 T | 50 T | [142] |
| MoS$_2$ | transition-metal dichalcogenide | FET interface | 0.49 UC (0.6 nm) | 10.8 K ($T_c^{mid}$, max) | 7 K (max) | 8.0 nm | 4 T | 52 T | [146] [147] |
| C$_6$CaC$_6$ | intercalated graphene | atomic sheet | 2 ML (+1 ML Ca) (0.68 nm) | 3 K ($T_c^{mid}$) | 11.5 K | 49 nm | 0.14 T | — | [160] |
| NbSe$_2$ | transition-metal dichalcogenide | atomic sheet | 1/2 UC (0.63 nm) | 3.1 K ($T_c^{mid}$) | 7.2 K | 7.8 nm (bulk) | 2 T | 35 T | [27] |
| FeSe/SrTrO$_3$ | iron chalcogenide | surface atomic layer | 1 UC (0.55 nm) | 32 K 109 K ($T_c^{mid}$) | 8 K | 4.5 nm (bulk) | > 52 T 116±12 T | — | [53] [29] |
| $\lambda$-(BETS)$_2$GaCl$_4$ | organic conductor | surface molecular layer | 1 UC | 9~10 K ($T_c^{gap}$) | 4.7 K | 1.6, 12.5 nm ($\xi_{GL\perp}(0)$, bulk) | — | — | [30] |
| CeCoIn$_5$/ YbCoIn$_5$ | rare-earth based compound | heteroepitaxial superlattice | 3 UC (2.27 nm) | 1 K ($T_c^{mid}$) | 2.3 K | 7.4 nm | 3.8 T | 9.0 T | [31] |

Table 1